\newif\ifmerci
\mercitrue

\documentclass[a4paper,twocolumn,eqsecnum,floatfix,preprintnumbers,aps,rmp,nofootinbib,amsmath,amssymb,amsfonts,superscriptaddress,twoside]{revtex4}

\usepackage[
top    = 2.80cm,
bottom = 2.80cm,
left   = 1.65cm,
right  = 1.65cm]{geometry}

\usepackage[dvipsnames]{xcolor}
\definecolor{airforceblue}{rgb}{0.12, 0.21,0.5}
\usepackage[colorlinks=true,linktocpage=true,linkcolor=airforceblue,citecolor=airforceblue,breaklinks=true,urlcolor=airforceblue]{hyperref}
\usepackage[utf8]{inputenc} 
\usepackage[T1]{fontenc}
\usepackage{bbold}

\usepackage{graphicx}

\usepackage{dsfont}
\usepackage{bm}
\usepackage{slashed}
\usepackage{subeqnarray}

\newcommand{\be}{\begin{equation}}
\newcommand{\ee}{\end{equation}}
\newcommand{\bea}{\begin{eqnarray}}
\newcommand{\eea}{\end{eqnarray}}
\newcommand{\beas}{\begin{subeqnarray}}
\newcommand{\eeas}{\end{subeqnarray}}

\newcommand{\stf}[1]{{\langle #1 \rangle}}
\newcommand{\dd}{{\rm d}}
\newcommand{\gr}[1]{{\bm #1}}
\newcommand{\nbb}{{\cal B}}
\newcommand{\cT}{{\cal T}}
\newcommand{\polar}{{\cal P}}
\newcommand{\mystar}[3]{{{\cal S}_{#1}[#2,#3]}}

 \newcommand{\coeffa}{\alpha}
 \newcommand{\Gap}{\Delta}
\newcommand{\Ee}{{\cal E}}
\newcommand{\dloge}{d_E}
\newcommand{\myk}{K}

\newcommand{\troisj}[6]{\left(\begin{array}{ccc}
      #1 & #2 & #3 \\
      #4 & #5 & #6\end{array}\right)}
\newcommand{\com}[1]{}

\newcommand{\myA}{{\cal R}}
\newcommand{\myB}{{\cal S}}
\newcommand{\myD}{{\cal U}}
\newcommand{\myZ}{{\cal Z}}
\newcommand{\ens}{{\delta_E}}

\def\signofmetric{1}
\newcommand{\hatM}{{\lambda}}

\newcommand{\vol}{{\cal V}}

\newcommand{\ii}{{\rm i}}

\newcommand{\sT}{{}\sigma_{\rm T}}

\newcommand{\deltarel}{\underline{\delta}}
\newcommand{\antipar}[1]{{\bar{#1}}}

\if0\signofmetric
\def\sgnzz{+}

\def\sgnii{-}

\def\sgnslash{}
\def\sgnminusslash{-}
\fi

\if1\signofmetric
\def\sgnzz{-}

\def\sgnii{+}

\def\sgnslash{-}
\def\sgnminusslash{}
\fi

\makeatletter
\def\l@subsubsection#1#2{}
\makeatother

\newcommand{\mypart}[2]{
    \part*{#2}
    \addcontentsline{toc}{part}{{\Large #2}}
}

\begin{document}

\title{Radiative transport of relativistic species in cosmology}

\author{Cyril Pitrou}
\email{pitrou@iap.fr}
\affiliation{Institut d'Astrophysique de Paris, CNRS UMR 7095,\\
Institut Lagrange de Paris, 98 bis Bd Arago 75014 Paris, France}

\date{\today}

%%%%%%%%%%%%%%%%%%%%%%%%%%%%%%%%%%%%%%%%%%%%%%%%%%%%%%%%%%%%%%%%%%%%%%%

%\thispagestyle{empty}

\begin{abstract}
  We review the general construction of distribution functions for gases
of fermions and bosons (photons), emphasizing the similarities and
differences between both cases. The central object which describes polarization for photons is a
tensor-valued distribution function, whereas for fermions it is a
vector-valued one. The collision terms of Boltzmann equations for
fermions and bosons also possess the same general structure and differ only in the quantum effects associated
with the final state of the reactions described. In particular, neutron-proton conversions in the early
universe, which set the primordial Helium abundance, enjoy many similarities with Compton scattering which
shapes the cosmic microwave background and we show that both can be
handled with a Fokker-Planck type expansion. For neutron-proton
conversions, this allows to obtain the finite nucleon mass corrections, required for precise theoretical predictions, whereas for Compton scattering it leads to the thermal
and recoil effects which enter the Kompaneets equation. We generalize the latter to the general case of anisotropic and polarized photon
distribution functions. Finally we discuss a parameterization of the
photon spectrum based on logarithmic moments which allows for a neat
separation between temperature shifts and spectral distortions.
\end{abstract}
%\eng

\maketitle

%%%%%%%%%%%%%%%%%%%%%%%%%%%%%%%%%%%%%%%%%%%%%%%%%%%%%%%%%%%%%%%%%%%%%%%
%\clearpage

\tableofcontents

\part*{Introduction}
\addcontentsline{toc}{section}{{\large Introduction}}

We review theoretical and practical aspects of the radiative transport
of relativistic species. Our emphasis is on cosmological applications, hence we focus on fermions (neutrons, protons, electrons, positrons and neutrinos) during
big-bang nucleosynthesis (BBN) and photons of the cosmic microwave
background (CMB). Relativistic species cannot be described by perfect
fluids and one must account for the distribution of particles using a distribution function $f(\gr{x},\gr{p},t)$, whose evolution is dictated by a Boltzmann
equation $L[f] = C[f]$. The left hand side is the Liouville term and
describes the free streaming of particles. In a curved space-time,
this requires the use of cosmological perturbation theory, that is general relativity. The
emphasis of this article is on the right hand side which is the
collision term, and describes the evolution of the distributions under
the influence of collisions, that is because of the micro-physics. 
Hence all the results presented here are independent of any
perturbation theory, as they are derived from the basic principles of
particle physics. From the equivalence principle, they are formulated
in a local orthonormal frame, that is in the context of special
relativity.

It is instructive to consider the cases of fermions and bosons side by
side as their description by distribution functions have numerous
similarities. In fact, the case of massless fermions is simpler than
the case of photons in many respects, essentially because the spin of
fermions ($1/2$) is smaller than the spin of photons ($1$). Hence the
paper is organized to allow for a detailed comparison of these two
cases. We show that the collision term for weak interactions during
BBN has a structure which is extremely similar to the structure of the
collision term for photons due to Compton scattering. Furthermore, we
can use common techniques to express in practice these collision terms in functions of the
distribution functions moments. Even though one could follow the
analogy for anisotropic distribution functions, it is not useful for
the case of fermions in the context of BBN. Hence we study in details
the angular structure only for Compton scattering and derive the {\it extended Kompaneets} equation, valid for anisotropic
and polarized photon distributions. Since the emphasis is on the
derivation and the structure of the equations, we only summarize how
the equations must be applied in the cosmological context, overviewing
briefly the main physical effects.

In \S~\ref{SecDistributionFunction} a general procedure to build a classical distribution
function out of the quantum number operator is summarized. We then
detail in \S~\ref{SecBoltzmann} how a classical Boltzmann equation can be derived, given a set
of suitable approximation and assumptions, from the quantum evolution
of the number operator. Sections~\ref{SecWeakInteraction} is then dedicated
to the collision terms of weak interactions  processes for fermions in
the early universe. In order to compute in
practice these collision terms, we review the Fokker-Planck expansion
in \S~\ref{SecFokkerPlanck} applied to BBN weak interactions. The similar treatment of Compton interactions between
electrons and photons and its Fokker-Planck expansion are reviewed in
\S~\ref{SecFokkerCompton}.  Applications for the evolution of isotropic photon distributions under Compton interactions are presented in
\S~\ref{SecIsotropicCMB}, with a brief discussion on its implications
for cosmology. The detailed form  of the collision term for anisotropic distributions, {\it
including polarization} is subsequently exposed in
\S~\ref{SecAnisotropicCMB}, using symmetric trace-free tensors to decompose the angular
dependence. It is the first general derivation of the Compton
collision term in the literature which includes thermal and recoil
effects while describing consistently polarization. Finally we present in
\S~\ref{SecSpectralDistortions} a parameterization for spectral
distortions and we collect in \S~\ref{SecDynStebbins} the equations governing the
generation of distortions from the Thomson part of the collision
term, with plots of the associated angular power spectrum generated during the reionization
era.

%%%%%%%%%%%%%%%%%%%%%%%%%%%%%%%%%%%%%
\mypart{1}{Theoretical framework}
%%%%%%%%%%%%%%%%%%%%%%%%%%%%%%%%%%%%%

\section{Distribution functions}\label{SecDistributionFunction}

In this section, based on \citet{FidlerPitrou}, we review how the distribution function is built
from the quantum expectations of the number operator, and how its
covariant components can be extracted. We also show
that for each spin there is an adapted expansion in spin-weighted
spherical harmonics for the dependence on the spatial momentum direction. The case of fermions is presented first, even
though it is less known, as it allows to understand better the photon case. 

\subsection{General construction}

\subsubsection{Notation}

Before considering the kinetic theory in curved spacetime, we build the formalism in a flat space-time (that is the Minkowski
space-time of special relativity) in which the quantum theory of
particles is very well established. An inertial frame is defined by a tetrad field, that is by a timelike vector field $e_0$ and three
spacelike vector fields $e_i$, together with the associated co-tetrad
$e^0,e^i$. Latin indices such as $i,j,\dots$ indicate
spatial components in the tetrad basis.  A four-vector is written as $V^\mu$ where Greek indices $\mu,\nu,\dots$ denote components in the tetrad basis. In particular, the components of the tetrad vectors and co-vectors in the tetrad basis are by definition $[e_\mu]^\nu = \delta_\mu^\nu$ and $[e^\mu]_\nu = \delta^\mu_\nu$. If gravity can be ignored, that is in the context of special relativity, the inertial frame is global. Later, when including the effect of gravity in the context of general relativity, the inertial frame is local and one must employ general coordinates whose indices are labelled by $\alpha,\beta,\dots$. For a given vector this implies $V^\alpha = V^\mu [e_\mu]^\alpha$. 

The momentum vector $p^\mu$ will often simply be denoted as $p$ and its spatial components $p^i$ allow to
build the spatial momentum $\gr{p} = p^i {\bf e}_i$. More generally, we
reserve boldface notation to spatial vectors. The energy associated
with the momentum is given by the time component 
\be
E = p^0\,,\quad \Rightarrow\quad E^2=m^2+|\gr{p}|^2\,.
\ee 
When a quantity depends on the spatial momentum, we use indifferently $\gr{p}$ or $p$ when no ambiguity can arise. The (special) relativistic (and Lorentz covariant) integration measure is defined as
\be
[\dd p] \equiv \frac{\dd^3 \gr{p}}{(2\pi)^3 2 p^0}\,,
\ee
and its associated (special) relativistic Dirac function is defined accordingly as
\be
\deltarel(p-p') = (2\pi)^3 2 p^0 \delta^3(\gr{p}-\gr{p}') \,,
\ee
such that $\int [\dd p] \deltarel(p-p') =1$. Our metric convention follows the standard notation employed in
cosmology, which is the opposite of the metric commonly used in
particle physics. In the tetrad basis, the metric $g_{\mu\nu}$ reduces to the Minkowski metric
\be
\eta = {\rm diag}(\sgnzz,\sgnii,\sgnii,\sgnii)\,.
\ee

The Levi-Civita tensor is fully antisymmetric and in the tetrad basis
all its components are deduced from the choice
\be
\epsilon_{0123} = -\epsilon^{0123} = 1\,.
\ee
We identify the time-like vector of a tetrad $e_0$ with the velocity
$u$ of an observer and its spatial Levi-Civita tensor is obtained from $\epsilon_{ijk}\equiv u^\mu\epsilon_{\mu ijk} $, such that $\epsilon_{123}= 1$.

\subsubsection{Number operator}

Creation and annihilation operators, $a^\dagger_r(\gr{p})$ and $a_r(\gr{p})$ respectively, where the index $r$ refers to a helicity basis and $\gr{p}$ to the particle momentum, are defined for each particle type from its corresponding quantum field. It allows to define a quantum number operator as
\be\label{defNrs}
N_{rs}(\gr{p},\gr{p}') \equiv a_r^\dagger(\gr{p}) a_s(\gr{p}')\,.
\ee
The total occupation operator is then obtained from a sum over all possible
momenta of the diagonal part as
\be
N_{rs} \equiv \int [\dd p] N_{rs}(p,p)\,.
\ee
When considering a given quantum state $| \Psi \rangle$, the average
of the number operators allows to define a distribution function with
helicity indices $f_{rs}$ as

\be\label{AverageNrs}
\langle \Psi |N_{rs}(\gr{p},\gr{p}')|\Psi \rangle =   \deltarel(p-p') f_{rs} (\gr{p})\,.
\ee
Hence the total number of particles is given by
\be
\langle \Psi | N_{rs}| \Psi \rangle  = \vol \int \frac{\dd^3 \gr{p}}{(2\pi)^3} f_{rs}(p)\,,
\ee
where we introduced the total volume $ (2\pi)^3 \delta^3(0)=
\vol$. In this expression, $f_{rs}(p)$ corresponds exactly to the definition of a classical
one-particle distribution function. By construction $N_{rs}$ and $f_{rs}$ are Hermitian, that is
\be\label{HermitianProperty}
N^\star_{rs}(p,p') =N_{sr}(p',p) \quad \Rightarrow \quad f^\star_{rs}(p) =f_{sr}(p)\,.
\ee
So far we have not specialized to particles nor antiparticles, not even
to a special spin type (fermions or bosons), and this construction is
very general. In the next two sections we study separately fermions
and bosons, and we show how the distribution function with helicity
indices ($f_{rs}(\gr{p})$) can be decomposed into covariant components.

\subsubsection{Adapted orthonormal basis}\label{SecAdaptedBasis}

\begin{figure}[htb!]
    \includegraphics[width=0.495\columnwidth,angle=0]{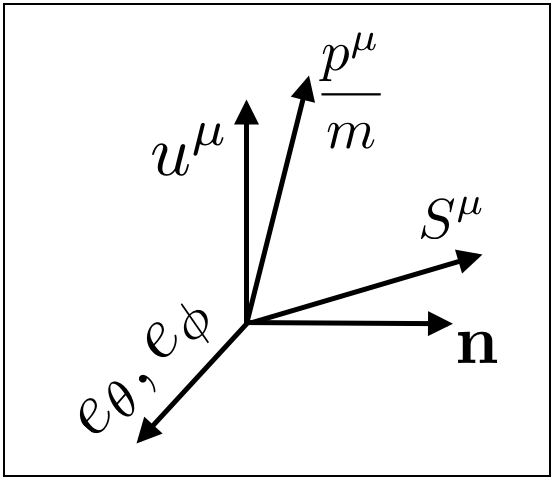}
\includegraphics[width=0.49\columnwidth,angle=0]{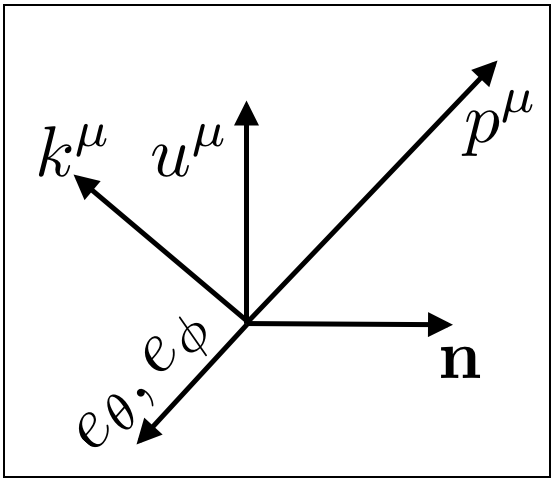}
     \caption{Left: In the massive case, $p^\mu/m$ and $e_\theta,e_\phi,S^\mu$ form an
       orthonormal basis. Right: In the massless case, $k^\mu,p^\mu$
       are null vectors orthogonal to $e_\theta,e_\phi$ such that
       $k^\mu p_\mu = \sgnslash 1$. In both cases, the polarization
       basis is formed by ${\epsilon}^\mu_{\pm} \equiv (e^\mu_\theta \mp \ii
e^\mu_\phi)/\sqrt{2}$.}
\label{fig1} 
  \end{figure}

For a given observer with four-velocity $u^\alpha$ which is chosen to
be aligned with the time-like tetrad vector $e_0$, we define the unit
spatial vector of momentum direction by 
\be\label{Defni}
\gr{n}\equiv \frac{\gr{p}}{|\gr{p}|}\,.
\ee
In spherical coordinates the momentum direction is given by $\theta,\phi$ and defines a radial unit vector. We then also consider
the usual basis in spherical coordinates $\gr{e}_\theta$ and
$\gr{e}_\phi$, which are purely spatial unit vectors. In tetrad components
these are given by
\bea\label{Explicitsphericalbasis}
n^i&=& \left( \begin{array}{c}\cos \phi \sin \theta\\\sin \phi \sin
                   \theta\\ \cos \theta \end{array} \right)\,,\\
               e_\theta^i &=&\left( \begin{array}{c}\cos \phi \cos \theta\\\sin \phi \cos \theta\\ -\sin \theta \end{array} \right) \,,\qquad
               e_\phi^i =\left( \begin{array}{c}-\sin \phi\\\cos \phi\\ 0\end{array} \right)\,.\nonumber
\eea
Let us introduce the helicity vector
\be\label{defSmu}
S^\mu(u^\nu,p^\nu) = -\frac{m}{|\gr{p}|} u^\mu +
\frac{E}{m|\gr{p}|}p^\mu\,, 
\ee
which is a unit vector in the direction of the spatial
momentum that is transverse to $p^\mu$ in the sense $S^\mu p_\mu =0$, and is thus
spacelike. Since the space of vectors orthogonal to $p^\mu$ is three-dimensional, the transverse property is
not enough to specify the helicity vector and the definition
(\ref{defSmu}) depends explicitly on the observer which is used to define the spatial part of the
momentum. When no ambiguity can arise we write simply $S^\mu$. In
components the helicity vector is given by
\be\label{DefS}
S^0 = m^{-1}|\gr{p}|\,,\qquad S^i = m^{-1} E n^i\,.
\ee
Geometrically (see Fig.~\ref{fig1}), the helicity vector corresponds to the spatial direction unit vector
${\gr{n}}$ boosted in its direction by the same boost needed to obtain $p^\mu/m$ from $u^\mu$. 

Finally, we define the polarization basis $\gr{\epsilon}_{\pm}(u,p)$
\bea\label{Explicitpolarbasis}
&&{\epsilon}^\mu_{\pm} \equiv\frac{1}{\sqrt{2}}(e^\mu_\theta \mp \ii
e^\mu_\phi)\,,\\
&&{\epsilon}^0_{\pm} =0\,,\quad {\epsilon}^i_{\pm}=\frac{1}{\sqrt{2}}\left( \begin{array}{c}
\text{cos}(\theta)\text{cos}(\phi) \pm i
\text{sin}(\phi)\\
\text{cos}(\theta)\text{sin}(\phi) \mp i
\text{cos}(\phi)\\
-\text{sin}(\theta) \end{array} \right)\,,\nonumber
\eea
where again the dependence on $(u,p)$ can be omitted whenever no ambiguity can arise.

The set of vectors $S^\mu,\epsilon_\pm^\mu$, and $p^\mu/m$ constitute
an adapted orthonormal basis to a given observer and a given momentum.

\subsubsection{Fermions}

The quantum fermion field is

\be\label{DefQuantumField}
\psi = \sum_{s=\pm\tfrac{1}{2}} \int [\dd p]\left[ {\rm e}^{\sgnslash \ii p \cdot x }
\antipar{a}^\dagger_s(p) v_s(p)+ {\rm e}^{\sgnminusslash \ii p \cdot x }
a_s(p) u_s(p)\right]\,,\nonumber
\ee
and satisfies the Dirac equation $(\ii \gamma^\mu \partial_\mu + m) \psi = 0$.
In this expression, the creation and annihilation operators of the particles ($a_s,a^\dagger_s$) and antiparticles ($\antipar{a}_s,\antipar{a}^\dagger_s$) satisfy the anti-commutation rules 
\beas\label{AntiCommuteRule}
\{a_r(p), a^\dagger_s(p')\} &=& \delta^{\rm K}_{rs}
\deltarel(p-p')\,,\\
 \{\antipar{a}_r(p), \antipar{a}^\dagger_s(p')\}&=& \delta^{\rm K}_{rs} \deltarel(p-p')\,,
\eeas
with all other anti-commutators vanishing and where $\delta^{\rm K}$ is the Kronecker function. 

We then define the operator in spinor space (beware of position of helicity indices for antiparticles)
\be\label{DefFparticles}
F_{\mathfrak{a}}^{\,\,\,\mathfrak{b}}(p) \equiv \begin{cases} \sum \limits_{rs} 
f_{rs}(p) u_{s,\mathfrak{a}}(p)  \bar u^{\mathfrak{b}}_r(p)\quad {\rm
  part.},\\
\sum \limits_{rs} 
f_{rs}(p) v_{r,\mathfrak{a}}(p) \bar v^{\mathfrak{b}}_s(p) \quad {\rm
antipart.}
\end{cases}
\ee
For the sake of clarity we use a notation where components of
operators in spinor space and Dirac spinors are explicit and are
denoted by indices of the type $\mathfrak{a},\mathfrak{b},\dots$. The plane waves solutions $u_{s,\mathfrak{a}}$ and
$v_{s,\mathfrak{a}}$ are the positive  and negative frequency solutions and satisfy
\be\label{eq:particledef}
(\slashed{p} \sgnii m) u_s(p) = 0\,,\qquad (\slashed{p}  \sgnzz  m) v_s(p) = 0\,,
\ee
with the standard Dirac slashed notation $\slashed{p} \equiv p_\mu
\gamma^\mu$ and the Dirac matrices satisfying the algebra
$\{\gamma^\mu,\gamma^\nu\}=\sgnzz 2 g^{\mu\nu}$. As detailed in appendix~\ref{SecSpinorOperators}, all spinor space operators
can be decomposed on the complete set
\be\label{SetO}
{\cal O} \equiv\{\mathds{1},\gamma^\mu,\Sigma^{\mu\nu},\gamma^\mu
\gamma^5,\gamma^5\}\,,
\ee 
and in particular  the operators (\ref{DefFparticles}) are decomposed as
\be\label{Fspinor}
 F_{\mathfrak{a}}^{\,\,\,\mathfrak{b}} = \begin{cases}\sum \limits_{X
     \in {\cal O}} c_X \sum \limits_{rs} f_{rs} (\bar u_r X u_s)X_{\mathfrak{a}}^{\,\,\mathfrak{b}} \com{= \sum
\limits_{X \in {\cal O}} c_X \sum_{ab }\left({}^ X{U_{ab}}  f_{ab} \right)
X_{\mathfrak{a}}^{\,\,\mathfrak{b}}}\,\,\, {\rm part.}\\
\sum \limits_{X \in {\cal O}} c_X \sum \limits_{rs} f_{sr} \,(\bar v_r X v_s) X_{\mathfrak{a}}^{\,\,\mathfrak{b}} \com{= \sum
\limits_{X \in {\cal O}} c_X \sum_{ab }\left( {}^ XV_{ab} f_{ba} \right)X_{\mathfrak{a}}^{\,\,\mathfrak{b}}}\,\,\, {\rm antipart.}
\end{cases}
\ee 
Note also how the indices are in reverse order for antiparticles
($f_{sr}$ instead of $f_{rs}$) echoing a similar placement of indices
in Eqs.~(\ref{DefFparticles}). 

The operators of the type (\ref{urus}) take a simple form in the
adapted orthonormal basis defined in \S~\ref{SecAdaptedBasis}. Using the helicity basis
\begin{eqnarray}\label{usvseasy}
u_{s} = \left(\begin{array}{c} \sqrt{E+2s|\vec{p}|}\chi_s \\ \sqrt{E-2s|\vec{p}|}\chi_s \end{array}\right) \quad 
v_{s} = \left(\begin{array}{c} 2s\sqrt{E-2s|\vec{p}|}\chi_{-s} \\ -2s\sqrt{E+2s|\vec{p}|}\chi_{-s} \end{array}\right),\nonumber
\end{eqnarray}
where the right and left chiral parts are defined as
\be
\chi_{\tfrac{1}{2}}= \left(\begin{array}{c}  {\rm e}^{-\ii \phi/2} {\rm
                            cos}(\theta/2) \\ {\rm e}^{\ii\phi/2}{\rm
                            sin}(\theta/2)\end{array}\right) \quad \chi_{-\tfrac{1}{2}}= \left(\begin{array}{c} - {\rm e}^{-\ii\phi/2}{\rm
                            sin}(\theta/2) \\  {\rm e}^{\ii\phi/2}{\rm cos}(\theta/2) \end{array}\right),\nonumber
\ee
we obtain
\begin{subeqnarray}\label{Closure}
u_s \bar{u}_s &=& \frac 1 2 \left(\mathds{1} \sgnzz 2 s \gamma_5
  \slashed{S}\right) \left(\sgnslash \slashed{p} + m \right) \,,\slabel{Closureuu}\\
v_s \bar{v}_s &=& \frac 1 2 \left(\mathds{1} \sgnzz 2 s \gamma_5
  \slashed{S}\right) \left(\sgnslash \slashed{p} - m \right)  \slabel{Closurevv}\,.
\end{subeqnarray}
In particular, we recover when summing on helicities the standard result
\be\label{Sumusus}
\sum_s u_s \bar{u}_s = \sgnslash \slashed{p}+m \,,\qquad \sum_s v_s \bar{v}_s = \sgnslash \slashed{p}-m\,.
\ee
Furthermore, when helicities are different, we obtain the so-called Bouchiat-Michel formulae~\citep{1958NucPh...5..416B} [see
also \citet[App. H.4]{BibleSpinors} or \citet[App. E.3]{Langenfeld:2007pf}]. Using the polarization
basis (\ref{Explicitpolarbasis}) we show \citep[App. D7]{FidlerPitrou}
that it is cast in the compact form
\beas\label{uruscov}
u_s \bar u_r &=& 
\frac{1}{\sqrt{2}} \gamma^5\slashed{\epsilon}_{r-s}\left(\slashed{p}\sgnzz
  m\right)\quad  {\rm if}\quad  r \neq s\,, \\
v_s \bar v_r &= &\frac{1}{\sqrt{2}} \gamma^5\slashed{\epsilon}_{s-r}\left(\slashed{p}\sgnii
  m\right)\quad  {\rm if}\quad  r \neq s \,.
\eeas
Let us define\footnote{We use the obvious abuse of notation $f_{++}$
   for e.g. $f_{+\tfrac{1}{2}\,+\tfrac{1}{2}}$.} from the distribution
 function with helicity indices
\bea\label{DefIVQ}
I&\equiv& f_{++}+f_{--}\,,\quad V\equiv f_{++}-f_{--}\,,\nonumber\\
Q_\pm&\equiv& \sqrt{2}f_{\pm\mp}\equiv Q\pm \ii U \,,
\eea
together with
\be\label{DefcalQ}
Q^\mu \equiv Q_+ \epsilon_+^\mu + Q_- \epsilon_-^\mu\,,\quad {\cal
  Q}^\mu \equiv Q^\mu+V S^\mu\,.
\ee
The functions $(I,Q,U,V)$ are the Stokes parameters. In detail, $I$ is the total intensity, $V$ the circular polarization, and $Q^\mu$
is the purely linear polarization vector. ${\cal Q}^\mu$ is the
total polarization vector, taking into account both circular and
linear polarization. By construction the total polarization ${\cal
  Q}^\mu$ is transverse to the momentum (${\cal Q}^\mu p_\mu=0$). The
linear polarization $Q^\mu$ is transverse both to the momentum and to
the observer velocity $u^\mu$, that is it is a purely spatial
vector.

The covariant parts are defined from the decomposition
\be\label{MasterFermionDecomposition}
\gr{F} = \frac{1}{2}\left(I \sgnzz \gamma^5 \gamma^\mu {\cal Q}_\mu
\right)\left(M\sgnzz \slashed{p}\right)\,,
\ee
\be\label{defMmass}
M=\begin{cases}
+m, \qquad {\rm particle}\\
-m, \,\,\quad {\rm antiparticle}.
\end{cases}
\ee
One degree of freedom corresponds to the total intensity $I(\gr{p})$ while the three remaining
degrees correspond to the state of polarization and are
covariantly contained in a vector ${\cal Q}^\mu(\gr{p})$ because of
its transverse property.

The decomposition (\ref{MasterFermionDecomposition}) can be understood
from group representations. Indeed, the total polarization vector is a
spin-$1$ representation of ${\rm SO}(3)\simeq {\rm SU}(2)$ and the intensity is a spin-$0$ representation. When forming the number operator (\ref{defNrs}), and thus $f_{rs}$, we are building the tensor product of spin-$1/2$ representations and what we have achieved is a decomposition of the reducible representation $\gr{2}\otimes \gr{2}$ in irreducible components $\gr{3} \oplus \gr{1}$, where we have denoted $\gr{1},\gr{2},\gr{3}$ the spin-$0,1/2,1$ representations of ${\rm SU}(2)$.

\subsubsection{Massless fermions}\label{SecMasslessFermions}

In the massless limit, the previous decomposition is slightly modified to
\be\label{BigDecompositionGoodMassless}
\gr{F} =\sgnslash\frac{1}{2}(I+ \hatM V \gamma^5)\slashed{p}+p^\mu \widetilde{\Sigma}_{\mu\nu} Q^\nu\,,\,\,\, \hatM \equiv \begin{cases}1\,\,\,\,\,\quad{\rm part}\\-1\quad{\rm antipart}\end{cases}
\ee
with $\widetilde{\Sigma}^{\mu\nu}$ defined in Eq.~(\ref{DefSigma}). Note that the linear polarization $Q^\mu$ and the circular polarization $V$ enter separately, and not as a total polarization vector ${\cal Q}^\mu$ as is the case for massive fermions. Using $\gamma^5 \slashed{Q} \slashed{p} = \slashed{p} \gamma^5 \slashed{Q} = 2 p^\mu \widetilde{\Sigma}_{\mu\nu} {Q}^\nu$ it can also be rewritten as
\be\label{BigDecompositionGoodMassless2}
\gr{F} =\sgnslash\frac{1}{2}(I+ \hatM V \gamma^5 \sgnzz \gamma^5 \slashed{Q})\slashed{p} \,.
\ee

In the massless case, the little group of the Lorentz group
\citep{Weinberg1} is not ${\rm SO}(3)$ but ${\rm SO}(2)\simeq {\rm U}(1)$. Hence the decomposition in irreducible representations is of the form $\gr{2}_1 \oplus \gr{1} \oplus \gr{1}$ where the purely linear polarization is in the spin-$1$ representation of ${\rm SO}(2)$ (noted $\gr{2}_1$) and circular polarization is in the representation $\gr{1}$. 

Also, it is no longer possible to overtake the
particles as they move at the speed of light in any coordinate
system. This leads to both, the circular and linear polarizations $V$
and $Q^\mu$ to be individually observer independent. More rigorously,
linear polarization is described by the coset of
\be\label{DefCoset}
[ Q^\mu ] \equiv \{ Q^\mu + \alpha p^\mu,\,\,\, \alpha \in \mathds{R}\}\,.
\ee
Indeed, since the polarization basis satisfies $\epsilon_\pm^\mu p_\mu
= 0$, but we also have $p_\mu p^\mu=0$, there is a gauge freedom in the definition of the
polarization basis. The choice (\ref{Explicitpolarbasis})
corresponds to the particular choice which is also transverse to the
observer velocity ($\epsilon_\pm^\mu u_\mu=0$), which selects unique
representatives of polarization vectors. Therefore the polarization vector
representative $\epsilon_\pm^\mu$ are observer dependent, but not the
associated cosets $[\epsilon_\pm^\mu]$. As a consequence $Q^\mu$ is
observer dependent but not its coset $[Q^\mu]$.

 Given a representative of the coset, the one associated with a given
 observer (that is such that it is transverse to that observer
 velocity) is obtained by projection with a screen projector ${\cal
   H}_{\mu\nu}(u,p)$, which is abbreviated as ${\cal H}_{\mu\nu}$ when no
 ambiguity can arise. Using the decomposition of the null momentum into energy and unit direction
\be\label{Decomposep}
p^\mu = E(u^\mu+n^\mu)\,,
\ee
where $E\equiv \sgnzz u_\mu p^\mu =  p^0$, the screen projector is built from the equivalent definitions
\beas\label{DefScreenH}
{\cal H^\mu}_\nu(u,p) &\equiv& \delta_\nu^\mu \sgnii p^\mu k_\nu \sgnii k^\mu
p_\nu\\
&=&\delta_\nu^\mu  \sgnii u_\nu u^\mu \sgnzz n_\nu n^\mu\slabel{DefScreenHb}\\
&=&\sgnminusslash \epsilon_-^{\star\mu} \epsilon_{-\,\nu} \sgnii
\epsilon_+^{\star\mu} \epsilon_{+\,\nu} \slabel{DefScreenHc}\\
&=&\delta^\mu_\nu-\frac{p^\mu p_\nu}{E^2}+\frac{p^\mu u_\nu}{E}+\frac{u^\mu p_\nu}{E}\,,
\eeas
with $k^\mu$ is a future directed null vector in the plane spanned by
$(u^\mu,p^\mu)$ such that $k^\mu p_\mu = \sgnslash 1$, $k_\mu \epsilon_\pm^\mu=0$. It can be checked that the screen projector satisfies the expected properties
${{\cal H}_{\mu}}^\sigma {\cal H}_{\sigma\nu} = {\cal H}_{\mu\nu}$ and
${\cal H}_{\mu\nu}p^\nu={\cal H}_{\mu\nu}u^\nu=0$.  If the observer
used in the definition is the natural observer associated with the
tetrad with which components are taken (that is if $u^\mu =
e_o^{\,\mu}$), the non-vanishing components of the screen projector
are only ${\cal H}_{ij} = \delta_{ij}- n_i n_j$.

For two screen projectors associated with two observers $u$ and
$\tilde u$ related by a boost
\be
\tilde u^\mu = \gamma(u^\mu + v^\mu)\,,\quad v_\mu u^\mu = 0\,,\quad
\gamma=\frac{1}{\sqrt{1-v_\mu v^\mu}}\,,
\ee 
but for the same momentum $p$, we find that they are related by
\bea\label{HtildeH}
{\cal H}_{\mu\nu}(\tilde u,p) &=& {\cal H}_{\mu\nu}(u,p) +2\frac{\gamma}{\tilde E} p_{(\mu} {\cal H}_{\nu) \sigma}(u,p)
v^\sigma \nonumber \\
&&+\left(\frac{\gamma}{\tilde E}\right)^2 p_\mu p_\nu {\cal
  H}_{\lambda \sigma}(u,p) v^\lambda v^\sigma\,,
\eea
where $\tilde E\equiv \sgnzz \tilde u_\mu p^\mu$. In particular this implies
\be\label{ProjectHcal}
{\cal H}_\mu^{\,\,\alpha}(u,p) {\cal H}_\nu^{\,\,\beta}(u,p) {\cal H}_{\alpha\beta}(\tilde{u},p) = {\cal H}_{\mu\nu}(u,p)\,.
\ee
Using the screen projector, another definition of the linear polarization coset is that two polarization vectors $Q_1^\mu$ and $Q_2^\mu$
describe the same state ($[Q_1^\mu] = [Q_2^\mu]$) if
\be\label{DefCosetFermions}
{\cal H}^\mu_{\,\,\nu}(Q_1^\nu-Q_2^\nu)=0\,.
\ee
Note that for a transverse vector ($X_\mu p^\mu = 0$) it is obvious
from the decomposition (\ref{DefScreenH}) or the transformation rule
(\ref{HtildeH}) that
\be\label{Independencecoset}
{\cal H^\mu}_\alpha(u,p){\cal H^\alpha}_\nu(\tilde u,p) X^\nu = {\cal H^\mu}_\nu(u,p) X^\nu\,,
\ee
implying that the definition~(\ref{DefCosetFermions}) is unambiguous.

For photons, that is massless bosons, on which we focus in
the next section, the structure is exactly similar and arises from the
electromagnetic gauge freedom.

\subsubsection{Massless bosons}

The null mass bosonic vector field of quantum electrodynamics is
\be\label{DefQuantumFieldA}
\hat A^\mu(p) = \sum_{s=\pm 1} \int [\dd p]\left[ {\rm e}^{\sgnslash \ii p \cdot x }
a^\dagger_s(\gr{p}) \epsilon_s^{\star \mu}(\gr{p})+ {\rm e}^{\sgnminusslash \ii p \cdot x }
a_s(\gr{p}) \epsilon_s^\mu(\gr{p})\right]\,\nonumber
\ee
where the creation and annihilation operators satisfy the commutation rule
\be
[a_r(p), a^\dagger_s(p')] = \delta^{\rm K}_{rs} \deltarel(p-p')\,.
\ee
If vectors are massive, then the null helicity ($s=0$) must also be
considered, see \citet[App. A]{FidlerPitrou}. 

A covariant distribution tensor is obtained by considering
\be\label{Deffmunumassless}
f^{\mu\nu}(u,p)\equiv\sum_{r,s=-1,1} f_{rs}(u,p) \epsilon_r^{\star \mu}(u,p)\epsilon_s^\nu(u,p)\,,
\ee
and by construction it is transverse to the momentum and the
observer's velocity ($f_{\mu\nu}p^\nu=f_{\mu\nu}u^\mu=0$). When no
ambiguity can arise, we omit the dependence on the observer's velocity
$u^\mu$ used in its definition.

We define  
\beas
I&\equiv&f_{++}+f_{--}\,, \\
V&\equiv&f_{++}-f_{--}\,,\\
\polar^{\pm\pm}&\equiv& \tfrac{1}{2}(Q\pm \ii U) = f_{\mp\pm}\,,
\eeas  
as the usual Stokes parameters\footnote{It is sometimes customary in the cosmic microwave background context to define the distribution function as~\cite{DurrerBook} $f^{\rm CMB}_{r\,s} \equiv f_{-r\,s}$. Accordingly, the tensor valued function (\ref{Deffmunumassless}) is defined as $f^{\mu\nu} = \sum_{rs}f^{\rm CMB}_{rs}\epsilon_r^{\mu}\epsilon_s^\nu$. With this definition the Stokes parameters are $I\equiv f^{\rm CMB}_{-+}+f^{\rm CMB}_{+-}$, $V \equiv f^{\rm CMB}_{-+} - f^{\rm CMB}_{+-}$ and $Q\pm \ii U \equiv 2 f^{\rm CMB}_{\pm\pm}$.} corresponding to intensity, circular polarization and linear polarization. 
For a given observer with four-velocity $u^\mu$, we use as in the massless fermion case the spatial momentum direction unit vector $\gr{n}$ defined in the decomposition~(\ref{Decomposep}). Let us also define the two-dimensional Levi-Civita tensor
\be\label{TwoDLeviCivita}
\epsilon_{\mu\nu}(u,p) \equiv u^\lambda\epsilon_{\lambda \mu \nu
  \sigma} n^{\sigma} =  \ii\left(\epsilon_+^{\star\mu} \epsilon_{+}^{\nu}-\epsilon_-^{\star\mu} \epsilon_{-}^{\nu} \right)\,.
\ee
We usually omit the dependence on $(u,p)$ and write simply
$\epsilon_{\mu\nu}$. The tensor-valued distribution function is
decomposed as
\be\label{fmunuphotons}
f_{\mu\nu}(u,p) = \polar_{\mu\nu}(u,p) \sgnii \frac{1}{2}{\cal
    H}_{\mu\nu} I(p)- \frac{\ii}{2} \epsilon_{\mu \nu}V(p) \,,
\ee
where the screen projector is defined exactly as for massless fermions
in Eqs.~(\ref{DefScreenH}). The distribution tensor is doubly transverse, that is transverse to the momentum $p^\mu$ and also  to the observer
velocity $u^\mu$. $\polar_{\mu\nu}$ is the linear polarization tensor and it is doubly transverse and traceless (it satisfies $\polar_{\mu\nu} u^\mu=\polar_{\mu\nu}p^\mu={\polar_{\mu}}^\mu=0$). It
is defined as
\be\label{DefPolar}
\polar^{\mu\nu}(u,p) \equiv \sum_{r=-1,1} f_{r\,-r}(p) \epsilon_r^{\star \mu}(u,p)\epsilon_{-r}^\nu(u,p)\,,
\ee
and its dependence on $u$ is often omitted. It can be extracted thanks to the transverse traceless projector
\beas
\polar_{\mu\nu}(u,p) &=& {\cal T}_{\mu\nu}^{\phantom{\mu\nu}\rho\sigma}(u,p)
f_{\rho\sigma}(u,p)\,,\\
 {\cal T}_{\mu\nu}^{\phantom{\mu}\,\,\rho\sigma}(u,p) &\equiv& {\cal
   H}_{\mu}^{(\rho} {\cal H}_{\nu}^{\sigma)} -\frac{1}{2}{\cal H}_{\mu\nu} {\cal H}^{\rho\sigma}\slabel{DefTTproj}\,.
\eeas

In the $\gr{e}_\theta,\gr{e}_\phi$ basis the components of the distribution tensor~(\ref{fmunuphotons}) form a $2\times 2$ Hermitian matrix~\cite{Hu:1997hp,Tsagas2007,DurrerBook}
\be\label{MatrixStokes}
\frac{1}{2}\left( \begin{array}{cc}
I+Q & U-\ii V  \\
U+\ii V & I-Q  \end{array} \right)\,,
\ee
whereas in the $\gr{\epsilon}_-,\gr{\epsilon}_+$ basis we obtain the Hermitian matrix
\be\label{MatrixStokes2}
\frac{1}{2}\left( \begin{array}{cc}
I+V & Q+\ii U  \\
Q-\ii U & I-V  \end{array} \right)\,.
\ee

For massless bosons, the structure of the decomposition can also be understood exactly like
in the discussion following
Eq.~(\ref{BigDecompositionGoodMassless2}) for massless fermions. The difference is that for
massless bosons, we decompose $\gr{2}_1  \otimes \gr{2}_1$ into
$\gr{2}_2 \oplus \gr{1} \oplus \gr{1}$, where $\gr{2}_1$
(resp. $\gr{2}_2$) is the spin-$1$ (resp. spin-$2$) representation of
${\rm SO}(2)$.

As in the case of massless fermions, the definition (\ref{Deffmunumassless}) and the
decomposition (\ref{fmunuphotons}) of the distribution tensor is observer
dependent, for exactly the same reasons that the polarization vectors
are defined up to factors of $p^\mu$. Hence, we should rather consider the
coset $[f_{\mu\nu}]$. Two polarization states $f^1_{\mu\nu}$ and
$f^2_{\mu\nu}$ are in the same coset if 
\be\label{DefCosetBosons}
{\cal H}_{\alpha}^{\,\,\mu} {\cal H}_{\beta}^{\,\,\nu} (f^1_{\mu\nu}- f^2_{\mu\nu})=0\,.
\ee
In particular, the linear polarization parts $\polar^1_{\mu\nu}$ and
$\polar^2_{\mu\nu}$ are equivalent if
\be
{{\cal T}_{\mu\nu}}^{\sigma\lambda}(\polar^1_{\sigma\lambda}-\polar^2_{\sigma\lambda})=0\,,
\ee
and one should rather consider the coset $[\polar_{\mu\nu}]$ of linear
polarization. With arguments similar to Eq.~(\ref{Independencecoset}),
this definition of equivalence (and its associated cosets) is observer independent.

\subsection{Multipolar decomposition}\label{SecMultipoles}

\subsubsection{Fermions}\label{SecMultipolesFermions}

The intensity $I(\gr{p})$ is easily decomposed into spherical
harmonics. Indeed, once an observer choice is made, that is its
four-velocity $u^\mu$ is identified with the time-like vector of the
tetrad $[e_0]^\mu$, we can define the
spatial momentum $\gr{p}$ and its direction unit vector
${\gr{n}}$ (see \S~\ref{SecAdaptedBasis}). We then perform the usual spherical harmonics decomposition 
\be\label{Ilm}
I(\gr{p}) = \sum_{\ell m} I_{\ell m}(|\gr{p}|) Y_{\ell m}(\gr{n})\,.
\ee
Using Eq. (\ref{OrthoYlm}) the multipoles are extracted as $I_{\ell m}= \{Y_{\ell m} | I\}$.

Alternatively one could use a decomposition based on symmetric
trace-free (STF) tensors $I_{i_1 \dots i_\ell}$ which is equivalent
~\cite{Thorne1980,BlanchetDamour1986,Pitrou2008,Pitrou2008GRG}
\be\label{STFscalar}
I(\gr{p}) = \sum_{\ell} I_{J_\ell}(|\gr{p}|) n^{J_\ell}\,,
\ee
where we use the tools and notation summarized in appendix~\ref{AppSTF}. From Eq.~(\ref{OrthoSTF}) the STF tensors are
extracted as 
\be
I_{J_\ell} = \Delta_\ell^{-1} \{ n_{\langle J_\ell \rangle} | I\}.
\ee
The relation between both expansions is obtained
from Eqs.~(\ref{MagicSTFYlm}) as
\beas\label{almtoSTF}
I_{J_\ell} &=& \sum_{m=-\ell}^\ell I_{\ell m} {\cal Y}^{\ell  m}_{J_\ell}\,,\\
I_{\ell m} &=& \Delta_\ell I_{J_\ell} {\cal Y}^{J_\ell}_{\ell m}\,.\slabel{lmfromSTF}
\eeas

For the polarization vector ${\cal Q}^\mu$ of fermions defined in
Eq.~(\ref{DefcalQ}), we have to pay attention to the transformation properties when performing a spatial rotation of the coordinate system around the direction of ${\gr{n}}$. The ordinary spherical harmonics, when evaluated at ${\gr{n}}$ do not transform under this rotation and are thus not suitable to decompose objects which have a non-trivial transformation under this rotation. 
The polarization vector $\cal Q$ transforms as an ordinary 4-vector (we have shown that it is observer-independent). 
However, this is not the case for the observer-dependent vectors and distribution functions used to build $\cal Q$.
The vector in direction of the spatial momentum $S^\mu$ is invariant
under this particular rotation as it points in the direction
${\gr{n}}$. Employing the observer-independence of $\cal Q$ which is discussed in the next section, we
therefore conclude that $V$ must be invariant under this rotation and
may be decomposed into ordinary spherical harmonics. 
\be\label{Vlm}
V(\gr{p}) = \sum_{\ell m} V_{\ell m}(|\gr{p}|) Y_{\ell m}(\gr{n})\,.
\ee
Again an expansion in STF tensors of the type (\ref{STFscalar}) is
possible and is obtained by relations exactly similar to Eqs.~(\ref{almtoSTF}).

The polarization vectors ${\epsilon}_{\pm}({\gr{n}})$ however transform with an additional spin $\mp 1$ complex rotation. To generate an observer-independent $\cal Q$ the corresponding $Q_{\pm}$ must transform with the opposite spin and they are decomposed into  spin-weighted spherical harmonics $Y^s_{lm}$~\cite{Goldberg1967} as
\beas
Q_+(\gr{p}) &\equiv& \sum_{\ell m} Q^+_{\ell m}(|\gr{p}|) Y_{\ell
  m}^+(\gr{n}) \,,\\
  Q_-(\gr{p}) &\equiv&\sum_{\ell m} Q^-_{\ell m}(|\gr{p}|) Y_{\ell
  m}^-(\gr{n})\,.
\eeas
Note that this discussion only concerns the observer dependence under
a specific spatial rotation and that due to the definition of helicity
an additional dependence mixing $V$ and $Q_{\pm}$ exists for more
general rotations and boosts. 
 ${\cal E}$ and ${\cal B}$ modes multipoles can be defined from 
\be
Q^\pm_{\ell m} \equiv \mp ({\cal E}_{\ell m} \pm \ii {\cal B}_{\ell
  m})\,.
\ee 
The ${\cal E}_{\ell m}$ have even parity (they get a factor $(-1)^\ell$
under parity transformation) whereas the ${\cal B}_{\ell
  m}$ have odd parity  (they get a factor $(-1)^{\ell+1}$
under parity transformation) since spin-weighted spherical harmonics
transform as $Y^s_{\ell m} \to (-1)^{\ell} Y^{-s}_{\ell m}$ and the polarization basis transforms as
$\gr{\epsilon}_\pm \to -\gr{\epsilon}_\mp $. Equivalently since $\gr{Q}(\gr{p})$ is a vector field
on the unit sphere in momentum space, it can be decomposed as the
gradient and the curl of two scalar functions as 
\be\label{twoPotentials}
Q_i(\gr{p}) = D_i
{E}(\gr{p}) + {\epsilon_{i}}^j D_j {B}(\gr{p})\,,
\ee
where $D_i$ is the covariant derivative on the unit sphere and
$\epsilon_{ij} \equiv {\epsilon_{ijk}} n^k$ is the Levi-Civita tensor on the unit sphere already defined in Eq.~(\ref{TwoDLeviCivita}). Decomposing the scalar functions ${E}$ and ${B}$ in
multipoles ${E}_{\ell m}$ and ${B}_{\ell m}$  as in the expansion (\ref{Ilm}) and using~\cite{DurrerBook}
\be
D^i Y_{\ell m} = \sqrt{\frac{\ell(\ell+1)}{2}}\left(-Y^+_{\ell m}
\epsilon_+^i + Y^-_{\ell m} \epsilon_-^i \right)\,,
\ee
the two possible definitions for the $E$ and $B$ modes multipoles are
related by ${\cal E}_{\ell m} = \sqrt{\ell(\ell+1)/2} {E}_{\ell m}$
and ${\cal B}_{\ell m} = \sqrt{\ell(\ell+1)/2} {B}_{\ell m}$.
Again a similar expansion can be obtained by using symmetric trace-free tensors to expand the scalar functions ${E}$ and
${B}$ directly in Eq. (\ref{twoPotentials}).

\subsubsection{Massless bosons}

The decomposition of intensity and circular polarization is performed
with spherical harmonics as in Eqs.~(\ref{Ilm}) and (\ref{Vlm}) or
with STF tensors as detailed in \S~\ref{SecMultipolesFermions} for
fermions. However, the linear polarization part $\polar_{\mu\nu}$
must be decomposed in spin-$2$ spherical harmonics. We decompose
polarization as
\be\label{DefPplusplus}
2\polar^{\mu\nu}(\gr{p}) =\polar_{++}(\gr{p}) { \epsilon}^\mu_+(\gr{p}) { \epsilon}^\nu_+(\gr{p}) +\polar_{--}(\gr{p}) { \epsilon}^\mu_-(\gr{p}) { \epsilon}^\nu_-(\gr{p}).
\ee
and the angular decomposition is 
\be\label{PplusplusYlms}
\polar_{\pm\pm}(\gr{p}) \equiv \sum_{\ell m} \polar^{\pm\pm}_{\ell m}(E) Y_{\ell
  m}^{\pm2}({\gr{n}}) \,.
\ee
Note that the factor $2$ in Eq. (\ref{DefPplusplus}) is purely
conventional. ${\cal E}$ and ${\cal B}$ modes are defined by 
\be\label{DefEBfromYlms}
\polar^{\pm\pm}_{\ell m} \equiv
({\cal E}_{\ell m} \pm \ii {\cal B}_{\ell m})\,.
\ee

Equivalently linear polarization can be decomposed with two potentials on
 the unit sphere in momentum space as \citep[Eq.~4.3.8]{Tsagas2007}
\be
\polar_{ij} = D_{\langle i} D_{i \rangle} {E}(\gr{p}) +
{\epsilon^k}_{\langle i} D_{j\rangle} D_k {B}(\gr{p})\,,
\ee
and the associated multipoles ${E}_{\ell m}$ and ${B}_{\ell
  m}$ can be related to the ${\cal E}_{\ell m}$ and ${\cal B}_{\ell m}$ by some
factors. Instead, if we use an expansion of ${E}$ and ${B}$
in STF tensors of the type (\ref{STFscalar}), we can decompose
$\polar_{ij}$ with them. However, it is customary to remove the $\ell(\ell-1)$
factors brought by the covariant derivatives $D_i$ and use the
expansion \citep{Dautcourt1978} [see also
\citet[Eq.~4.3.9]{Tsagas2007} or \citet[Eq.~1.33]{Pitrou2008}]
\be\label{DefEBTsagas}
\polar_{ij}(\gr{p}) = \left[\sum_{\ell} {E}_{ij K_{\ell}}(E) n^{K_\ell} -
{\epsilon^p}_{(i} {B}_{j) p K_\ell}(E) n^{K_\ell} \right]^{\cal T}.
\ee
The exponent ${\cal T}$ indicates that free indices are to be
projected on the transverse traceless part with the operator
(\ref{DefTTproj}). From the definition (\ref{DefPplusplus}) of
$\polar_{\pm\pm}$, and using the notation~(\ref{Defns}), this expansion is equivalent to
\be\label{PolarplusplusEB}
\polar_{\pm\pm}(\gr{p}) = \sum_\ell \left[E_{I_\ell}(E) \mp \ii B_{I_\ell}(E)\right]
n_{\mp 2}^{\langle  I_\ell \rangle}\,.
\ee
The STF tensors of the decomposition (\ref{DefEBTsagas}) are extracted
thanks to \citep{Tsagas2007,Pitrou2008}
\beas
E_{I_\ell} &=& M_\ell^2 \Delta_\ell^{-1}\{ n_{\langle I_{\ell-2}} |
\polar_{i_{\ell-1} i_\ell \rangle } \}\,,\\
B_{I_\ell} &=& M_\ell^2 \Delta_\ell^{-1}\{ n_j
\epsilon^{jk}_{\phantom{jk} \langle i_\ell } n_{I_{\ell-2}} | \polar_{i_{\ell-1} \rangle k} \}\,,
\eeas
where
\be
M_{\ell} \equiv \sqrt{\frac{2 \ell (\ell-1)}{(\ell+1)(\ell+2)}}\,.
\ee

If we now associate to these STF tensors
$E_{K_\ell}$ and $B_{K_\ell}$ the ${E}_{\ell m}$ and ${B}_{\ell m}$,
using a relation of the type (\ref{lmfromSTF}), these are related to the
${\cal E}_{\ell m}$ and ${\cal B}_{\ell m}$ defined in (\ref{DefEBfromYlms}) by
\be
{\cal E}_{\ell m} \pm \ii {\cal B}_{\ell m} =
\frac{\sqrt{2}}{M_\ell} \left({E}_{\ell m} \mp \ii {B}_{\ell m} \right)\,.
\ee
This is obtained using Eqs.~(\ref{Magiceplusns}) in Eq.~(\ref{PolarplusplusEB}) and comparing with Eq.~(\ref{PplusplusYlms}).

\subsection{Observer independence}\label{SecObserverIndependence}

In this section, we detail the transformation property of the
distribution function under a general Lorentz transformation $\Lambda \in {\rm SO}^+(1,3)$. 
It is more appropriate to take the passive point of view and consider
a transformed tetrad basis related to the initial one by
\be\label{DefLorentz} 
\tilde{\gr{e}}^{\tilde\nu} = {\Lambda^{\tilde\nu}}_{\nu'} {\gr{e}}^{\nu'}\qquad \tilde{\gr{e}}_{\tilde\nu} = {\gr{e}}_{\nu'} (\Lambda^{-1})^{\nu'}_{\,\,\,\tilde\nu}= {\Lambda_{\tilde\nu}}^{\nu'} {\gr{e}}_{\nu'}
\ee
The new observer's velocity $\tilde{\gr{u}}$ is identified with the
time-like vector of the new tetrad $\tilde{\gr{e}}_{\tilde0}$. That is we take the point of view
that when considering a change of frame we also consider the
associated change of observer, such a that any observer is not moving in
its own frame.  In that sense, the observer's velocity is not observer
independent. 

The new components of the momentum $p^{\tilde\mu} \equiv p \cdot
\tilde{\gr{e}}^{\tilde\mu}$ are related to the previous ones $p^\mu \equiv p \cdot \gr{e}^\mu$ by
\be
p^{\tilde\mu} = \Lambda^{\tilde\mu}_{\phantom{\mu}\nu} p^\nu\,,
\ee
and we abbreviate $p^{\tilde\mu}$ as $\tilde p$.

\subsubsection{Massive fermions}

In \citet{FidlerPitrou}, we showed that the spinor valued operator
transforms under the Lorentz transformation  $\Lambda \in {\rm
  SO}^+(1,3)$ defined by Eqs.~(\ref{DefLorentz}) as
\be \label{WonderfulTransformation}
{\widetilde{F}_\mathfrak{a}}^{\,\,\mathfrak{b}}(\tilde p) = {D_\mathfrak{a}}^{\mathfrak{a}'}(\Lambda){F_{\mathfrak{a}'}}^{\mathfrak{b}'}( p){D_{\mathfrak{b}'}}^\mathfrak{b}(\Lambda^{-1})\,,
\ee
where $D(\Lambda)$ is the spinor-space representation of $\Lambda$.
Using the property for Dirac matrices
\be
D(\Lambda) \gamma^\mu D(\Lambda^{-1}) = \gamma^\nu {\Lambda_\nu}^\mu\,,
\ee
it implies that the covariant components for massive fermions transform as
\be
\widetilde{I}(\tilde p ) = I(p)\,,\qquad \widetilde{\cal Q}^{\tilde\mu}(\tilde p) = {\Lambda^{\tilde\mu}}_\nu
{\cal Q}^\nu(p)\,.
\ee
This means that they transform exactly as a scalar and vector field, and they are therefore observer independent.

The observer independence is important as it allows to build a
statistical description of the fluid without the need to specify an
observer first. This is particularly useful for deriving simple
transport equations in general relativity.   

The scalar $I$ describes the total intensity of the field and is observer independent since the local number of particles is identical for each observer. The information of the polarization of the fluid is contained in the observer independent vector ${\cal Q}^\mu$.

On the other hand the parameters $V$ and $Q_\pm$, describing
individually the circular and linear polarizations are not observer
independent. The circular polarization $V$, for example, changes if
the observer is boosted and overtakes the momentum considered. We have defined
\be
 {\cal Q}^\mu = Q^\mu + V S^\mu\,,
\ee
where ${\cal Q}^\mu$ combines multiple observer dependent quantities
into one observer independent vector. In the example of the observer
overtaking a particle momentum, we change all left-helical $-\tfrac{1}{2}$
states into right-helical $+\tfrac{1}{2}$ states. This means that the
boosted observer will find $\widetilde{V} = - V$. At the same time the
vector $S^\mu$ is also observer dependent and the new observer will
define the spatial momentum of the particles with the opposite
sign. Therefore the combination $V S^\mu$ is invariant under this
boost. At the same time the off-diagonal distributions are swapped:
$\widetilde{f}_{+-} = f_{-+}$. However these are combined with the polarization vectors $\epsilon_{\pm}$ to form $Q^\mu$, which are also interchanged for the new observer, leading to $Q^\mu$ being invariant. 

In a more general case $Q^\mu$ and $V$ cannot be disentangled in an
observer independent manner and there always exists a subset of
observers, all related by boosts along the momentum direction and
rotations around the momentum direction, that will perceive the field
to be entirely circularly polarised without any linear
polarization. For this reason we will work with the observer
independent polarization vector ${\cal Q}^\mu$ and only refer to the
circular and linear polarizations when we have specified an
observer. Only in the case of massless fermions, considered in
\S~(\ref{SecMasslessFermions}), the linear and circular polarization
can be disentangled, and are observer independent, the latter in the
sense of the polarization coset (\ref{DefCoset}) as detailed in the
next section.

\subsubsection{Massless fermions}

Using the decomposition (\ref{BigDecompositionGoodMassless}) for massless fermions, we deduce that the covariant components transform as
\be
\widetilde{I}(\tilde p ) = I(p)\,,\quad\widetilde{V}(\tilde p ) = V(p)\,,\quad \widetilde{Q}^{\tilde\mu}(\tilde p) = {\widetilde{\cal H}^{\tilde\mu}}_{\,\,\tilde\sigma}{\Lambda^{\tilde\sigma}}_\nu
{Q}^\nu(p)\,.
\ee
The screen projector [see Def. (\ref{DefScreenH})] associated with the
new observer and the new momentum components, 
\be\label{DefHtilde}
\widetilde{\cal H}^{\tilde\mu}_{\,\,\tilde\nu} \equiv
\delta^{\tilde\mu}_{\tilde\nu}-\frac{p^{\tilde\mu}
  p_{\tilde\nu}}{\tilde E^2}+\frac{p^{\tilde\mu} \tilde{u}_{\tilde
    \nu}}{\tilde E}+\frac{\tilde{u}^{\tilde \mu} p_{\tilde\nu}}{\tilde
  E},
\ee 
(with $\tilde E \equiv \sgnzz \tilde{u}_{\tilde \mu} p^{\tilde \mu} =  p^{\tilde 0}$) ensures that the linear polarization
remains spatial for the new observer. Hence in the massless case, the
linear polarization part is not strictly observer independent, but
since this dependence introduced by the screen projector is there only
as the result of a choice to remove a non physical degree of freedom,
we can still conclude that in that sense the covariant components are
observer independent. More rigorously, it is the coset of linear
polarization [see definition (\ref{DefCoset})] which is observer independent
and only the special choice of its representative element is observer
dependent. Hence we should rather write the transformation rule of linear
polarization cosets which is 
\be
\left[\widetilde{Q}^{\tilde\mu}(\tilde p)
\right]=\left[ {\Lambda^{\tilde\mu}}_\nu{Q}^\nu(p)\right],
\ee 
for which the observer independence is manifest.

\subsubsection{Massless bosons}\label{SecIndependenceBosons}

For massless bosons, the tensor-valued distribution function transforms as
\be
\widetilde{f}^{\tilde\mu\tilde\nu}(\tilde p) = (\widetilde{\cal
    H} \Lambda)^{\tilde\mu}_{\,\,\sigma}   (\widetilde{\cal H} \Lambda)^{\tilde\nu}_{\,\,\tau} f^{\sigma\tau}(p)\,,
\ee
with the definition $(\widetilde{\cal  H} \Lambda)^{\tilde\mu}_{\,\,\sigma} \equiv {{\widetilde{\cal
    H}}^{\tilde\mu}}_{\,\,\tilde\nu} \Lambda^{\tilde\nu}_{\,\,\sigma}$. 
Since the screen projector satisfies\footnote{This is exactly Eq. (\ref{HtildeH}) but expressed with components associated to different tetrads.}
\be
 (\widetilde{\cal
    H} \Lambda)^{\tilde\mu}_{\,\,\sigma}   (\widetilde{\cal H}
  \Lambda)^{\tilde\nu}_{\,\,\tau} {\cal H}^{\sigma\tau} =
  \widetilde{\cal H}^{\tilde \mu \tilde \nu} \,,
\ee
and the two-dimensional Levi-Civita tensor (\ref{TwoDLeviCivita}) satisfies a
similar property, then we deduce that the covariant components transform as
\beas
\widetilde{I}(\tilde p ) &=& I(p)\,,\quad \widetilde{V}(\tilde p ) =
V(p)\,,\\
\widetilde{\polar}^{\tilde\rho\tilde\lambda}(\tilde p) &=&
\widetilde{\cal T}^{\tilde\rho\tilde \lambda}_{\phantom{\alpha\beta}\tilde\mu\tilde\nu}\Lambda^{\tilde\mu}_{\,\,\sigma}   \Lambda^{\tilde\nu}_{\,\,\tau}\polar^{\sigma\tau}(p)\,,
\eeas
where $\widetilde{\cal T}$ is the transverse-traceless projector associated with $\widetilde{\cal H}_{\tilde\mu\tilde\nu}$, using the definitions
(\ref{DefTTproj}) and (\ref{DefHtilde}). As in the case of massless fermions, it is the coset of linear
polarization [see Eq. (\ref{DefCosetBosons})] which is observer independent, and only the special choice of
its representative element is observer dependent. Hence we should rather
write the transformation rule as
$\left[\widetilde{f}^{\tilde\mu\tilde\nu}(\tilde p) \right]=\left[
  {\Lambda^{\tilde\mu}}_\sigma
  {\Lambda^{\tilde\nu}}_\tau{f}^{\sigma\tau}(p)\right]$, 
with the cosets defined by the equivalence relation
(\ref{DefCosetBosons}), and for which the observer independence is manifest.

\subsubsection{Relation to abstract tensor indices}

Since we have shown that all components of the vectors or tensors
associated to fermions and photons have the expected transformation
properties, we could decide to work with abstract indices as in
\citet{Challinor1999,Challinor2000,Tsagas2007} instead of
working with indices referring to a particular tetrad. In most cases
this reinterpretation is straightforward. However both approaches
differ when it comes to expressing in practice the transformation of
the STF tensors presented in \S~\ref{SecMultipoles} for the angular
decomposition of the distribution functions. With abstract indices,
projectors still appear in the transformation rules, as
e.g. in  Eqs.~(4.3.31-4.3.33) of \citet{Tsagas2007}, whereas with indices referring to components in
tetrads the transformations relate STF tensors which all are purely
spatial in their associated tetrad, that is we relate only spatial
indices, as in Eqs.~(1.56-1.58) of \citet{Pitrou2008}. However, this subtlety
only shows when the transformation of the multipoles is performed at
least at second order in the boost velocity. In the remainder of this
article, we use a method where no change of frame is needed, hence we do
not detail any further the procedure to obtain the multipoles transformation rules. More details can be found in \citet{Pitrou2008}.

\section{Boltzmann equation}\label{SecBoltzmann}

\subsection{Liouville equation in curved space-time}\label{SecLiouvilleGR}

The previous construction was restricted to a homogeneous system,
hence the functions appearing ($I$ and $Q^\mu$) depended only on
$(t,p)$. In order to describe a gas of particles classically, one
must assume that this construction is in fact valid only locally. That is
we assume that there is a mesoscopic scale and that our previous
construction was restricted to scales much smaller. The functions,
which were dependent on $(t,p)$ must depend now on $(t,\gr{x},p)$.
In order to derive a Liouville equation in curved space-time which
describes the evolution of the covariant components, we must
also distinguish between the massive and the massless cases.

\subsubsection{Massive fermions}

In the previous sections we have shown that $I$ and ${\cal Q}$ are observer independent. In addition, in the local Minkowski frame, they are also parallel transported in the absence of collisions. The helicity of particles does not change in free propagation and, considering that the momentum $p^\mu$ is conserved, the vectors $\epsilon_\pm^\mu$ and $S^\mu$ used to build the quantities $I$ and ${\cal Q}$ remain unchanged. Hence, in the local Minkowski space we obtain the equations of motion
\be\label{eq:Minkowskievo}
\frac{\dd I}{\dd t} = 0\,, \qquad \frac{\dd {\cal Q}^\mu}{\dd t} = 0\,.
\ee

From the point of view of general relativity, these equations are only valid locally and neglect entirely the impact of the relativistic space-time. The intensity $I$ describes the total number of particles. The conservation of $I$ in the absence of collisions in Eq.~(\ref{eq:Minkowskievo}) is equivalent to mass or particle number conservation. The geometrical impact of general relativity does not change the number of particles and we may generalise the equation of motion by requiring the conservation of $I$ along a full geodesic
\be\label{DIDlambda}
\frac{DI}{D\lambda} = 0 \,,
\ee
where $\frac{D}{D\lambda}$ is the derivative along the particle trajectory parameterized by $\lambda$.

The vector ${\cal Q}$ is parallel transported in the local space-time and describes the polarization of particles in an observer-independent way. Again, the geometrical nature of general relativity does not change the polarization of particles and we require that ${\cal Q}$ is parallel transported along the non-trivial trajectory of the particles. Note that the observer dependent linear and circular polarization may change non-trivially during the transport and require a specification of the dynamics of the observer.

Using the observer-independence, we are able to uniquely define the vector ${\cal Q}$ on our full space-time by employing the tetrads
\be\label{QalphaQmu}
{\cal Q}^\alpha = {\cal Q}^\mu [e_\mu]^\alpha  \,,
\ee
where we remind that the index $\mu$ is a tetrad component index, but the index $\alpha$ is a general coordinate index. Assuming parallel transport, we obtain the equation of motion
\be\label{DQDlambda}
\frac{D{\cal Q}^\alpha}{D\lambda} = 0 \,.
\ee
Note that ${\cal Q}$ is by definition orthogonal to the momentum. This property is automatically conserved in the relativistic
evolution as both the momentum and $\cal{Q}$ are parallel-transported along the geodesic of a free particle. 

\subsubsection{Massless fermions}

In the massless case, linear polarization and circular polarization
must be considered separately. Circular polarization $V$ is
transported exactly like the intensity $I$ in Eqs. (\ref{DIDlambda})
because the direction of the helicity vector is identical to the
momentum and therefore parallel-transported. However the linear
polarization vector (considered in general coordinates with $Q^\alpha
= Q^\mu [e_\mu]^\alpha$) cannot be parallel transported because it is
transverse to both the momentum and the observer velocity
$u^\alpha$, and the latter is not (necessarily) parallel
transported. However, in the process of free streaming, any variation
of $Q^\alpha$ in the direction of the momentum is not physical. Hence
this unphysical degree of freedom must be eliminated by an appropriate
projection so as to obtain an unambiguous equation for parallel
transport. To that purpose, we use the screen projector
(\ref{DefScreenH}) in general coordinates and write
\be\label{LiouvilleQmassless}
{{\cal H}^\beta}_\alpha \frac{D Q^\alpha}{D \lambda} = 0\,.
\ee
The transport of linear polarization in the massless case is the same
as the transport of the full polarization vector in the massive case
[Eq. (\ref{DQDlambda})], up to an additional
screen projection which ensures that the double transverse property
holds. It can be equivalently formulated by saying that the coset
$[Q^\alpha]$ is parallel transported, that is
\be
\left[\frac{D Q^\alpha}{D \lambda} \right]= [0]\,.
\ee
%Notation for coset?

\subsubsection{Massless bosons}

The parallel transport of linear polarization for massless bosons,
that is photons is very similar to massless fermions, except that instead of
projecting a vector we must project a tensor~\citep{Challinor:1999zj,Challinor2000,Tsagas2007,Pitrou2008,Pitrou2008GRG}. We define polarization
on the full spacetime as in Eq. (\ref{QalphaQmu}), that is
\be\label{PalphaPmu}
{\polar}^{\alpha\beta} = {\polar}^{\mu\nu} [e_\mu]^\alpha  [e_\nu]^\beta  \,.
\ee
Similarly, the non-physical degree of freedom must be projected and
the evolution of linear polarization is dictated by
\be\label{LiouvillePolarPhoton}
{{\cal T}^{\beta \beta'}}_{\alpha\alpha'} \frac{D
  \polar^{\alpha \alpha'}}{D \lambda} = 0,\,\,\,\,\,\Rightarrow\,\,\,\, \left[\frac{D
  \polar^{\alpha \alpha'}}{D \lambda} \right]= [0]\,.
\ee
Note that for massless bosons, we need not postulate this equation as
it is obtained from the eikonal approximation of electromagnetism, see
e.g. \citet{FleuryPhD} for a detailed account on the procedure.

\subsection{Quantum evolution in the interaction picture}

So far we have discussed the free propagation of fermions or bosons. When in addition considering collisions, we will employ a separation of scales. We assume that the relativistic evolution is dominant on macroscopic scales, while individual collisions act on microscopic scales. We therefore may compute the collision term in the local tangent space corresponding to special relativity. Then averaging over the local Minkowski space-time of the observer we will provide an effective collision term for the relativistic evolution of the distribution functions.

\begin{figure}[!htb]
  \center
  \includegraphics[width=\columnwidth]{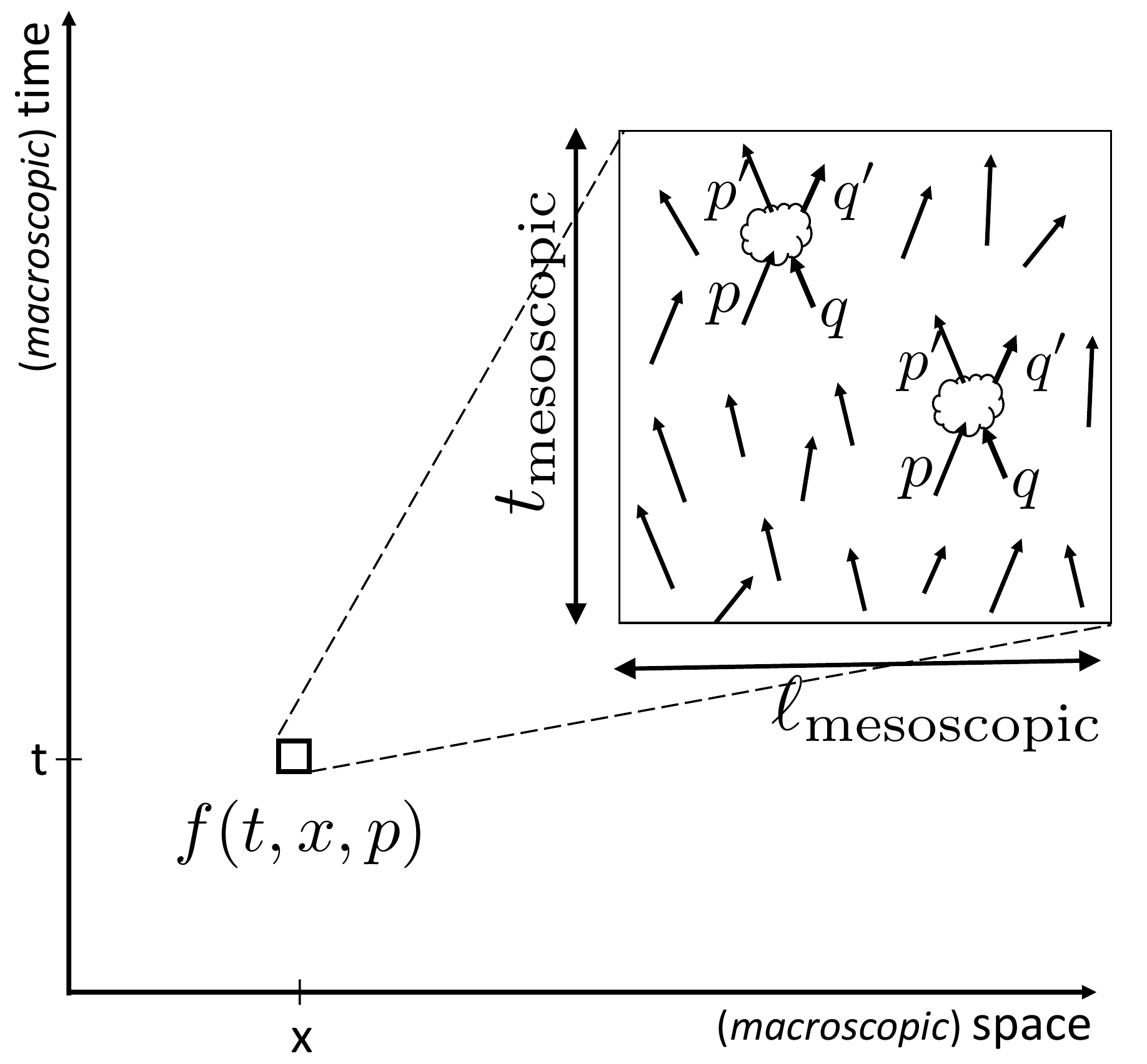}
      \caption{Illustration of the scale separation needed to derive
        the Boltzmann equation from quantum evolution. The mesoscopic
        scale (for both time and space) is much larger than the microscopic
        one, hence allowing to compute the expression of the
        collisions in a Minkowski space-time, as is standard in particle
        physics. Conversely the mesoscopic scale is much smaller than
        the macroscopic scale of general relativity, allowing to
        define a local distribution function and the associated local collision term governing its evolution.} 
      \label{FigMeso}
\end{figure}

We therefore introduce three separate scales, the microscopic scale of
individual interactions, typically the Compton timescale of
interacting particles. Then a mesoscopic scale over which we average
the individual collisions, define our local distribution functions and
describe the impact of the collisions on the averaged fluid. Finally,
the macroscopic scale on which particles free stream on general
relativistic geodesics. This separation of scales is illustrated in Fig.~\ref{FigMeso}.

We begin with the description of collisions in the local frame of our observer. The full Hamiltonian $H$ can be separated into a free part $H_0$ and an interaction part $ H_{\rm I}$. We employ the Heisenberg picture in which the states are time-independent. The time evolution of our distribution function is given by (omitting to specify the momentum dependence of $f_{rs}$ and $N_{rs}$ for simplicity)
\be\label{dfdtfromdNdt}
\deltarel(0)\frac{\dd}{\dd t}f_{rs} = \langle \Psi | \frac{\dd N_{rs}}{\dd t}  | \Psi \rangle = \ii \langle \Psi |  [H_{\rm I}, N_{rs}] |\Psi\rangle \,.
\ee 
We find a differential equation for the operator $N_{rs}$ and are able
to write an approximate solution as closed integration if we restrict
ourselves to a given order in the interaction Hamiltonian. The details
can be found in \cite{FidlerPitrou} and are also summarized in
appendix \ref{SecConstructionCollision}. They require to separate
between the microscopic scales of the quantum collisions and the
macroscopic scales of the classical Boltzmann transport description.

Eventually, defining the collision term as
\bea\label{eq:defcoll}
&&\deltarel(0) C[f_{rs}(t)]\equiv \\
&&-\frac{1}{2} \langle \Psi(t) | \int_{-\infty}^{\infty} \dd t_{\rm mic} [H_I(t),[H_I(t+t_{\rm mic}), N^{(0)}_{rs}]] |\Psi(t)\rangle\nonumber
\eea
the evolution of $f_{rs}$ is then ruled by the Boltzmann equation
\be\label{QuantumBoltzmann2}
\deltarel(0) \frac{\dd f_{ss'}(t,p)}{ \dd t } = \deltarel(0) C[f_{ss'}(t,p)]\,.
\ee

In the case of fermions, a spinor space operator associated with this collision term is obtained by contraction with $u_{s'}(p) \bar u_s(p)$ (or $v_{s}(p) \bar v_{s'}(p)$ for antiparticles) as in Eq. (\ref{Fspinor}), and we define
\be\label{DefCollisionOperator}
C[\gr{F}(t,p)] \equiv \begin{cases}\sum_{s\,s' }C[f_{ss'}(t,p)]u_{s'}(p) \bar u_s(p)\,,\,\,\, {\rm part.}\\\sum_{s\,s' }C[f_{ss'}(t,p)]v_{s}(p) \bar v_{s'}(p)\,,\,\,\,\, {\rm antipart}\,.\end{cases}
\ee
The covariant parts of this spinor space collision operator, $I_C(p)$
and ${\cal Q}_C^\mu(p)$ are obtained exactly like in
Eq.~(\ref{MasterFermionDecomposition}). In the massless case the covariant
parts are $I_C(p)$, $V_C(p)$ and ${Q}_C^\mu(p)$ and are obtained as in
Eq.~(\ref{BigDecompositionGoodMassless}). For massless bosons
(photons), a tensor valued collision function is built as in Eq.~(\ref{Deffmunumassless}).

The classical Boltzmann equation is obtained when considering that
this derivation, which has been made for a homogeneous system is in
fact valid locally. That is in the derivation
we assumed that the distribution function depends on time and momentum
only $f_{rs}(t,p)$, but we now assume that it also depends on the
position and employ $f_{rs}(t,\gr{x},p)$. This amounts to considering
that under the mesoscopic length scale the system can be
considered as homogeneous (see also Fig.~\ref{FigMeso}), such that the volume integral in the
Hamiltonian $H_I=\int \dd^3 \gr{x} {\cal H}_I$ can be extended to
infinity in the computation of the local collision term
$C[f_{ss'}(t,\gr{x},p)]$. Expressed in terms of spinor valued or
tensor valued operators the classical Boltzmann equation reads 
\be\label{dgrFdt}
\frac{\dd \gr{F}(t,\gr{x},p)}{ \dd t } =C[\gr{F}(t,\gr{x},p)]\,.
\ee
Finally, in order to include this collision term in the right hand side
of the Liouville equation in curved space-time discussed in
\S~\ref{SecLiouvilleGR}, one must multiply it by $\dd t /\dd \lambda =
p^0=E$. This converts the collision term, seen as a rate of change of
the distribution function per unit of proper time of the observer in
the tetrad frame, to a collision term which is a change of the
distribution function per unit of the affine parameter $\lambda$. In
practice the Boltzmann equation obtained needs to be converted again
to an equation giving the change of the distribution function per unit of a generalized time
coordinate, and this final step requires a specific form of the
metric.

\subsection{Molecular chaos}\label{SecChaos}

In principle, when considering an interacting system, the one-particle
distribution function is not enough to describe it statistically,
because $n$-particle correlation functions are generated by
collisions. In order to obtain a description only in terms of a one-particle
distribution function, we must assume that the connected part of the
$n$-particle functions vanishes and thus that $n$-particle functions
are expressed only in terms of one-particle functions, corresponding
to the assumption of molecular chaos. We review how this assumption is
implemented in this section.

Let us introduce a multi-index notation which encodes
both the helicity index and the momentum, and which consists in using
$s$ for $(s,p)$ or $s'$ for $(s',p')$. With this notation we write for
instance $a_{s'}$ instead of $a_{s'}(p')$. We also introduce a
generalized delta function on both helicities and momenta which is
\be
\deltarel_{s s'} \equiv \delta^{\rm K}_{s s'} \deltarel(p-p')\,.
\ee 
In particular the number operator (\ref{defNrs}) is noted
\be
N_{s s'} \equiv N_{s s'}(p,p') = a^\dagger_{s}(p) a_{s'}(p')\,.  \nonumber
\ee
For fermions, we get from anticommuting rules
\be\label{Magic4}
a_{s'}a^\dagger_{s} = \left(\deltarel_{s s'}-
  N_{s s'}\right) \equiv \widehat{N}_{s s'}\qquad 
a^\dagger_{s} a_{s'}=N_{s s'}\,,
\ee
which defines the Pauli blocking operator $\widehat{N}_{s s'}$.
Similarly for bosons, we get from commutation rules
\be\label{Magic4Boson}
a_{s'}a^\dagger_{s} = \left(\deltarel_{s s'}+
  N_{s s'}\right) \equiv \widehat{N}_{s s'}\qquad 
a^\dagger_{s} a_{s'}=N_{s s'}\,,
\ee
which defines the stimulated emission operator $\widehat{N}_{s s'}$.
The $n$-particle number operators for species $a$ are defined as
\be
N^{(n)}_{r_1 \dots r_n s_1 \dots s_n} \equiv a^\dagger_{r_1}\dots a^\dagger_{r_n}a_{s_1} \dots a_{s_n}\,.
\ee
Under the molecular chaos assumption, their expectation
value for fermions in a quantum state is related to the expectation value of the number operator as\footnote{For bosons we remove all minus signs, that is the factor $(-1)^{n+1}\epsilon(\sigma)$  so we would for instance get
$
\langle N^{(2)}_{r_1 r_2 s_1 s_2} \rangle = \langle N^{(1)}_{r_1 s_1}
\rangle \langle N^{(1)}_{r_2 s_2} \rangle + \langle N^{(1)}_{r_1 s_2}
\rangle \langle N^{(1)}_{r_2 s_1} \rangle
$.
}
\be
\langle N^{(n)}_{r_1 \dots r_n s_1 \dots s_n} \rangle = \sum_{\sigma \in S_n}(-1)^{n+1} \epsilon(\sigma) \langle N^{(1)}_{r_1 s_{\sigma(1)}} \rangle \dots \langle N^{(1)}_{r_n s_{\sigma(n)}}\rangle\nonumber
\ee
where the sum is on the group of permutation $S_n$ and $\epsilon(\sigma)$ is the signature of the permutation.
This approximation is exactly similar to the Boltzmann approximation of the BBGKY hierarchy~\citep{Volpe:2015rla}. In practice this assumption of molecular chaos is used to obtain the following property for the expectation in a quantum state of a product of one-particle number operators
\be\label{MagicChaos}
\langle N^{(1)}_{r_1 s_1}\dots N^{(1)}_{r_n s_n} \rangle =
\sum_{\sigma \in S_n}\langle \{N^{(1)}_{r_1 s_{\sigma(1)}}\}\rangle
\dots \langle \{N^{(1)}_{r_n s_{\sigma(n)}}\}\rangle \nonumber
\ee
\be
\{N^{(1)}_{r_a s_b}\}= \begin{cases}N^{(1)}_{r_a s_b}\quad a\leq b\\\widehat{N}^{(1)}_{r_a s_b}\quad a > b\end{cases}\,,\nonumber
\ee
where $\widehat{N}_{rs}$ is either the Pauli blocking operator (for fermions) or the
stimulated emission operator (for bosons) defined by Eqs.~(\ref{Magic4}) and (\ref{Magic4Boson}). The expectation value of a product of one-particle number operators is the sum of products of expectation values of all possible pairings between creation and annihilation operators. For each pair, if the indices $(r_a,s_b)$ correspond to operators which were initially in the creation-annihilation order, that is with $a\leq b$ (resp. annihilation-creation order, that is $a>b$) we use $N_{r_a s_b}$ (resp. $\widehat{N}_{r_a s_b}$). For instance the expectation value for a product of two one-particle number operators is simply
\be
\langle N^{(1)}_{r_1 s_1}N^{(1)}_{r_2 s_2} \rangle  = \langle N^{(1)}_{r_1 s_1}\rangle \langle N^{(1)}_{r_2 s_2} \rangle + \langle N^{(1)}_{r_1 s_2}\rangle \langle \widehat{N}^{(1)}_{r_2 s_1} \rangle\,.
\ee
Finally, we also assume that species are uncorrelated such that the expectation value for operators of various species is the product of expectation values of the operators of each species. For instance for two species $a$ and $b$ we assume $\langle a_r^\dagger a_s b_p^\dagger b_q \rangle=\langle a_r^\dagger a_s \rangle \langle b_p^\dagger b_q \rangle$.

%%%%%%%%%%%%%%%%%%%%%%%%%%%%%%%%%%%%%%%
\mypart{2}{Weak interactions}
%%%%%%%%%%%%%%%%%%%%%%%%%%%%%%%%%%%%%%%

The Fermi theory of weak interactions is a contact interaction between
four fermions. All reactions in that approximation are of the type
\be\label{Eqabcd}
a+b \leftrightarrow c+d\,,
\ee
with other reactions involving antiparticles ($\bar a, \bar b ,\bar
c,\bar d$) deduced from charge conjugation or crossing symmetry. In the next section we derive the
general collision term for general weak currents and apply it to the
case of neutrino interactions which is relevant for the early
universe. In \S~\ref{SecFokkerPlanck} we apply it to the case of neutron-proton
conversions by weak interactions which controls the primordial Helium abundance.

\section{General collision term}\label{SecWeakInteraction}

\subsection{Fermi theory of weak interactions}\label{SecFermiTheory}

All weak interaction take the form of current-current interactions~\cite{Nachtmann1991} at low energy (low compared to the $W^\pm$ and $Z$ masses), that is they are given by
\be\label{StructureHI}
{\cal H}_I = -{\cal L}_I = \sgnslash\frac{G_F}{\sqrt{2}}\left[J^{\rm NC}_\mu
  J_{\rm NC}^\mu + J^{{\rm CC} \dagger}_\mu J_{\rm CC}^\mu
\right]\,,
\ee
where $G_F\simeq 1.1663787 \,\times\, 10^{-5}\,{\rm GeV}^{-2}$ is the Fermi constant of weak interactions.

\subsubsection{Neutral currents}

Neutral currents describe the exchange of $Z$ bosons and as these are not charged they mediate elastic scatterings that do not alter the involved types of particles and only transfer momentum, spin and energy. 

The neutral current is simply the sum of the neutral currents of all
particles undergoing weak interactions
\be\label{StructureJNC}
J_{\rm NC}^\mu = J^\mu_{ee} + J^\mu_{\nu\nu} +\dots\,.
\ee
For neutrinos, the neutral current couples only the left chiralities and, noting ${\bm \nu}$ the neutrino quantum field, it is simply
\be\label{JNCneutrinos}
 J^\mu_{\nu\nu} \equiv e^\nu_- \,\bar{\gr{\nu}} \gamma^\mu(\mathds{1}-\gamma_5)\gr{\nu}\,,\qquad e^\nu_- \equiv \frac{1}{2}\,.
\ee
\com{\CF{we have not defined $\gr{\nu}$, $e^\nu_-$ etc.. do you think it is standard enough so that everyone knows what we mean? }}
with similar expressions for other flavors. However for electrons and
(similarly pions and taus) the neutral currents must be further
decomposed into left and right chiral interactions as
\beas\label{JNCelectrons}
J^\mu_{ee} &=& \epsilon^e_-  J^{-\,\mu}_{ee}+ \epsilon^e_+
J^{+\,\mu}_{ee}\,,\\
J^{-\,\mu}_{ee} &\equiv& \bar{\gr{e}}
\gamma^\mu(\mathds{1}-\gamma_5)\gr{e} \,,\\
 J^{+\,\mu}_{ee} &\equiv& \bar{\gr{e}} \gamma^\mu(\mathds{1}+\gamma_5)\gr{e} \,,
\eeas
where we noted ${\bm e}$ the electronic quantum field. The chiral coupling constants are for electrons
\be\label{ChiralCouplings}
\epsilon^e_- \equiv -\frac{1}{2} + \sin^2\theta_W \,,\qquad \epsilon^e_+ \equiv \sin^2\theta_W \,,
\ee
with $\theta_W$ the Weinberg angle ($\sin^2 \theta_W \simeq 0.23$).

\subsubsection{Charged currents}

Opposed to the neutral currents, the charged currents describe the exchange of charged $W$-bosons and therefore are inelastic.  
The structure of the charged current is more complex since it couples eigenmass states of different flavors, thanks to the
Cabbibo-Kobayashi-Maskawa (CKM) matrix for quarks or the
Pontecorvo-Maki-Nakagawa-Sakata (PMNS) matrix for massive neutrinos. We ignore these complications for the examples that we shall consider and employ effective charged currents for the neutron/proton pair
which is involved in beta decays and related processes, and the
charged currents of the first two lepton flavors, that is of the
electron/neutrino and muon/muon neutrino pairs. We use
\be
J_{CC}^\sigma = V_{ud}  J^\sigma_{pn} +  J^\sigma_{e \nu} +
J^\sigma_{\mu \nu_\mu }\,,
\ee
where $V_{ud} =  0.97420(20)$ is a Cabbibo-Kobayashi-Maskawa (CKM) angle~\cite{PDG17}.

The charged currents for electron/neutrino and muon/muon neutrino pairs
are coupling only the left chiralities
\be\label{JCCdetail}
J^\sigma_{e \nu} \equiv  \bar{\bm \nu}\gamma^\sigma
(\mathds{1}-\gamma^5) {\gr{e}}\,,\qquad  J^\sigma_{\mu \nu_\mu} \equiv
\bar{\bm \nu}_\mu\gamma^\sigma (\mathds{1}-\gamma^5) {\gr{\mu}}\,.
\ee
However, due to internal QCD effects, the coupling in the
neutron/proton pair is not purely left chiral. The deviation from left chirality of the
coupling is parameterized by the $g_A$ parameter whose measured value
is approximately $1.2723(23)$~\citep{PDG17} and the corresponding charged current reads
\be\label{JCCpn}
J^\mu_{pn} \equiv \bar{\gr{p}} \gamma^\mu (\mathds{1}- g_A \gamma^5) \gr{n}\,.
\ee

When considering the cumulative effect of neutral currents and
charged currents, we can use the Fierz identities which for anticommuting fields give~\cite{Sarantakos:1982bp,1993NuPhB.406..423S}
\be\label{CrossingSymmetry}
 J^{\dagger\mu}_{e\nu} {(J_{e\nu})}_\mu\ = J^{-\,\mu}_{ee} {(J_{\nu\nu}^{-})}_\mu \,.
\ee
This means that the effect of multiple charged currents can be replaced by equivalent neutral currents. In the collision term we may therefore replace the charged currents by modifying the neutral chiral coupling factors  (\ref{ChiralCouplings}), yielding
\be\label{epsilonmoinsplusun}
\epsilon^e_- \to \epsilon^e_- +1\,. 
\ee

\subsection{Two-body processes} \label{SecGeneralabcd}

\subsubsection{Notation}

We use the compact notation introduced in \S~\ref{SecChaos} that we
adapt to also account for the fact that we have several different
species in the reaction (\ref{Eqabcd}). We introduce
\be\label{DefMultiIndex}
\alpha \equiv (s_\alpha, p_\alpha) \qquad \alpha' \equiv (s'_\alpha, p'_\alpha)\,.
\ee
These multi-indices contain all information characterising one single particle (its momentum and helicity). We will typically label ingoing states as unprimed and outgoing states with primed indices. For species $a$ we employ the multi-index $\alpha$ and 
similarly for species $b$ (resp. $c$ and $d$) we use the multi-indices $\beta$ (resp. $\gamma$ and $\delta$). The plane wave solutions are written in a compact form in this
notation. For instance for the species $a$ we write $u_\alpha \equiv u_{s_\alpha}(p_\alpha)$ and $v_\alpha \equiv v_{s_\alpha}(p_\alpha)$.
Furthermore this allows to write a compact relativistic Dirac delta function
which acts both on helicities and momenta as
\be
\deltarel_{\alpha \alpha'} \equiv \delta^{\rm K}_{s_\alpha s'_\alpha} \deltarel(p_\alpha-p'_\alpha)\,.
\ee
We denote the number operator associated with species $a$ as
\be
A_{\alpha \alpha'} \equiv N_{s_\alpha s'_\alpha}(p_\alpha,p'_\alpha) = a^\dagger_{s_\alpha}(p_\alpha) a_{s'_\alpha}(p'_\alpha)\,. 
\ee
We also define the Pauli blocking operator
\be\label{DefHatted}
\widehat{A}_{\alpha \alpha'} \equiv \deltarel_{\alpha \alpha'}-{A}_{\alpha \alpha'}\,.
\ee
The expectation value of these operators is denoted as
\beas
\langle A_{\alpha \alpha'}\rangle &=& \deltarel(p_\alpha-p'_\alpha)
  {\cal A}_{\alpha \alpha'}(p_\alpha)\,,\\
\langle \widehat{A}_{\alpha \alpha'}\rangle &=& \deltarel(p_\alpha-p'_\alpha) \widehat{\cal A}_{\alpha \alpha'}(p_\alpha)\,,
\eeas
where we introduce the short-hand notation ${\cal A}_{\alpha
  \alpha'}(p_\alpha) = {\cal A}_{s_\alpha
  s'_\alpha}(p_\alpha)$. We recall that this quantity is exactly the one-particle distribution
function associated with species $a$ [see Def.~\ref{AverageNrs}]. Note that for the Pauli blocking factor, $\widehat{\cal A}_{\alpha
  \alpha'}(p_\alpha)$ is a shorthand notation for $\delta^{\rm K}_{s_\alpha
  s'_\alpha} - {\cal A}_{s_\alpha  s'_\alpha}(p_\alpha)$. We
associate to the one-particle distribution function (resp. the Pauli blocking function) a spinor valued operator following the
procedure (\ref{Fspinor}) that we note ${A_\mathfrak{a}}^\mathfrak{b}$ (resp. ${\widehat{A}_\mathfrak{a}}^{\,\,\mathfrak{b}}$) in component notation or simply $\gr{A}$ (resp. $\widehat{\gr{A}}$) in operator notation. Having defined for species $a$ the number operator $A_{\alpha \alpha'}$, the distribution function ${\cal A}_{\alpha \alpha'}$ and the spinor-valued (observer-independent) operator $\gr{A}$, we proceed identically for species $b$ (resp. $c$ , $d$) and we use $B_{\beta \beta'}$, ${\cal B}_{\beta \beta'}$ and $\gr{B}$ (resp. $C_{\gamma  \gamma'}$, ${\cal
  C}_{\gamma \gamma}'$ and $\gr{C}$, $D_{\delta \delta'}$, ${\cal
  D}_{\delta \delta'}$ and $\gr{D}$), and associated hatted notations for Pauli blocking factors.
Furthermore, for the antiparticles species $\bar a,\bar b,\bar c,\bar
d$ related to the species $a,b,c,d$, we use barred notation for number
operators (e.g. $\overline{A}_{\alpha \alpha'}$), distribution function
(e.g. $\overline{\cal A}_{\alpha \alpha'}$) and spinor valued operators (e.g. $\overline{\gr{A}}$), along with their hatted versions for Pauli blocking terms. Finally we define the collision term as in Eq. (\ref{eq:defcoll}), that is
\be\label{DefCollisionTerm}
\deltarel(0) C[{\cal A}_{ss'}(p)]\equiv -\frac{1}{2} \langle \int_{-\infty}^{\infty} \dd t' [H_I(0),[H_I(t'), A_{ss'}(p)]]
\rangle 
\ee
such that the quantum Boltzmann equation (\ref{QuantumBoltzmann}) for species $a$ is written as (when neglecting forward scattering)
\be\label{QuantumBoltzmann3}
\deltarel(0) \frac{\dd {\cal A}_{ss'}(p)}{ \dd t } = \deltarel(0) C[{\cal
A}_{ss'}(p)]\,.
\ee

\subsubsection{Collision term structure}\label{SecCollStructure}

Following the previous discussion, our goal is to compute the collision term $C[{\cal
A}_{ss'}(p)]$ corresponding to the reaction (\ref{Eqabcd}) due to weak
interactions. It is mediated by an Hamiltonian density of the form
\be\label{HIabcd}
{\cal H}_I = -{\cal L}_I =\sgnslash g\left( J^{ac}_\mu  J^\mu_{bd}+ {\rm cc} \right)\,,
\ee
where, depending on the interaction, the same species may be represented by multiple indices. The chiral contributions of
these currents are parameterized by $\epsilon^{ac}_\pm$ and $\epsilon^{bd}_\pm$ as
\be\label{GeneralCurrents}
 J^\mu_{ac} =\overline{\gr{\psi}}_c {\cal \chi}^\mu_{(ac)}\gr{\psi}_a \,, \qquad\qquad 
J^\mu_{bd} =\overline{\gr{\psi}}_d {\cal \chi}^\mu_{(bd)} \gr{\psi}_b\,,
\ee
with the notation
\bea\label{Omu}
{\cal \chi}^\mu_{(ac)} &\equiv& \epsilon^{ac}_+ \gamma^\mu(\mathds{1}+\gamma^5) + \epsilon^{ac}_- \gamma^\mu(\mathds{1}-\gamma^5)\,,\\
{\cal \chi}^\mu_{(bd)} &\equiv& \epsilon^{bd}_+ \gamma^\mu(\mathds{1}+\gamma^5) + \epsilon^{bd}_- \gamma^\mu(\mathds{1}-\gamma^5)\,.
\eea

The interaction Hamiltonian associated to the Hamiltonian density (\ref{HIabcd}) is explicitly given by
\bea\label{HIofM}
&&{H}_I^{a+b\leftrightarrow c+d} \equiv \int [\dd p_\alpha] [\dd p_\beta] [\dd p_\gamma] [\dd p_\delta]
(2\pi)^3 {\cal M}^{a+b\leftrightarrow c+d} \nonumber\\
&&\qquad\times\delta(\gr{p}_\alpha+\gr{p}_\beta-\gr{p}_\gamma-\gr{p}_\delta){\rm e}^{-\ii
  (p_\alpha^0+p_\beta^0-p_\gamma^0-p_\delta^0) t}
\eea
where we used the scattering operator for this reaction
\bea
&&{\cal M}^{a+b\leftrightarrow c+d}  \\
&&\equiv \sum_{\rm spins}\left(d^\dagger_\delta c^\dagger_\gamma b_\beta a_\alpha M_{\alpha\,\beta\to \gamma\,\delta}
  +b^\dagger_\beta a^\dagger_\alpha d_\delta c_\gamma M_{\gamma\,\delta\to \alpha\,\beta} \right)\,.\nonumber
\eea
The $M$ matrices are defined with the multi-index notation (\ref{DefMultiIndex})
\be
M_{\alpha\,\beta\to \gamma\,\delta}  \equiv  M[(s_\alpha,p_\alpha)\,(s_\beta,p_\beta)\to
  (s_\gamma,p_\gamma)\,(s_\delta,p_\delta)] 
\ee
and for weak interactions they are of the general form
\bea
M_{\alpha\,\beta\to \gamma\,\delta}  &\equiv& g  [\bar u_\gamma{\cal \chi}^\mu_{(ac)} u_\alpha ][ \bar u_\delta {\cal \chi}^{(bd)}_\mu
u_\beta] \\
M_{\gamma\,\delta\to \alpha\,\beta}  &=&M^\star_{\alpha\,\beta\to
  \gamma\,\delta}  =g  [\bar u_\alpha {\cal \chi}^\mu_{(ac)} u_\gamma ][ \bar u_\beta {\cal \chi}_\mu^{(bd)} u_\delta]\,.\nonumber
\eea

To compute the collision term we first need to compute the operator $
[{\cal M},[{\cal M},A_{ss'}]] $. Using the commutation rules of
Appendix B in \citet{FidlerPitrou}, and using the molecular chaos assumption described in \S~\ref{SecChaos}, we get
\bea\label{MMAresult}
&&
 [{\cal M},[{\cal M},A_{ss'}]]  = M^\star_{\alpha\,\beta\to
   \gamma'\,\delta'}M_{\alpha'\,\beta'\to \gamma\,\delta}\\
 &&\left\{\deltarel_{s \alpha'} \left[B_{\beta \beta'}A_{\alpha s'} 
\widehat D_{\delta \delta'} \widehat C_{\gamma \gamma'}  -D_{\delta
  \delta'} C_{\gamma \gamma'}
\widehat B_{\beta \beta'} \widehat A_{\alpha
  s'}\right]\right.\nonumber\\
&&\left.+\deltarel_{\alpha s'} \left[B_{\beta \beta'}A_{s \alpha'} 
\widehat D_{\delta \delta'} \widehat C_{\gamma \gamma'}  -D_{\delta
  \delta'} C_{\gamma \gamma'}
\widehat B_{\beta \beta'} \widehat A_{s\alpha'}\right]\right\}\,.\nonumber
\eea
We now employ this result in Eqs. (\ref{HIofM}) and (\ref{DefCollisionTerm}).
We integrate a total of five momentum integrals (each one being itself three-dimensional in
momentum space) using the Dirac distributions. Of these, four Dirac functions are contained in the expectation values
of the number operators associated to the four species, and there is
an extra Dirac function ($\deltarel_{s \alpha'}$ or $\deltarel_{\alpha
  s'}$ in Eq.~(\ref{MMAresult})) from the collision
term ensuring local energy and momentum conservation. Eventually,
taking the expectation in the quantum state, we get
\begin{widetext}
\bea\label{ColAss}
 2 E C[{\cal A}_{ss'}(p)]&=& {\cal K} M^\star[(s_\alpha,p)\,(s_\beta,p_\beta)\to (s'_\gamma,p_\gamma)\,(s'_\delta,p_\delta)]M[(s'_\alpha,p)\,(s'_\beta,p_\beta)\to (s_\gamma,p_\gamma)\,(s_\delta,p_\delta)]\nonumber\\
&&\left\{\delta^{\rm K}_{s\alpha'}\left[-{\cal B}_{\beta \beta'}(p_\beta){\cal A}_{\alpha s'}(p) 
\widehat{\cal D}_{\delta \delta'}(p_\delta) \widehat{\cal
  C}_{\gamma\gamma'}(p_\gamma) +{\cal D}_{\delta \delta'}(p_\delta)
{\cal C}_{\gamma \gamma'}(p_\gamma)
\widehat{\cal B}_{\beta \beta'}(p_\beta) \widehat{\cal A}_{\alpha s'}(p)\right]\right.\nonumber\\
&&\left. +\delta^{\rm K}_{\alpha s'}
\left[-{\cal B}_{\beta \beta'}(p_\beta){\cal A}_{s \alpha'}(p) 
\widehat{\cal D}_{\delta \delta'}(p_\delta) \widehat{\cal C}_{\gamma
  \gamma'}(p_\gamma)  +{\cal D}_{\delta \delta'}(p_\delta) {\cal
  C}_{\gamma \gamma'}(p_\gamma)
\widehat{\cal B}_{\beta \beta'}(p_\beta) \widehat{\cal A}_{s\alpha'}(p)\right]\right\}\,,
\eea
\end{widetext}
with the integration on momenta
\be\label{calK}
{\cal K} \equiv \frac{1}{2} \int [\dd p_\delta][\dd p_\gamma][\dd
p_\beta](2\pi)^4\delta^{(4)}(p_\delta+p_\gamma-p_\beta-p) \,.
\ee
We note that:
\begin{itemize}
\item The collision term is made of two types of terms. The first terms on
the second and the third line of Eq.~(\ref{ColAss}) correspond to
\emph{scattering out} processes, that is collisions which due to the minus sign deplete the
distribution function associated with species $a$ and they correspond
to $a+b \to c+d $. The second term on the second and third line
correspond conversely to \emph{scattering in} processes, which increase the
distribution function of species $a$, and they are due to the reaction $c+d \to
a+b$. 

\item For scattering out processes, the collision term is proportional
to the distribution function of the initial states (species $a$ and
$b$), but also to the Pauli blocking function of the final states
(species $c$ and $d$), and the reverse is true for the scattering in processes.

\item The distribution functions are Hermitian, that is ${\cal  A}^\star_{s s'}(p) = {\cal A}_{s' s}(p)$ as in
  Eq.~(\ref{HermitianProperty}). Let us now consider
  $C[{\cal A}_{ss'}(p)]^\star$. Given the Hermiticity of the distribution functions and thus
  of the Pauli blocking functions, with a simple renaming of all
  primed indices as unprimed indices (and also of unprimed
  indices as primed indices), it is straightforward to show that this is equal to $C[{\cal A}_{s's}(p)]$, hence the collision term is
  also Hermitian as expected.

\item In the previous computation when checking the Hermiticity, the second and third line of
  Eq. (\ref{ColAss}) are interchanged. Terms of the second line are
  proportional to $\delta^{\rm K}_{s\alpha'}$ and correspond
  physically to the scattering of the helicity index $s'$, and
  conversely in the third line the terms are proportional to
  $\delta^{\rm K}_{\alpha s'}$ and it corresponds to the scattering of
  the helicity index $s$. Hence we see that the collision term
  possesses four terms corresponding to the \emph{in/out} contributions and
  the $s/s'$ contributions.

\item Finally even though we computed the collision term for a
  homogenous system in a Minkowski space-time, the total volume, which appears as
  $\delta^{(3)}(0)$, drops out from both the left and the right hand
  side of Eq.~(\ref{QuantumBoltzmann3}). Hence, as argued before Eq.~(\ref{dgrFdt}), we can consider that this collision
  term is valid locally, allowing us to consider in a classical
  macroscopic description that all distribution functions should be
  considered with a dependence on the point of space-time. We started
  a computation with total number of particles in a quantum system,
  but we end up using it with \emph{number densities} of particles,
  considering that the collisions are point-like.

\end{itemize}

The procedure to follow is now transparent. The helicity indices of
the distribution functions (or the related Pauli blocking functions)
are contracted with the plane waves solutions contained in the $M$
matrices. From Eqs.~(\ref{DefFparticles}) this is exactly what is needed to build the
spinor space operators related to each species. Since only the indices
$s$ and $s'$ remain uncontracted in Eq.~(\ref{ColAss}), we contract them
with $u_{s'}(p) \bar u_s(p)$ (or $v_{s}(p) \bar v_{s'}(p)$ for antiparticles) so as to form a spinor space collision operator $C[\gr{A}(p)] $ as specified in the definition~(\ref{DefCollisionOperator}). 
\com{\be
C[\gr{A}(p)] \equiv \sum_{s\,s' }C[{\cal A}_{ss'}(p)]u_{s'}(p) \bar u_s(p)\,.
\ee}
Note that the contraction of $\delta^{\rm
  K}_{ss'}$ with $u_{s'}(p) \bar u_s(p)$ or $v_{s}(p) \bar
v_{s'}(p)$ gives simply
\be\label{IdentityFermions}
\mathds{1}(p) \equiv \sgnslash\slashed{p}+M\,,
\ee 
with the notation (\ref{defMmass}), as can be seen from Eqs.~(\ref{Sumusus}).  We finally obtain the structure of the collision term
\bea\label{GenCA}
&&E \,C[\gr{A}(p)] = -[\mathbb{1}(p) \cdot{\bm K} \cdot
\gr{A}(p)+\gr{A}(p)\cdot{\bm K} \cdot \mathbb{1}(p) ]
\nonumber\\
&&\qquad \qquad+ [\mathbb{1}(p) \cdot\widehat{\bm K} \cdot \widehat{\gr{A}}(p)+\widehat{\gr{A}}(p)\cdot\widehat{\bm K} \cdot \mathbb{1}(p) ],
\eea
where ${\bm K}={\bm K}[\gr{B},\widehat{\gr{C}},\widehat{\gr{D}}]$ is
an operator depending on other species distribution functions
integrated over momenta, and $\widehat{\bm K}={\bm
  K}[\widehat{\gr{B}},{\gr{C}},{\gr{D}}]$ is its hatted version. Its
expression is
\bea
&&\gr{K}(\gr{B}, \widehat{\gr{C}},\widehat{\gr{D}}) \equiv \\
&& \frac{g^2}{2} {\cal K}\left\{{\rm Tr}[\gr{B} \cdot {\cal \chi}_\mu^{(bd)}  \cdot \hat{\gr{D}} \cdot
  {\cal \chi}_\nu^{(bd)} ]{\cal \chi}^\mu_{(ac)}  \cdot \widehat{\gr{C}} \cdot {\cal
    \chi}^\nu_{(ac)}\right\}\,,\nonumber
\eea
where the momentum dependence $\gr{A}(p)$ and $\gr{B}(p_\beta)$,
$\gr{C}(p_\gamma)$,  $\gr{D}(p_\delta)$ (and similarly for Pauli
blocking operators) are omitted for a more compact
notation. Since$\gr{K}$, $\gr{A}$ and  $\mathds{1}$ are all Hermitian, it
is obvious from Eq.~(\ref{GenCA}) that so is the collision
term. Furthermore, its structure is again manifest. The first line
corresponds to scattering out processes. As for the second line, it
corresponds to the scattering in processes, and differs only by an
overall sign and the exchange of the distribution and Pauli blocking functions.

This collision term $C[\gr{A}]$, being itself an operator in spinor
space, can be decomposed into its covariant parts $I_{C[A]}$ and ${\cal Q}_{C[A]}^\mu$ as in the decomposition~(\ref{MasterFermionDecomposition}).  These components can be found by
multiplying by the appropriate $X \in {\cal O}$ and taking the trace,
that is using the extraction~(\ref{DecompositionFormula}).  Since all operators involved in the collision term are made of $\gamma^\mu$ or $\gamma^5$ matrices,
the problem is reduced to taking traces of products of these
operators \citep[App.~C]{FidlerPitrou}.  This systematic computation can be handled by a computer algebra
package such as {\it xAct}~\cite{xAct} and this is particularly powerful
since it also takes care of all simplifications involving space-time indices.

In particular, when using Eqs.~(\ref{DecompositionFormula}) to extract
the intensity part of the collision term (\ref{GenCA}), we find
\bea\label{ICA}
&&E \, I_{C[A]}(p)\\ 
&&\quad=  -g^2 {\cal K}\left\{ {\rm Tr}[\gr{B} \cdot {\cal \chi}_\mu^{bd}  \cdot \hat{\gr{D}} \cdot
  {\cal \chi}_\nu^{bd} ] {\rm Tr}\left[\gr{A} \cdot {\cal \chi}^\mu_{ac}  \cdot \widehat{\gr{C}} \cdot {\cal
    \chi}^\nu_{ac}\right]\right.\nonumber\\
&&\quad\qquad\qquad\,\left.-(\gr{A} \leftrightarrow \widehat{\gr{A}},\gr{B} \leftrightarrow
\widehat{\gr{B}},\widehat{\gr{C}} \leftrightarrow
{\gr{C}},\widehat{\gr{D}} \leftrightarrow {\gr{D}})\right\},\nonumber
\eea
which is compactly written as
\be
E\,I_{C[A]}(p) = - 2{\rm Tr}[{\bm K}.\gr{A}(p)]+2{\rm Tr}[\widehat{\bm K}.\widehat{\gr{A}}(p)]\,.
\ee
Reactions related to the reaction~(\ref{Eqabcd}) by crossing symmetry are deduced by replacing
the operators describing the distributions by those of the
antiparticle, and changing distribution operators for Pauli-blocking
operator. For instance the collision term for $a+\bar c
\leftrightarrow \bar b + d$ is deduced by ${\gr{B}} \to
\widehat{\overline{\gr{B}}}$ and $\widehat{\gr{C}} \to {\overline{\gr{C}}}$, where the bar indicates that we consider
the operator associated to the antiparticles [see
Eq.~(\ref{MasterFermionDecomposition})]. Similarly the reaction $\bar a + \bar b
\leftrightarrow \bar c + \bar d$ is obtained by a global charge
conjugation, where all operators are replaced by the one associated to the antiparticle. From the decomposition
(\ref{MasterFermionDecomposition}) it is obviously equivalent to $m
\to -m$ for all masses. Finally the intensity part of the collision
term for the species $a$ in the reaction~(\ref{Eqabcd})  is the same as
the intensity part fo the species $b$~\footnote{When focusing on the polarization part of the collision term, this is no
  longer the case~\citep{FidlerPitrou}.}, and if we are to compute the
collision term for $c$ or $d$ we need only to change the global sign.

\subsection{General collision term}\label{SecGenCollTerm}

Let us now restrict to the case where all particles are
unpolarized, the general case being detailed in
\citet{FidlerPitrou}. 
More specifically, we assume that massive particles (such as electrons, positrons neutrons or protons) are unpolarized, that is
for these species ${\cal Q}^\mu=0$. For these particles we
define\footnote{The notation $f_\mu^\pm$ is obviously useless but we
  keep it as it is a particular case of the general case when species
  are polarized, which is considered in detail in
  \citet{FidlerPitrou}. Furthermore it allows to write the general
  collision term (\ref{GeneralI}).}
\bea
{f}&=&\frac{1}{2}I\,,\qquad\quad {f}^\pm_\mu = p_\mu {f}\,, \\
\widehat{f}&=&1-f\,,\qquad \widehat{f}^\pm_\mu =  p_\mu \widehat{f}\,.
\eea
However, for neutrinos, when considered as strictly massless, circular
and linear polarization are separate concepts. We still assume that
they do not have linear polarization. However, neutrinos have circular
  polarization since there are only left-helical neutrino and
  right-helical antineutrino states. We define for neutrinos
\be
{f}^\pm_\mu \equiv \frac{I \pm \lambda V}{2} p_\mu,\qquad
\widehat{f}^\pm_\mu = p_\mu - {f}^\pm_\mu\,,
\ee
where $\lambda = 1 $ for particles (neutrinos) and $\lambda = -1$ for
antiparticles (antineutrinos). In fact given the left-chirality of
weak interactions for neutrinos, we have  $V=-\lambda I$ such that the
previous definition reduces to
\be
{f}^+_\mu = 0,\,\,\, {f}^-_\mu = I p_\mu,\,\,\,\widehat{f}^+_\mu = p_\mu,\,\,\, \widehat{f}^-_\mu = p_\mu(1-I).
\ee
Hence for neutrinos we also define
\be
{f} = I \,,\qquad \widehat{f} = (1-I)\,.
\ee
$f$ is the distribution function {\it per helicity state}, which has a
clear meaning if the distribution is unpolarized, and in thermal
equilibrium it reduces to a Fermi-Dirac distribution. Since massless neutrinos exist only with left chiralities, that is left
helicities $f=I$, whereas for other fermionic massive species, $f=I/2$
since they exist in two different helicities.

Under all these restrictions and with these definitions, the intensity
part of the collision term is reduced to

\bea\label{CollisionTermIFermions}
&&E\, I_{C[A]} =\\
&&\quad 2^8 g^2 {\cal K}\left[{\cal T}_I(\widehat{\gr{A}},\widehat{\gr{B}},\gr{C},\gr{D}) - {\cal T}_I(\gr{A},\gr{B},\widehat{\gr{C}},\widehat{\gr{D}})\right]\,,\nonumber
\eea
where the Kernel takes the general form (using the generic notation (\ref{defMmass})
for masses)
\bea\label{GeneralI}
&&{\cal  T}_I(\gr{A},\gr{B},\gr{C},\gr{D})=\\
&&\phantom{+}\qquad\sum_{r=\pm}(\epsilon_r^{ac})^2(\epsilon_r^{bd})^2 \left({f}_a^r \cdot {f}_b^r\right) \left({f}_c^r \cdot  {f}_d^r\right)\nonumber\\
&&\qquad+\sum_{r=\pm}(\epsilon_r^{ac})^2(\epsilon_{-r}^{bd})^2 \left({f}_a^r \cdot {f}_d^{-r}\right) \left({f}_c^r \cdot
  {f}_b^{-r}\right)\nonumber\\
&&\qquad-\sum_{r=\pm}(\epsilon_r^{ac})^2(\epsilon_{-}^{bd}\epsilon_{+}^{bd})
\left({f}_a^r \cdot {f}_c^{r}\right) M_b M_d {f}_b {f}_d\nonumber\\
&&\qquad-\sum_{r=\pm}(\epsilon_r^{bd})^2(\epsilon_{-}^{ac}\epsilon_{+}^{ac})
\left({f}_b^r \cdot {f}_d^{r}\right) M_a M_c {f}_a {f}_c\nonumber\\
&&\qquad+4 (\epsilon_+^{ac}\epsilon_-^{ac})(\epsilon_{-}^{bd}\epsilon_{+}^{bd})M_a  M_b M_c M_d {f}_a {f}_b {f}_c {f}_d\,.\nonumber
\eea
The Kernel can be separated into a squared amplitude and a phase
space in the form
\be
{\cal  T}_I(\gr{A},\gr{B},\gr{C},\gr{D}) \equiv {f}_a  {f}_b
{f}_c {f}_d |M|^2\,,
\ee
such that the collision term (\ref{CollisionTermIFermions}) is reduced to
\bea\label{CollisionTermIFermionsUsual}
E I_{C[A]} &=& 2^8 g^2 {\cal K}\left[\widehat{f}_a \widehat{f}_b {f}_c {f}_d- {f}_a  {f}_b \widehat{f}_c \widehat{f}_d\right]|M|^2.
\eea
\subsection{Standard reactions with neutrinos}

Let us review the standard two-body reactions for neutrinos. These are
required to describe the decoupling of neutrinos in the early
universe~\citep{Dolgov1997,Mangano2005,Grohs:2015tfy,Froustey:2019owm}. We consider the
various type of reactions one by one, and we summarize the results in table~\ref{Table2}.

In the particular case that the species $a$ and $c$ are neutrinos or
antineutrinos, that is can be considered as massless, and their coupling is only
left-chiral ($\epsilon_+^{ac}=0$), we find 
\bea\label{Mgeneralparticular}
|M|^2 &=& (\epsilon_-^{ac})^2(\epsilon_-^{bd})^2 (p_a \cdot p_b) (p_c
\cdot p_d) \\
&+&(\epsilon_-^{ac})^2(\epsilon_{+}^{bd})^2 \left(p_a \cdot p_d\right)
\left(p_c \cdot p_b\right)\nonumber\\
&-&(\epsilon_-^{ac})^2(\epsilon_{-}^{bd}\epsilon_{+}^{bd})\left(p_a
  \cdot p_c\right) M_b M_d\,. \nonumber
\eea

\subsubsection{Muon decay}

The muon decay is due to the interaction between the muon ($\mu^-$)/muon
neutrino ($\nu_\mu$) charged current and the electron ($e^-$)/neutrino ($\nu$) charged current. Furthermore it involves only left-chiral
couplings. It thus corresponds to the case
\bea\label{abcd1}
&&a=\mu^-,\quad c= \nu_\mu,\quad b = \nu,\quad d = e^-\,,\\
&&\epsilon_{+}^{ac}=\epsilon_{+}^{bd}= 0\,,\quad \epsilon_{-}^{ac}=\epsilon_{-}^{bd}= 1\,,\quad g = \frac{G_F}{\sqrt{2}}\,.\nonumber
\eea
We remind that for the decay reaction $a \leftrightarrow \bar b +c
+d$, the collision term is deduced from the reaction $a+b
\leftrightarrow c+d$ by crossing symmetry. The collision term deduced
from the general form (\ref{CollisionTermIFermionsUsual}) is therefore
\bea\label{CollisionTermMuon}
E\,I_{C[A]} &=&2^7 G_F^2 {\cal K}
\left[\widehat{f}_a \overline{f}_b {f}_c {f}_d- {f}_a  \widehat{\overline{f}}_b \widehat{f}_c \widehat{f}_d\right]|M|^2,\nonumber\\
|M|^2 &=& (p_a \cdot p_b)( p_c \cdot p_d)\,,
\eea
and where it is stressed by a barred notation that the covariant
quantities related to the species $a$, $c$ and $d$ are those of
particles, and those for the species $\bar b$ are those of
antiparticles. 
The muon lifetime is recovered from this collision term
evaluated at null spatial momentum of  ($\gr{p}=0$), and ignoring
Pauli blocking effects, thanks to the definition $I_{C[A]}(\gr{p}=0) = \dd I_a/\dd t (\gr{p}=0) \equiv -\Gamma_a I_a(\gr{p}=0)$.
We get
\bea\label{Gammamuon}
\Gamma_a&=&32 G_F^2 \int [\dd p_b][\dd
p_c][\dd
p_d](2\pi)^4\delta^{(3)}(\gr{p}_d+\gr{p}_c-\gr{p}_b)\nonumber\\
&&\times\delta^{(1)}(E_d+E_c-E_b-m_a)\,E_b\, (\sgnslash p_c \cdot p_d)\,,
\eea
and this is exactly the expression that would be obtained from the Fermi golden rule.

\subsubsection{neutrino/muon neutrino scattering}\label{Secnunumuon}

The interactions between neutrinos of different types (e.g. electronic
neutrinos and muonic neutrinos) are only due to neutral currents
with a pure left chiral coupling. The effect of the reaction $\nu +
\nu_\mu \leftrightarrow \nu +
\nu_\mu$ thus corresponds to the case
\bea\label{abab1}
&&a=\nu\,,\quad b = \nu_\mu\,,\quad c=\nu\,,\quad d =
\nu_\mu\,, \quad g = 2 \frac{G_F}{\sqrt{2}}\,,\nonumber\\
&&\epsilon_-^{ac} = \epsilon_-^{bd} = e^\nu_-=\frac{1}{2}\,,\quad \epsilon_+^{ac} = \epsilon_+^{bd} = 0\,.
\eea
Using Eq.~(\ref{Mgeneralparticular}), the covariant parts of the collision term take the form
\bea\label{CollisionTermNuNumu}
E I_{C[A]} &=& 2^9 G_F^2 {\cal K} \left(\widehat{f}_a \widehat{f}_b {f}_c {f}_d-{f}_a {f}_b \widehat{f}_c \widehat{f}_d\right)|M|^2\,,\nonumber\\
|M|^2 &=& \frac{1}{4}(p_a \cdot p_b)(p_c \cdot p_d)\,.
\eea

The effect of the reaction $\nu +
\bar \nu_\mu \leftrightarrow \nu + \bar \nu_\mu$, which in our general
notation is $a+ \bar d \leftrightarrow c + \bar b$, is obtained by a simple crossing symmetry. For instance the intensity part of the
Kernel would be for that process
\bea
E I_{C[A]} &=& 2^9 G_F^2 {\cal K} \left(\widehat{f}_a \overline{f}_b {f}_c \widehat{\overline{f}}_d-{f}_a \widehat{\overline{f}}_b \widehat{f}_c \overline{f}_d\right)|M|^2\,,\nonumber
\eea
and $|M|^2$ given by (\ref{CollisionTermNuNumu}).

For completeness, we must stress again that the effect of
antineutrino-muonic antineutrino reactions ($\bar \nu +
\bar \nu_\mu \leftrightarrow \bar \nu +\bar \nu_\mu$) on
antineutrinos is obtained by charge conjugation, that is by
considering the case
\bea\label{abab1bar}
&&a=\bar\nu\,,\quad b = \bar \nu_\mu\,,\quad c=\bar \nu\,,\quad d = \bar
\nu_\mu\,, \quad g = 2 \frac{G_F}{\sqrt{2}}\,,\nonumber\\
&&
\epsilon_-^{ac} = \epsilon_-^{bd} = e^\nu_-=\frac{1}{2}\,,\quad \epsilon_+^{ac} = \epsilon_+^{bd} = 0\,.
\eea
This means that the collision term takes the same form as Eqs.~(\ref{CollisionTermNuNumu}) but where
all covariant components should now refer to antiparticle species. For
instance the intensity part takes the form
\bea
E I_{C[A]} &=& 2^9 G_F^2 {\cal K} \left(\widehat{\overline{f}}_a \widehat{\overline{f}}_b \overline{f}_c \overline{f}_d-\overline{f}_a \overline{f}_b \widehat{\overline{f}}_c \widehat{\overline{f}}_d\right)|M|^2\,.\nonumber
\eea

\subsubsection{neutrino/neutrino scattering}\label{Secnunu}

Neutrino-neutrino scattering ($\nu +\nu \leftrightarrow \nu+\nu$) and
neutrino-antineutrino scattering  ($\nu +\bar \nu \leftrightarrow
\nu+\bar \nu$) are special cases of the previous electronic
neutrino-muonic neutrino scattering but there are a few crucial
differences in the derivation of the collision term which are detailed
in \citet{FidlerPitrou}.

To summarize, when considering interactions between neutrinos
($\nu+\nu \leftrightarrow \nu+\nu$) one must consider the two-body
case (\ref{abab1}) in the particular case $a=b=c=d=\nu$ and multiply
the result by a factor $2$ [this point was omitted in \citet{Hannestad:1995rs}]. And when considering interactions between neutrinos and antineutrinos of the same flavor ($\nu+\bar \nu \leftrightarrow \nu+\bar \nu$) one must consider the two-body interaction in the particular case $a=c=\nu$, $b=d=\bar \nu$ and multiply the result by a factor $4$ in agreement with \citet{Dolgov1997}. In particular, a simple crossing symmetry allows to get the former reactions $\nu+\nu \leftrightarrow \nu+\nu$ from the $\nu+\bar \nu \leftrightarrow \nu+\bar \nu$ only up to a factor $1/2$. We can interpret this reduction by a factor two using the fact that outgoing particles are identical and one must not double count the outgoing states.

\subsubsection{neutrino/electron scattering}

Contrary to neutrino-neutrino scattering, electron-neutrino
scattering is due to both charged and neutral currents. However the
Fierz reordering reduces the problem to an interaction of neutral
currents with modified chiral couplings. 
Using Eqs.~(\ref{ChiralCouplings}) and (\ref{epsilonmoinsplusun}), the effect of $\nu+e^-
\leftrightarrow \nu+e^-$ on neutrinos corresponds to the case
\bea\label{abab3}
&&a= c= \nu\,,\quad b = d = e^-\,,\quad \epsilon_{-}^{bd} =
\epsilon^e_- +1\,,\\
&& \epsilon_{+}^{bd} = \epsilon^e_+\,,\quad
\epsilon_-^{ac} = e^\nu_-=\frac{1}{2}\,,\quad \epsilon_+^{ac} = 0\,,\quad g=2 \frac{G_F}{\sqrt{2}}\,,\nonumber
\eea
which must be used in Eq.~(\ref{Mgeneralparticular}).

The effect of $\nu+e^+ \leftrightarrow \nu+e^+$ is obtained by a crossing symmetry.
The effect of $\bar \nu+e^+ \leftrightarrow \bar
\nu+e^+$ on antineutrinos is obtained from charge conjugation of
(\ref{abab3}), that is it corresponds to the case
\bea\label{abab4}
&&a= c= \bar \nu\,,\quad b = d = e^+\,,\quad \epsilon_{-}^{bd} =
\epsilon^e_- +1\,,\\
&& \epsilon_{+}^{bd} = \epsilon^e_+\,,\quad
\epsilon_-^{ac} = e^\nu_-=\frac{1}{2}\,,\quad \epsilon_+^{ac} = 0\,,\quad g=2 \frac{G_F}{\sqrt{2}}\,,\nonumber
\eea
and the effect of $\bar \nu+e^- \leftrightarrow \bar
\nu+e^-$ is obtained from crossing symmetry.

Finally, we can check that in the unpolarized case, these results for neutrino/electrons interaction and those for neutrino/neutrinos interactions obtained in \S~\ref{Secnunumuon} and \ref{Secnunu} are exactly the results of \citet{Grohs:2015tfy}. However note that as mentionned in this reference, there is a typo in the annihilation of neutrino and antineutrinos into electrons and positron in tables $1$ and $2$ of \citet{Dolgov1997}, and thus Tables 1.5 and 1.6 of \citet{NeutrinoBook}. The process described in these tables should be of the form $\nu+\bar \nu \leftrightarrow e^-+e^+$ and not $\nu+\bar \nu \leftrightarrow e^++e^-$. Up to this typographical correction our results agree also with \citet{Dolgov1997,NeutrinoBook} and we gather all reactions in Table~\ref{Table2}.

{\renewcommand{\arraystretch}{2}%
\begin{table*}[!htb]
\begin{tabular}{|c|c|c|c|}
\hline
{\rm Reaction} & {\rm Particles names} & {\rm Chiral couplings} &
                                                                  $2^{-9}\,G_F^{-2}
                                                                  \,E
                                                                  \,I_{C[A]}
                                                                  =
                                                                  {\cal
                                                                  K}
                                                                  |M|^2
                                                                   \times$ \\[0.1cm]
\hline
   $\nu+\nu_\mu \leftrightarrow \nu+\nu_\mu$& $a+b \leftrightarrow c+d$ &  $\epsilon_-^{ac}=\tfrac{1}{2}\quad \epsilon_+^{ac}=0\quad \epsilon_-^{bd}=\tfrac{1}{2}\quad\epsilon_+^{bd}=0$ & $\left(\widehat{f}_a \widehat{f}_b {f}_c {f}_d-{f}_a {f}_b \widehat{f}_c \widehat{f}_d\right)$\\
   \hline
   $\nu+\bar\nu_\mu \leftrightarrow \nu+\bar \nu_\mu$& $a+\bar d \leftrightarrow c+\bar b$ &  $\epsilon_-^{ac}=\tfrac{1}{2}\quad \epsilon_+^{ac}=0\quad \epsilon_-^{bd}=\tfrac{1}{2}\quad\epsilon_+^{bd}=0$ & $\left(\widehat{f}_a \overline{f}_b {f}_c \widehat{\overline{f}}_d-{f}_a \widehat{\overline{f}}_b \widehat{f}_c \overline{f}_d\right)$\\
   \hline
   $\nu+\bar \nu \leftrightarrow \bar \nu_\mu+\nu_\mu$& $a+\bar c \leftrightarrow \bar b+d$ &  $\epsilon_-^{ac}=\tfrac{1}{2}\quad \epsilon_+^{ac}=0\quad \epsilon_-^{bd}=\tfrac{1}{2}\quad\epsilon_+^{bd}=0$ & $\left(\widehat{f}_a \overline{f}_b \widehat{\overline{f}}_c {f}_d-{f}_a \widehat{\overline{f}}_b \overline{f}_c \widehat{f}_d\right)$\\
   \hline
   $\nu+\nu \leftrightarrow \nu+\nu$& $a+b \leftrightarrow c+d$ &  $\epsilon_-^{ac}=\tfrac{1}{2}\quad \epsilon_+^{ac}=0\quad \epsilon_-^{bd}=\tfrac{1}{2}\quad\epsilon_+^{bd}=0$ &  $2\left(\widehat{f}_a \widehat{f}_b {f}_c {f}_d-{f}_a {f}_b \widehat{f}_c \widehat{f}_d\right)$\\
   \hline
   $\nu+\bar\nu\leftrightarrow \nu+\bar \nu$& $a+\bar d \leftrightarrow c+\bar b$ &  $\epsilon_-^{ac}=\tfrac{1}{2}\quad \epsilon_+^{ac}=0\quad \epsilon_-^{bd}=\tfrac{1}{2}\quad\epsilon_+^{bd}=0$ & $4\left(\widehat{f}_a \overline{f}_b {f}_c \widehat{\overline{f}}_d-{f}_a \widehat{\overline{f}}_b \widehat{f}_c \overline{f}_d\right)$\\
   \hline
   $\nu+e^- \leftrightarrow \nu+e^-$& $a+b \leftrightarrow c+d$ &  $\epsilon_-^{ac}=\tfrac{1}{2}\quad \epsilon_+^{ac}=0\quad \epsilon_-^{bd}=\epsilon^e_-+1\quad\epsilon_+^{bd}=\epsilon_+^e$ &$\left(\widehat{f}_a \widehat{f}_b {f}_c {f}_d-{f}_a {f}_b \widehat{f}_c \widehat{f}_d\right)$\\
   \hline
   $\nu+e^+ \leftrightarrow \nu+e^+$& $a+\bar d \leftrightarrow c+\bar b$ &  $\epsilon_-^{ac}=\tfrac{1}{2}\quad \epsilon_+^{ac}=0\quad \epsilon_-^{bd}=\epsilon^e_-+1\quad\epsilon_+^{bd}=\epsilon_+^e$ &$\left(\widehat{f}_a \overline{f}_b {f}_c \widehat{\overline{f}}_d-{f}_a \widehat{\overline{f}}_b \widehat{f}_c \overline{f}_d\right)$\\
   \hline
   $\nu+\bar \nu \leftrightarrow e^+ +e^-$& $a+\bar c \leftrightarrow \bar b +d$ &  $\epsilon_-^{ac}=\tfrac{1}{2}\quad \epsilon_+^{ac}=0\quad \epsilon_-^{bd}=\epsilon^e_-+1\quad\epsilon_+^{bd}=\epsilon_+^e$ &$\left(\widehat{f}_a \overline{f}_b \widehat{\overline{f}}_c {f}_d-{f}_a \widehat{\overline{f}}_b \overline{f}_c \widehat{f}_d\right)$\\
   \hline
\end{tabular}
\caption{Main two-body reactions for the collision term  of
  neutrinos. The electronic neutrino is noted $\nu$ and the muonic
  neutrino is noted $\nu_\mu$. Similar reactions for antineutrinos can
  be deduced with a global charge conjugation on all these reactions,
  and thus on all the collision Kernels. The squared amplitudes
  $|M|^2$ are expressed with Eq. (\ref{Mgeneralparticular}). The
  integration on momenta is defined in Eq.~(\ref{calK}).}
\label{Table2}
\end{table*}}

\section{Neutrons-protons conversions}\label{SecFokkerPlanck}

Neutron-proton conversions are controlled by weak interactions in the
early universe. As they enforce statistical equilibrium, and since the
neutron is more massive and thus less likely statistically, the frozen
neutron abundance depends directly on the reaction rates. For larger
reaction rates, the frozen abundance is smaller and thus it leads to
less primordial Helium production~\citep{PitrouBBN}. We now review
the general form of these rates and we detail how a Fokker-Planck
expansion can be used to compute them in practice.

\subsection{General expression of the rates}

Let us first consider the reactions
\beas\label{AllWeakReactions}
n+\nu &\leftrightarrow& p + e^-\,,\slabel{EqWeak1}\\
n&\leftrightarrow& p + e^-+\bar \nu\,, \slabel{EqWeak2}\\
n+e^+ &\leftrightarrow& p + \bar \nu \,.\slabel{EqWeak3}
\eeas
They are mediated by the coupling of the neutron ($n$)/proton
($p$) charged current and the neutrino/electron charged current. While
the latter is purely left chiral, the former has both chiral couplings
due to the effective constant $g_A$ defined in
Eq.~(\ref{JCCpn}). Hence we must consider the case
\bea\label{abcd2}
&&a=n,\quad c=p,\quad b = \nu,\quad d = e^-\,,\quad g = \frac{G_F V_{ud}}{\sqrt{2}}\,,\nonumber\\
&& 
\epsilon_{+}^{ac}=\frac{1-g_A}{2}\,,\quad\epsilon_{+}^{bd}= 0\,,\quad \epsilon_{-}^{ac}=\frac{1+g_A}{2}\,,\quad \epsilon_{-}^{bd}= 1\,.\nonumber
\eea
With unpolarized species, the collision term for the forward reaction (\ref{EqWeak1}) takes the simpler form
\bea\label{CollNeutronProton}
E_a I_{C[A]} &=& 2^7 G_F^2 {\cal K} \left(\widehat{f}_a \widehat{f}_b {f}_c {f}_d-{f}_a {f}_b \widehat{f}_c \widehat{f}_d\right)|M|^2\,,\\
|M|^2 &=& c_{LL}{\cal M}_{LL} + c_{RR}{\cal M}_{RR} + c_{LR}{\cal M}_{LR} ,
\eea
where the coupling constants are
\beas
c_{LL}&\equiv& \left(\frac{1+g_A}{2}\right)^2\,,\\
c_{RR}&\equiv& \left(\frac{1-g_A}{2}\right)^2\,,\\
c_{LR}&\equiv&\left(\frac{g_A^2-1}{4}\right)\,,
\eeas
and the left-left right-right and left-right chiral couplings are
\beas\label{papbpcpd}
{\cal M}_{LL}  &=&  (p_a \cdot p_b)  (p_c \cdot p_d)\,,\\
{\cal M}_{RR}  &=&  (p_a \cdot p_d)  (p_b \cdot p_c)\,,\\
{\cal M}_{LR}  &=& m_a m_c (p_b \cdot p_d)\,.
\eeas
All other reactions are related by crossing symmetry or time reversal,
which affect only the phase space, but not $|M|^2$, that is we only
need to make sure to put the distribution function $f$ for initial
particles and the Pauli-blocking factor $1-f$ for final particles.

The number density of nucleons $N=n,p$ is related to the
distribution function $f_N$ by
\be
2\int f_N(\gr{p}) \frac{\dd^3 \gr{p}}{(2\pi)^3} = n_N\,.\label{Nucleonrule1}
\ee
Hence from Eq.~(\ref{CollNeutronProton}) we can define reaction rates
for the densities of neutrons and protons. The forward rates $\Gamma_{n \to p}$ are of the form
\bea\label{GeneralGamma2}
&&n_n \Gamma_{n \to p} \nonumber\\
&&= \int \frac{ \dd^3 \gr{p}_n \dd^3 \gr{p}_e \dd^3
  \gr{p}_\nu }{2^4 (2\pi)^8}\delta\left({E}_n-E_p + \coeffa_e E_e +
  \coeffa_\nu E_\nu\right) \nonumber\\
&&\qquad\times\frac{2^7 G_F^2\left| M\right|^2}{{E}_n {E}_p {E}_e
  {E}_\nu}  f_n(E_n)
f_\nu(\coeffa_\nu {E}_\nu) f_e(\coeffa_e {E}_e) \,,
\eea
where $ \coeffa_e=1$ (resp. $ \coeffa_e=-1$) if the electron/positron is in the initial (resp. final) state, and with a similar definition
for the neutrino/antineutrino coefficient $\coeffa_\nu$. Hence,
Eq.~(\ref{GeneralGamma2}) describes all reactions (\ref{AllWeakReactions}).
Note, that we have neglected Pauli-blocking effects of the final
proton, since the baryon-to-photon ratio is very low.
%, which is very well
%justified since the rest mass energy of nucleons is much larger than
%the typical energy during BBN. 
However we have correctly included Pauli-blocking effects of electrons/positrons and
neutrinos/antineutrinos since for a Fermi-Dirac (FD) distribution
without chemical potential
\be
f_{\rm FD}(-E) = 1-f_{\rm FD}(E)\,.
\ee
The vanishing of the electron/positron chemical potential is enforced
by the very low baryon-to-photon number ratio~\citep[App.~A.2]{PitrouBBN}. However, if we want to investigate
the possibility of non-vanishing neutrino chemical potentials $\mu$
\citep{PitrouBBN,SerpicoRaffelt,IoccoReport,Simha:2008mt}, once must
use instead
\be
f_{\rm FD}(-E,-\mu) = 1-f_{\rm FD}(E,\mu)\,.
\ee

\subsection{Isotropy of distributions}

At low temperature, it is enough to assume that nucleons follow an isotropic
Maxwellian distribution of velocities at the plasma temperature $T$. Hence the following integrals are obtained
\beas\label{IntMaxwellian}
2\int f_N(\gr{p}) \frac{p^i}{m_N}\frac{\dd^3 \gr{p}}{(2\pi)^3} &=&0\,, \slabel{Nucleonrule2}\\ 
2\int f_N(\gr{p}) \frac{p^i p^j}{m_N^2}\frac{\dd^3 \gr{p}}{(2\pi)^3} &=& 
\frac{T}{m_N} \delta^{ij} n_N\,. \slabel{Nucleonrule3}
\eeas
In particular contracting with $\delta_{ij}$ we recover the expression for the pressure of nucleons
in the low temperature limit
\be
P_N = 2 \int f_N(\gr{p}) \frac{p^2}{3m_N}\frac{\dd^3 \gr{p}}{(2\pi)^3}
= T n_N\,.
\ee
For electron or neutrino distributions, since we have assumed
isotropy, we deduce the property
\be
\int g(E) p^\alpha E^\beta p^i p^j \frac{\dd^3 \gr{p}}{(2\pi)^3} = \frac{\delta^{ij}}{3}\int g(E) p^{\alpha+2} E^\beta \frac{\dd^3 \gr{p}}{(2\pi)^3}
\ee
where $\alpha$ and $\beta$ are some numbers. From isotropy we also find that
\be
\int g(E) p^\alpha E^\beta p^i \frac{\dd^3 \gr{p}}{(2\pi)^3} =0\,.
\ee
Hence for all practical purposes, we can perform the replacements 
\be\label{CutRule}
p^i p^j  \to  p^2 \delta^{ij}/3\,,\qquad p^i \to 0\,.
\ee 
on all species, resulting in great simplifications.

\subsection{Expansion in the energy transfer}\label{FokkerFermions}

The integral in (\ref{GeneralGamma2}) is $8$-dimensional when on
removes the Dirac function. Due to the isotropy of all distributions,
this can be reduced to a $5$-dimensional integral. This is the method
followed by \citet{Lopez1997}. Here we follow a much simpler route
by performing a Fokker-Planck expansion, that is an expansion in the momentum transferred to the nucleons. It consists in expanding the energy difference between the
nucleons, $E_n-E_p$ around the lowest order value 
\be
\Gap=m_n-m_p\simeq 1.29333\,{\rm MeV}\,.
\ee
As we shall see, this results in one-dimensional integrals which are much faster to evaluate. 

We evaluate the rates by performing an expansion in powers of $\epsilon\equiv \sqrt{T/m_N}$. To evaluate the order of each term, we consider that the momentum or energies of neutrinos are of order $T
\sim \Gap$, that is factors of the type $E_e/m_N$ or $E_\nu/m_N$ are
of order $\epsilon^2$. Furthermore, from (\ref{IntMaxwellian}) a factor $\gr{p}_n/m_n$ is of order $\sqrt{T/M}
\sim \sqrt{\Gap/M}$ and thus $\epsilon$. However since only
even powers of the spatial momentum of nucleons must appear [see
Eqs.~(\ref{IntMaxwellian})], we shall encounter terms of the type $|(\gr{p}_p/m_n)^2|$ which are of order $\epsilon^2$.

Keeping only the lowest corrections this expansion reads
\be\label{DiffEnEp}
E_n-E_p = \Gap+\delta Q_1 +\delta Q_2 +\delta Q_3
\ee
\beas\label{dQnp}
\delta Q_1&\equiv&-\frac{\gr{p}_n \cdot \gr{q}}{m_N}\\
\delta Q_2 &\equiv & - \frac{|\gr{q}|^2}{2 m_N}\\
\delta Q_3 &\equiv&
\frac{|\gr{p}_n|^2}{2}\left(\frac{1}{m_n}-\frac{1}{m_p}\right)\simeq 
-\frac{|\gr{p}_n|^2 \Gap}{2 m_N^2}\,.\slabel{dQnp3}
\eeas
where $\gr{q}\equiv \gr{p}_p-\gr{p}_n = \coeffa_\nu
\gr{p}_\nu+\coeffa_e \gr{p}_e$ is the
spatial momentum transfered. The first term in (\ref{DiffEnEp}) is the
lowest order, or Born approximation, that is the only appearing when
considering the infinite nucleon mass approximation. The second term is an order $\epsilon$ 
correction, and the third term is an order $\epsilon^2$ correction. Finally the last term is of order $T \Gap/m_N$ so it is an
order $\epsilon^2$ correction as well. It is the only corrective term
for which it is crucial to take into account the difference of mass between
neutrons and protons.  
Using Eq. (\ref{DiffEnEp}), we expand the Dirac delta function on
energies as
\bea\label{DiracExpansion}
&&\delta\left(E_n-E_p+\coeffa_e E_e
  +\coeffa_\nu E_\nu\right)  \simeq \\
&&\qquad \delta(\Sigma) + \delta'(\Sigma)\left(\sum_{i=1}^3\delta
  Q_i\right)+\frac{1}{2}\delta''(\Sigma) (\delta Q_1)^2\,,\nonumber
\eea
where $\Sigma \equiv \Gap +\coeffa_e E_e +\coeffa_\nu E_\nu$.

We must then expand the matrix element and the energies appearing in
Eq.~(\ref{GeneralGamma2}). It proves much easier to expand all these
contributions together. Furthermore, whenever a term is already of
order $\epsilon^2$, we know that it should multiply only the Born term
of the expansion (\ref{DiracExpansion}), so we can apply the
simplification rule (\ref{CutRule}). With this method we find
\beas\label{RulesMEEEE}
\frac{{\cal M}_{LL}}{\Pi_i E_i} &\to& 1 -\frac{\gr{p}_n}{m_N} \cdot
\left(\frac{\gr{p}_e}{E_e}+\frac{\gr{p}_\nu}{E_\nu}\right)
-\frac{\coeffa_\nu|\gr{p}_\nu|^2}{m_N E_\nu}\slabel{Expansion1}\\
\frac{{\cal M}_{RR}}{\Pi_i E_i} &\to& 1 -\frac{\gr{p}_n}{m_N} \cdot
\left(\frac{\gr{p}_e}{E_e}+\frac{\gr{p}_\nu}{E_\nu}\right)
-\frac{\coeffa_e|\gr{p}_e|^2}{m_N E_e}\slabel{Expansion2}\\
\frac{{\cal M}_{LR}}{\Pi_i E_i} &\to&
\left(1-\frac{|\gr{p}_n|^2}{m_N^2}\right)\left(1-\frac{\gr{p}_e \cdot \gr{p}_\nu}{E_e E_\nu}\right) \slabel{Expansion3}\,.
\eeas
The second term in Eqs.~(\ref{Expansion1}) and (\ref{Expansion2}) is
of order $\epsilon$ and the last term in these equations is of
order $\epsilon^2$. Hence the second term needs to be coupled with the
order $\epsilon$ term in the Dirac delta expansion
(\ref{DiracExpansion}) which is $\delta'(\Sigma) \delta Q_1$, and
simplified with the rules (\ref{CutRule}). 

There are four steps to complete this Fokker-Planck
expansion. 
\begin{enumerate}
\item First, using Eqs.~(\ref{RulesMEEEE}) and (\ref{DiracExpansion}) in
the reaction rates (\ref{GeneralGamma2}) we perform the integral on
the initial neutron momentum with the rules
(\ref{IntMaxwellian}). 
\item Second, we can replace the differential elements
for the integral on electron and neutrino momenta with $\dd^3 p \to
4\pi p^2 \dd p$ because we have already performed all angular
averages. 
\item We are left with a two dimensional integral
on the electron and neutrino momentum magnitudes $p_e=|\gr{p}_e|$ and
$p_\nu=|\gr{p}_\nu|$. Let us note $E_\nu=p_\nu$ in order to write the
result in a easily readable form. Third, we perform the integral on  $E_\nu$ using the Dirac delta and its derivatives. Whenever a
Dirac delta derivative appears, it means that we have to perform
integration by parts to convert it into a normal Dirac delta. This
will introduce derivatives with respect to the $E_\nu$ applied on the
neutrino distribution function or Pauli-blocking factor. Also for a
given reaction it might appear that the value of $E_\nu$ constrained
by the Dirac delta is not physical for that reaction if
$\coeffa_\nu=1$ and physical if $\coeffa_\nu=-1$, or vice-versa. This is the
reason why we consider the total reaction rate of the reactions (\ref{EqWeak1}) and (\ref{EqWeak2}). Once their rates are added, the
Dirac delta automatically selects either the neutrino in the initial
state, with the corresponding distribution function, or the antineutrino
in the final state, with the associated Pauli-blocking
factor. Eventually once the rates (\ref{EqWeak1}) and (\ref{EqWeak2})
are added, we might forget about $\coeffa_\nu$, that is about the
position of the neutrino. We need only to compute two rates, one where
the electrons is in the initial state [reaction~(\ref{EqWeak3})], and
one where it is a positron which is in the final state [the sum of reactions~(\ref{EqWeak1}) and (\ref{EqWeak2})].
\item Finally, we need to determine the procedure to convert the rate with a neutron in the initial state into the reverse rate with a proton in the initial state. Even if the matrix
  element is the same for all reactions, the method to perform a
  finite mass expansion is not symmetric under the interchange $p
  \leftrightarrow n$. Indeed we chose to expand the momentum of the
  final nucleon around the initial one, and we remove the integral on
  the final nucleon momenta. It is apparent on Eqs.~(\ref{papbpcpd})
  that the electron (resp. neutrino) momentum is contracted with the
  neutron (resp. proton) in the   $LL$ term but this is the opposite
  in the $RR$ term. Since the coupling factors of these terms are
  interchanged by the replacement $g_A \to -g_A$, we can deduce the
  rates with an initial proton from those with an initial neutron
  using the rule $g_A \to -g_A$. Obviously the argument of the Dirac
  delta contains now $E_p - E_n= -\Gap + \dots$ instead of $E_n-E_p=
  \Gap + \dots$ so we must also apply the rule $\Gap \to
  -\Gap$. Finally when considering a reverse reaction, the electron in
  the initial state turns into a positron in the final state so we
  must also apply the rule $E_e \to -E_e$, that is change the electron
  distribution function to a Pauli-blocking factor or
  vice-versa. 
\end{enumerate}
Having sketched the details of the procedure, we are in position to
give the results. 
%However, it only takes a very simple form because we
%can use that for a homogeneous and isotropic universe the distribution
%must respect these symmetries. 
In the next section, we report the lowest order reaction rates in
\S~\ref{SecBorn}, also called Born approximation rates. The first
corrections, that we call finite nucleon mass corrections, are reported in appendix~\ref{AppFM}.

\subsection{Lowest order $n \leftrightarrow p$ reaction rate}\label{SecBorn}

Let us note $g(E)$ the Fermi-Dirac distribution at
temperature of electrons $T$ and $g_\nu(E)$ the Fermi-Dirac
distribution at the neutrino temperature $T_\nu$, that is
\be\label{DefgnuANDg}
g(E) \equiv \frac{1}{\left({\rm e}^{\frac{E}{T}}+1\right)}\qquad g_\nu(E) \equiv  \frac{1}{\left({\rm e}^{\frac{E}{T_\nu}}+1\right)}\,.
\ee
At lowest order in the Fokker-Planck expansion, the reaction rates take simple forms. First, the
factors entering the matrix element reduce to
\be
\frac{{\cal M}_{LL}}{E_n E_p E_\nu E_e}=\frac{{\cal M}_{RR}}{E_n E_p E_\nu E_e}=\frac{{\cal M}_{LR}}{E_n E_p E_\nu E_e}=1\,,
\ee
as seen from Eqs.~(\ref{RulesMEEEE}). The last equality is correct only if it is understood that an angular
average either on electrons momentum or neutrino momentum is
performed, that is using the rule (\ref{CutRule}). Hence from Eq.~(\ref{GeneralGamma2}), we find the Born rates~\cite{BrownSawyer,LopezTurner1998,Weinberg1972,Bernstein1989,PitrouBBN}
\beas\label{GammanBorn}
\overline{\Gamma}_{n \to p}&=&\overline{\Gamma}_{n \to
  p+e}+\overline{\Gamma}_{n +e \to p}\\
&=&\myk \int_0^\infty p^2 \dd p [\chi_+(E) + \chi_+(-E)]\,,
\eeas
with $E=\sqrt{p^2+m_e^2}$ and
\bea\label{Defchipm}
\chi_\pm(E) &\equiv& (E_\nu^\mp)^2 g_\nu(E^\mp_\nu) g(-E)\,,\\
E^\mp_\nu &\equiv& E\mp\Gap\,,\\
\myk &\equiv& \frac{4 G_F^2 V_{ud}^2 (1+3 g_A^2)}{(2\pi)^3} \label{Defk}\,.
\eea
The first contribution in Eq.~(\ref{GammanBorn}) corresponds
to the $n\to p$ processes (\ref{EqWeak1}) and (\ref{EqWeak2}) added, that is
for all processes where the electron is in the final state. It can be
checked indeed that the electron distribution is evaluated as $g(-E)=1-g(E)$. Furthermore, if the neutrino is in
the initial state (when $E>\Gap$) its energy is $E_\nu = E-\Gap$ and its
distribution function appears as $g_\nu(E_\nu)$, but if it is in the final state (when $E<\Gap$) its energy is $E_\nu =
\Gap-E$ and the neutrino distribution function is evaluated as
$g_\nu(E-\Gap)=1-g_\nu(\Gap-E)$. 

The second term of Eq.~(\ref{GammanBorn})  corresponds to the reaction
(\ref{EqWeak3}), that is to the process where the positron is in the
initial state. The energy of the positron is $E$ and
its distribution function appears as an initial state [$g(E)$], whereas the neutrinos in the final state have energy $E_\nu = \Gap+E$ and
their distribution function appear thus as Pauli-blocking factor
$g_\nu(-E-\Gap)=1-g_\nu(E+\Gap)$.

The reaction rate for protons, that is $\Gamma_{p \to n}$, is obtained
by the simple replacement $\Gap \to -\Gap$, which amounts to $\chi_+
\to \chi_-$. We give it for completeness
\beas\label{GammanBornp}
\overline{\Gamma}_{p \to n} &=& \overline{\Gamma}_{p \to n+e} +
\overline{\Gamma}_{p +e \to n}  \\
&=&\myk \int_0^\infty
p^2 \dd p [\chi_-(E) + \chi_-(-E)]\,.
\eeas
Similarly the second term corresponds to the reverse processes
(\ref{EqWeak1}) and (\ref{EqWeak2}) added since the electron
distribution function is always in an initial state [$g(E)$], and the
neutrino is in the initial or final state depending on the sign of
$E_\nu=-E+\Gap$. The first term corresponds to the reverse process
(\ref{EqWeak3}) with the positron always in the final state
[$g(-E)=1-g(E)$] and the neutrino always in the initial state [$g_\nu(E+\Gap)$].

Finally, note that using
\be\label{BasicLawDetailedBalance}
g(-E) = 1-g(E) = {\rm e}^{E/T} g(E) \,,
\ee
we get in the case of thermal equilibrium between neutrinos and the plasma (that is when $T_\nu = T$)
\be\label{BasicLawDetailedBalance2}
\chi_+(E) ={\rm  e}^{\Gap/T} \chi_-(-E)\,.
\ee
This implies that if neutrinos have the same temperature as the
plasma, the reaction rates satisfy the Born approximation detailed
balance relation \citep{BrownSawyer,PitrouBBN}
\be\label{MagicDetailedBalanceBorn}
\overline{\Gamma}_{p \to n} = {\rm e}^{-\Gap/T} \overline{\Gamma}_{n \to p}\,.
\ee

%%%%%%%%%%%%%%%%%%%%%%%%%%%%%%%%%%%%
\mypart{3}{Compton scattering}
%%%%%%%%%%%%%%%%%%%%%%%%%%%%%%%%%%%%

The Fokker-Planck expansion exposed in \S~\ref{FokkerFermions} was inspired
from a similar expansion often used for Compton scattering. We have
already stressed the numerous similarities between the construction of
distribution functions for fermions and bosons. We now turn to the
computation of the collision term for photons associated with Compton
scattering onto electrons. As we detail in the next section, the structure
of the Compton collision term is nearly identical to the weak
interaction collision term except that stimulated
emission factors replace Pauli-blocking ones. The collision term
obtained is exposed in \S~\ref{SecIsotropicCMB} for isotropic
distributions (along with a discussion on its cosmological
implications) and in \S~\ref{SecAnisotropicCMB} for the general case of anisotropic distributions.

\section{Compton collision term}\label{SecFokkerCompton}

\subsection{Extended Klein-Nishina formula}

We consider the Compton reaction
\be\label{DefComptonreaction}
\gamma(p)+ e^{-}(q)\leftrightarrow\gamma(p')+ e^{-}(q')\,.
\ee
The initial photon and electron momenta are decomposed as
\bea\label{Momenta}
&&p^0 = E,\qquad\quad\,\,\,\,p^i = E n^i\,,\\
&&q^0 = \Ee=m\Gamma,\quad q^i = \Ee\beta^i,\quad\Gamma\equiv\frac{1}{\sqrt{1-\beta_i  \beta^i}}\,,\nonumber
\eea
with similar decompositions for the final particles. Throughout this part, the electron mass is noted $m$.

Even though the Hamiltonian of QED accounts for a vertex between the
electronic current and a single photon, that is it is a three-leg vertex, it
is more adapted to consider an effective QED Hamiltonian assuming that the electron propagates freely between two
interactions with photons \citep[\S.~II.D]{BenekeFidler}. This is
essentially similar to our treatment of weak interactions in the Fermi
theory of \S~\ref{SecFermiTheory}, except that we do not use that the propagator
of the internal electron line is dominated by the electron mass. The effective
interaction Hamiltonian takes the form (\ref{HIofM}) with $a,c=\gamma$
and $b,d=e^-$ and a matrix element
\be\label{EffectiveHIQED}
M[(r,p)\,(h,q)\to(r',p')\,(h',q')]   = e^2 \bar u(q',h'){\cal O} u(q,h)\,,
\ee
\bea\label{DefCalO}
{\cal O}
&\equiv&\slashed{\epsilon}^\star_{r'}(p')\frac{\slashed{q}+\slashed{p}-m}{(q+p)^2+m^2}\slashed{\epsilon}_r(p)\nonumber\\
&+&\slashed{\epsilon}_{r}(p)\frac{\slashed{q}-\slashed{p}'-m}{(q-p')^2+m^2}\slashed{\epsilon}^\star_{r'}(p')\,.\nonumber
\eea
The two terms of $\cal O$ correspond to the two possible Feynman
diagrams associated with the reaction (\ref{DefComptonreaction}).
Then, the procedure to obtain a collision operator exactly follows our
derivation in \S~\ref{SecCollStructure}, that is we also obtain
Eq.~(\ref{ColAss}), with the only difference that the hatted notation
on photons now refers to stimulated emission factors instead of
Pauli-blocking factors, because we use Eq.~(\ref{Magic4Boson}) instead of Eq.~(\ref{Magic4}) when ordering operators. Once expressed as a
collision term for an operator by contraction with
$\gr{\epsilon}^\star_s \gr{\epsilon_{s'}}$ [as in Eq. (\ref{Deffmunumassless})], it takes a form fully similar to
Eq. (\ref{GenCA}). We prefer to report it with explicit indices for
all operators. Furthermore, we assume that electrons are unpolarized and they
are thus described by their distribution function (per helicity) $g(q)=
I_{e^-}(q)/2$. We also assume that we can neglect the associated
Pauli-blocking factors as the baryon-to-photon ration is very low. The equivalent of
Eq.~(\ref{GenCA}) for photons under Compton scattering finally reads
\bea\label{GeneralCompton1}
E C^\mu_{\,\,\nu}[f(p)] &=& -{\cal H}^\mu_{\,\,\alpha} K^\alpha_{\,\,\beta}
f^\beta_{\,\,\nu}(p)-f^\mu_{\,\,\alpha}(p) K^\alpha_{\,\,\beta}{\cal
    H}^\beta_{\,\,\nu}\\
&&+{\cal H}^\mu_{\,\,\alpha}\widehat{K}^\alpha_{\,\,\beta}
\widehat{f}^\beta_{\,\,\nu}(p)+\widehat{f}^\mu_{\,\,\alpha}(p) \widehat{K}^\alpha_{\,\,\beta}{\cal
    H}^\beta_{\,\,\nu}, \nonumber
\eea
where the dependence of ${\cal H}^\mu_{\,\,\nu}$ and $K^\mu_{\,\,\nu}$ on $p$ has been omitted. The operators involved are
\beas
K^\alpha_{\,\,\beta}(p) &=& \frac{1}{2}{\cal K}\,{\cal  M}^{\alpha\phantom{\beta\gamma}\delta}_{\phantom{\alpha}\beta\gamma}(p,q,p',q')g(q) \widehat{f}^\gamma_{\,\,\delta}(p'),\\
\widehat{K}^\alpha_{\,\,\beta}(p) &=& \frac{1}{2}{\cal K}\,{\cal  M}^{\alpha\phantom{\beta\gamma}\delta}_{\phantom{\alpha}\beta\gamma}(p,q,p',q')g(q'){f}^\gamma_{\,\,\delta}(p').
\eeas
where as in Eq.~(\ref{calK}) we defined
\be\label{defKCompton}
{\cal K} \equiv \frac{1}{2}\int[\dd p'][\dd q][\dd q'](2\pi)^4 \delta^4(p+q-p'-q')\,,
\ee
and where the detailed form of ${\cal M}$ is reported below in Eq.~(\ref{CovariantM}). Note that for photons, the equivalent of the identity operator
(\ref{IdentityFermions}) is ${\cal H}^\mu_{\,\nu} = \sum_{s s'}
\delta^K_{s s'} \epsilon^{\star\mu}_{s} (\epsilon_{s'})_\nu$ as seen
on Eq.~(\ref{DefScreenH}), and the structure is totally similar to Eq.~(\ref{GenCA}), but
with the stimulated emission operators
\be
\widehat{f}_{\mu\nu}(p) \equiv {\cal H}_{\mu\nu}(p)+ f_{\mu\nu}(p)\,.
\ee
If $g(q')=g(q)$ was satisfied, the general collision term
(\ref{GeneralCompton1}) would take an even simpler form. It is thus
convenient to define
\be\label{Defdg}
\delta_g(q,q') \equiv g(q') - g(q) \,,
\ee
so as to recast it under the sum of two contributions as
\be
C^\mu_{\,\,\nu}[f(p)] \equiv C^\mu_{\,\,\nu}[f(p)]_{\rm G-L} +C^\mu_{\,\,\nu}[f(p)]_{\rm G}\,.
\ee

The first contribution is the {\it gain minus loss}
contribution~\citep[\S~IV.B]{BenekeFidler}
\bea\label{GainMinusLoss}
&&E C^\mu_{\,\,\nu}[f(p)]_{\rm G-L} ={\cal K} g(q) \left\{{\cal H}^\mu_{\,\,\alpha} {\cal
    H}^\beta_{\,\,\nu}\,{\cal
    M}^{\alpha\phantom{\beta\gamma}\delta}_{\phantom{\alpha}\beta\gamma}{f}^\gamma_{\,\,\delta}(p')\right.\nonumber\\
&&\left.-\frac{1}{2}\,{\cal  M}^{\alpha\phantom{\beta\gamma}\delta}_{\phantom{\alpha}\beta\gamma}\left[{\cal H}^\mu_{\,\,\alpha} 
f^\beta_{\,\,\nu}(p)+f^\mu_{\,\,\alpha}(p) {\cal
  H}^\beta_{\,\,\nu}\right]{\cal H}^\gamma_{\,\,\delta}(p') \right\}
\eea
where we omitted to write the $(p,q,p',q')$ dependence on $\cal M$ and
with the dependence of ${\cal H}^\mu_{\,\,\nu}$ omitted when it is on
$p$. This part of the collision term is linear in the distribution function and is not impacted by stimulated
effects.

The second contribution is the pure {\it gain} term
\bea\label{Gain}
E C^\mu_{\,\,\nu}[f(p)]_{\rm G} &=& {\cal H}^\mu_{\,\,\alpha} \delta K^\alpha_{\,\,\beta}
\widehat{f}^\beta_{\,\,\,\nu}(p)+\widehat{f}^\mu_{\,\,\,\alpha}(p) \delta K^\alpha_{\,\,\beta}{\cal
    H}^\beta_{\,\,\nu}\,,\nonumber\\
\delta K^\alpha_{\,\,\beta} &\equiv&\frac{1}{2}{\cal K}\,{\cal
  M}^{\alpha\phantom{\beta\gamma}\delta}_{\phantom{\alpha}\beta\gamma}(p,q,p',q')
\delta_g(q,q') {f}^\gamma_{\,\,\delta}(p')\,.\nonumber
\eea
This gain term is however affected by stimulated effects, and it requires $\delta_g \neq 0$. Hence we already see that for very heavy
electrons (in the sense that their rest mass energy is much
larger than the typical energy of photons) corresponding to the
Thomson limit of Compton scattering, we would have
$\delta_g(q,q') \simeq 0$ and no stimulated emission effects.

The squared amplitude of the QED process associated to Compton collisions is 
\bea
{\cal M}^{rsr's'}(p,q,p',q')&=&\sum_{h\,h'}M[(r,p)\,(h,q)\to(r',p')\,(h',q')]\nonumber\\
&\times&M^{\star}[(s,p)\,(h,q)\to(s',p')\,(h',q')].\nonumber
\eea
Using (\ref{EffectiveHIQED}), and using the initial electron frame to
define polarization vectors, we checked that [using {\it
    xAct}~\cite{xAct} to handle products of Dirac matrices and
  contractions of vectors]
\begin{widetext}
\bea\label{KleinNishina}
\frac{{\cal M}^{rsr's'}(p,q,p',q')}{48 \pi \sT m^2} &=& [\epsilon_r(p) \cdot \epsilon^\star_{r'}(p')]
[\epsilon^\star_s(p) \cdot
\epsilon_{s'}(p')]\\
&+&\frac{1}{4}\left(\frac{p \cdot q}{p' \cdot
    q}+\frac{p' \cdot q}{p \cdot q}-2\right)\Big\{1+ [\epsilon_r(p) \cdot \epsilon^\star_{r'}(p')]
[\epsilon^\star_s(p) \cdot \epsilon_{s'}(p')]- [\epsilon_r(p) \cdot
\epsilon_{s'}(p')] [\epsilon^\star_s(p) \cdot\epsilon^\star_{r'}(p')]\Big\}\,,\nonumber
\eea
\end{widetext}
in agreement with \citet[Eq.~187]{Portsmouth:2004ee} or
\citet{Stedman1982}. Eq.~(\ref{KleinNishina}) is the extended
Klein-Nishina formula for Compton scattering. The Thomson cross-section $\sT = 8 \pi
\alpha_{\rm FS}^2/(3 m^2) $,  where $\alpha_{\rm FS} \equiv e^2/(4\pi)$,  has been introduced and it is related to
the electron charge by $6\pi \sT m^2 = e^4$. This is covariantized by
contraction with photon polarization vectors as
\bea
{\cal M}^{\alpha\beta\mu\nu}(p,q,p',q') &\equiv& \sum_{rsr's'}{\cal
  M}^{rsr's'}(p,q,p',q')\\
&&\times\epsilon_r^\alpha(p)\epsilon_{s}^\beta(p) \epsilon_{r'}^\mu(p') \epsilon_{s'}^\nu(p')\,, \nonumber
\eea
and it then takes the form \citep[Eq.~189]{Portsmouth:2004ee}
\bea\label{CovariantM}
&&\frac{{\cal M}^{\alpha\beta\mu\nu}(p,q,p',q')}{48 \pi \sT m^2} = {\cal P}^{\alpha\beta\mu\nu} +\frac{1}{4}\left(\frac{p \cdot q}{p' \cdot
    q}+\frac{p' \cdot q}{p \cdot q}-2\right)\nonumber\\
&&\qquad\times\left[{\cal
    H}^{\alpha\beta}(q,p) {\cal H}^{\mu\nu}(q,p')+{\cal P}^{\alpha\beta\mu\nu}-{\cal P}^{\alpha\beta\nu\mu}\right],
\eea
where we use the combinations of the screen projectors (note that
these are screen projectors in the frame of the initial electron)
\beas\label{DefPH}
{\cal P}^{\mu\nu}(q,p,p') &\equiv& {\cal H}^{\mu\alpha}(q,p) {\cal H}_\alpha^{\,\,\nu}(q,p')\,,\\
{\cal P}^{\mu\nu\alpha\beta}(q,p,p') &\equiv& {\cal P}^{\mu\alpha}(q,p,p') {\cal P}^{\nu\beta}(q,p,p').
\eeas

\subsection{Kinematics}

Let us detail the kinematics of the reaction
(\ref{DefComptonreaction}) enforced by the Dirac function on
momenta. From $(p+q)^2 = (p'+q')^2$, which leads to $p \cdot q = p'
\cdot q'$, and expressing the energy and spatial momentum of the
final electron using energy and momentum conservation, we find \citep[Eq.~C7c]{Chluba2012}
\bea\label{EpoverE}
\frac{E'}{E} &=& \frac{1-{\bm \beta}\cdot \gr{n}}{1-\gr{\beta}\cdot
  \gr{n}' +E/(m \Gamma) (1-\gr{n} \cdot \gr{n}')}\nonumber\\
&\equiv& \frac{\phi(E,\gr{n},\gr{n}',\gr{\beta})}{E}\,.
\eea
We define a fractional photon energy shift 
\be
E' \equiv E(1+ \ens)\,,
\ee
which is deduced from Eq.~(\ref{EpoverE}). Note that in the initial electron frame $\Gamma=1$, $\beta^i=0$ and 
\be
\ens|_{\gr{\beta}=0} = -\frac{E}{m}(1-\gr{n} \cdot \gr{n}') + {\cal O}(E/m)^2\,.
\ee
In particular, when considering the initial electron frame, we find
for the prefactor of the last term of Eq. (\ref{CovariantM})
\be\label{MagicEpeeEp}
\frac{p \cdot q}{p' \cdot
    q}+\frac{p' \cdot q}{p \cdot q}-2= \left.\frac{E'}{E} + \frac{E}{E'}-2\right|_{\gr{\beta}=0} = {\cal O}(E/m)^2\,.
\ee

\subsection{Electron velocity distribution}

Given that we neglect the degeneracy of electrons, the electron
distribution function is well described by a Boltzmann distribution,
and Pauli-blocking effects can be ignored. In the frame comoving with the bulk
velocity we have \citep{Challinor1999}
\be\label{gq}
g(q) = \frac{n_e \pi^2}{m^2 T_e K_2(m/T_e)} {\rm e}^{-{\cal E}(q)/T_e}\,,
\ee
where $K_2$ is a modified Bessel function. The energy of a given
electron with momentum $q^\mu$ in the bulk electron frame is given by
\be
{\cal E}(q) \equiv \sgnzz u_{\rm b}^\mu q_\mu =  \gamma({\cal E} - \gr{v} \cdot \gr{q})\,.
\ee
where $u^\mu_{\rm b} = \gamma(1,\gr{v})$ (with $\gamma\equiv 1/\sqrt{1-v^2}$) is
the electron bulk velocity. The distribution of the final electron
$g(q')$ is related to $g(q)$ using \citep{Challinor1998,Chluba2012}
\be\label{EprimeE}
{\cal E}(q') = {\cal E}(q) - \gamma E \ens (1-\gr{v} \cdot \gr{n}')-\gamma E \gr{v}\cdot(\gr{n}-\gr{n}')\,.
\ee

However for our purpose it is enough to consider that electrons follow
a Maxwell-Boltzmann distribution
\be\label{DefMaxwell}
g(q) = \frac{n_e}{2}\left(\frac{2\pi}{m T_e}\right)^{3/2}{\rm
  e}^{-\frac{(\gr{q}-m \gr{v})^2}{2 m T_e}}\,.
\ee
The first few moments of this distribution are simply
\beas\label{MainIntegrations}
2 \int \frac{\dd^3 \gr{q}}{(2\pi)^3} g(\gr{q}) &=& n_e\,,\\
2 \int \frac{\dd^3 \gr{q}}{(2\pi)^3} g(\gr{q}) \frac{q^i}{m} &=& n_e v^i\,,\slabel{Moment1}\\
2 \int \frac{\dd^3 \gr{q}}{(2\pi)^3} g(\gr{q}) \frac{q^i q^j}{m^2} &=&
n_e \left(\frac{T_e}{m} \delta^{ij}+ v^i v^j\right).\slabel{Moment2}
\eeas
As long as we restrict our expansion to second order contributions in the electron
velocities (see next section), using the distribution
(\ref{DefMaxwell}) and the integrations (\ref{MainIntegrations}) is
valid. In the more general case, considered in e.g. 
\citet{Challinor1997,Challinor1998,Challinor1999}, one must use the
distribution (\ref{gq}) and Eq.~(\ref{EprimeE}) to expand $g(q')$
around $g(q)$.

\subsection{Fokker-Planck expansion}

The Fokker-Planck expansion is an expansion in the momentum
transferred to the heavy species. As in the case of weak interactions detailed in
\S~\ref{FokkerFermions}, we consider that terms of order $\gr{v}$ or $\gr{\beta}$
are of order $\epsilon$, whereas terms of order $T_e/m$ or $E/m$ are
of order $\epsilon^2$. We list all steps required to perform such expansion.

\subsubsection{Energy shift}

There are two equivalent methods to perform this expansion. The first
method, as used by \citet{Chluba2012} [see also \citet[\S~5.5]{Peskin}] consists in simplifying the integrations~(\ref{defKCompton}) using 
\bea\label{PeskinMethod}
&&\int \frac{\dd^3 \gr{p}'}{E'}\frac{\dd^3 \gr{q}'}{{\cal E}'}
\delta^4(p+q-p'-q')\\
&&\quad=\int \frac{E' \dd E' \dd^2 \Omega'}{{\cal E}'} \delta(E+{\cal E}-E'-{\cal
  E}')\nonumber\\
&&\quad=\int E' \dd E' \dd^2 \Omega'
\frac{\delta(E'-\phi(E,\gr{n},\gr{n}',\gr{\beta}))}{{\cal E}+E(1-\gr{n}\cdot \gr{n}')-m \Gamma \gr{\beta} \cdot
  \gr{n}'}\nonumber\\
&&\quad=\int \frac{(E')^2  \dd E' \dd^2 \Omega' }{m E \Gamma(1-\gr{\beta} \cdot \gr{n})}\delta(E'-\phi(E,\gr{n},\gr{n}',\gr{\beta}))\,,\nonumber
\eea
where Eq.~(\ref{EpoverE}) was used in the last step. The integration on $E'$ is then trivially performed and it amounts to
the replacement $E' \to \phi(E,\gr{n},\gr{n}',\gr{\beta})$. In
practice all quantities which depend on $E'$ (among which the
distribution functions) are expanded in the small parameter
$\phi-E$. 

The second method is the long-standing method for the CMB computations~\citep{Hu:1993tc},
among which \citet{Dodelson1993,Bartolo2006,Pitrou2008,BenekeFidler}. After
integration on the spatial momentum of the final electron $\gr{q}'$,
the remaining Dirac function ensuring conservation of energy is
handled via a Taylor expansion. Eq.~(\ref{EpoverE}) cannot be used as it results precisely from the Dirac function, and the final photon energy
$E'$ must remain unknown at this stage. The Taylor expansion reads
\bea\label{TaylorDirac}
&&\delta(E+{\cal E}-E'-{\cal E}') =\delta(E-E')\\
&& +\frac{(\gr{p}-\gr{p}')\cdot
  \gr{q}}{m}\partial_{E'}\delta(E-E')\nonumber\\
&&+\left\{\frac{(\gr{p}-\gr{p}')^2}{2 m}\partial_{E'}+
\frac{1}{2} \left[\frac{(\gr{p}-\gr{p}')\cdot \gr{q}}{m}\right]^2\partial^2_{E'}\right\}\delta(E-E'),\nonumber
\eea
where the second line is of order $\epsilon$ and the last
line of order $\epsilon^2$. The expression of the energy transfer that
we have used to write this Taylor expansion is essentially similar to Eqs.~(\ref{dQnp}), except that the equivalent
of the term (\ref{dQnp3}) does not exist since the final massive particle is the
same as the initial massive particle (both are electrons).

It must be understood that integration by parts must be performed to remove
derivatives on Dirac functions. Since $\gr{p}' = E' \gr{n}'$, these
integration by parts also act on the expansion (\ref{TaylorDirac})
itself, in addition to acting on the distribution function. In both
methods, it is assumed that the spectrum is smooth enough such that
the Taylor expansion of the spectrum is meaningful. In the case of a
spectrum containing narrow lines, the Fokker-Planck expansion might
not be applicable \citep{Sazonov2000}. Furthermore, for the Taylor
expansion to remain meaningful, one must ensure that the various
moments of the scattering Kernel remain
small~\citep{Sarkar:2019har}. This is equivalent to asking the
conditions $E/m \ll 1$ and $p \ll 1$ or equivalently $T_e/m \ll 1$.

\subsubsection{Electron distribution}

Furthermore, in both methods, one also expands $\delta_g(q,q')$ and
${\cal M}$ [Eq.~(\ref{CovariantM})]. The first expansion is given by
\bea
&&\frac{\delta_g(q,q')}{g(q)}-1 = -\frac{(\gr{p}-\gr{p}')\cdot (\gr{q}-m\gr{v})}{m
  T_e}\\
&&\qquad-\frac{(\gr{p}-\gr{p}')^2}{2 m T_e}+\frac{1}{2}\left[\frac{(\gr{p}-\gr{p}')\cdot (\gr{q}-m\gr{v})}{m T_e}\right]^2 +\dots\nonumber
\eea
In the first method, it is understood that $\gr{p}'=\phi \gr{n}'$ and one must further expand in $\phi-E$.
In the second method we use $\gr{p}' = E' \gr{n}'$ so that $\gr{p}'$
is acted upon by the integrations by parts as $\partial_{E'} \gr{p}' =
\gr{n}'$.

\subsubsection{Squared amplitude}

The expansion of Eq.~(\ref{CovariantM}) is simplified when
restricting to orders smaller or equal to $\epsilon^3$, as we need only consider the first term
in its r.h.s given the property (\ref{MagicEpeeEp}). Its expansion is
deduced from its relation to the screen projector in the
definitions (\ref{DefPH}). The expansion of the screen projector
in the initial electron frame  ${\cal H}_{\mu\nu}(q/m,p)$ is obtained
from the transformation rule (\ref{HtildeH}) considered for $\tilde u^\mu = q^\mu/m$,
and using  $\tilde E = \Gamma E(1- \gr{n}\cdot \gr{\beta})$. For instance
restricting to order $\epsilon$, we find 
\be
{\cal H}_{ij}(q/m,p) \simeq {\cal H}_{ij}(p) + 2 n_{(i} {\cal H}_{j) k}(p) \beta^k\,.
\ee 
While the expansion of Eq.~(\ref{CovariantM}) up to order $\epsilon^2$
is rather sizable, it is instructive to consider the case in which the
distribution function is unpolarized, restricting also to the
intensity part of the collision term. One must thus consider
\be
{\cal M}(p,q,p',q') \equiv {\cal H}_{\alpha \beta}(p){\cal H}_{\mu\nu}(p'){\cal M}^{\alpha\beta\mu\nu}(p,q,p',q')\,.
\ee
Considering only the contribution of the first term in the r.h.s of
Eq.~(\ref{CovariantM}), that is neglecting terms of order
$\epsilon^4$, we find using property (\ref{ProjectHcal}) that
\bea\label{Unpluscos}
\frac{{\cal M}(p,q,p',q')}{48 \pi \sT m^2} &=& {\cal H}_{\mu\nu}(q/m,p) {\cal
  H}^{\mu\nu}(q/m,p') \nonumber\\
&=& 1+ (\tilde{\gr{n}}\cdot \tilde{\gr{n}}')^2\,,
\eea
where the direction in the initial electron frame of a momentum $p$ is
defined as in Eq.~(\ref{Decomposep}), that is $p^\mu= \tilde E(\tilde u ^\mu + \tilde{n}^\mu)$. This is the usual
Thomson squared amplitude in the electron frame $\propto 1+ \cos^2
\theta$ where $\theta$ is the deflection angle. 
Eventually, we must expand $\tilde{\gr{n}} \cdot \tilde{\gr{n}}'$
around of $\gr{n} \cdot \gr{n}'$ in order to obtain its expansion. From the invariance of $p'_\mu p^\mu$ we get
\be
(1- \tilde{\gr{n}} \cdot \tilde{\gr{n}}') = \frac{(1- \gr{n} \cdot \gr{n}')}{\gamma^2 (1- \gr{n} \cdot \gr{v}) (1-
  \gr{n}' \cdot \gr{v})} \,.
\ee
This is the procedure followed by \citet{Dodelson1993} or \citet[\S 2.2]{HuThesis}.

\subsubsection{Integration measure}

Finally, the expansion is completed by expanding the energies of
electrons which appear in the relativistic integration elements
contained in (\ref{defKCompton}). Up to order $\epsilon^2$ we find
\be
\frac{m^2}{{\cal E}{\cal E}'} = 1- \frac{\gr{q}\cdot \gr{q}}{m^2} + \dots
\ee
Furthermore, when using the first method for handling the energy
shift, the factor $1/\Gamma/(1-\gr{\beta} \cdot \gr{n})$ appearing in
(\ref{PeskinMethod}) must be expanded in powers of $\gr{\beta}$.

\subsection{Structure of the expansion}

We checked that both methods for the energy shifts lead to the same
result\footnote{The computations were performed with {\it
    xAct}~\cite{xAct}.} and we present and analyze them in the
subsequent sections. Let us briefly comment on the prominent features of
the method.

In the integration (\ref{defKCompton}), we had nine integrals (on $\gr{q}$, $\gr{p}'$ and $\gr{q}'$). Those on $\gr{q}'$ were removed
using the spatial part of the Dirac function. The integration on
$\gr{q}$ is subsequently performed using the moments
(\ref{MainIntegrations}), once the expansion in the momentum
transferred is written. Finally, the integration on $E'$ is performed either
as in (\ref{PeskinMethod}) with the first method, or from the Taylor
expansion (\ref{TaylorDirac}) in the second method. Eventually we are
left with an integration on $\gr{n}'$, that is the direction of the
final photon. As we shall detail, this is handled from the multipolar
decomposition of the distribution function.

The {\it gain minus loss} terms are always linear in the distribution function. The
loss terms are particularly simple to compute because they do not
depend on the distribution of the final photon, hence the residual
integration on $\gr{n}'$ is simple.

The stimulated emission factors appear only as a result of $\delta_g
\neq 0$. As a consequence, they do not appear in the Thomson limit of
the collision term, even for a general non-isotropic distribution function.

The various contributions of the collision term can be classified as follows.
\begin{itemize}
\item {\it Thomson} terms are the lowest order ones (order
  $\epsilon^0$). They are entirely due to the structure of ${\cal
    P}^{\mu\nu\alpha\beta}(q,p,p')$ (when the velocity of the initial
  electron is the same as the one of the observer) coupled with the
  structure of the {\it gain minus loss} term (\ref{GainMinusLoss}).
\item {\it Thermal terms} are of order $\epsilon^2$, and they appear whenever we average over the
  electron distribution products of the type $q^i q^j$, by using Eq.~(\ref{Moment2}).
\item {\it Kinetic terms} are proportional to $v^i$ and arise from the
  order  $\epsilon$ terms using Eq.~(\ref{Moment1}). {\it Non-linear
    kinetic} terms arise in the same condition as the thermal terms,
  that is they are of order $\epsilon^2$,  and are proportional to
  $v^i v^j$ or $v^2$. One can conveniently check thermal terms by
  replacing $v^i v^j \to \delta^{ij}T_e/m $.
\item Recoil terms arise from the fact that the
  electron mass is not infinite. They are of order $\epsilon^2$ and are proportional to $E/m$. They
  originate from the energy shift $\ens$, and also from $\delta_g$
  since it is also affected by $\ens$.
\end{itemize}

Finally, the factor $n_e \sT$ appears as a prefactor to all terms,
since collisions are proportional to the number densities of electrons
and to the Thomson cross section. It is customary to define the optical depth $\tau$ by
\be\label{DefTau}
 \frac{\dd \tau }{\dd t}  \equiv n_e \sT\,.
\ee
Once divided by $\dd \tau /\dd t$, the collision term is reinterpreted as a rate of variation per unit of $\tau$ instead of per unit
of time.

\section{Evolution of isotropic distributions}\label{SecIsotropicCMB}

In this section we restrict to the case of an isotropic distribution
function. Given the absence of preferred directions, there is no linear
polarization ($\polar_{ij}=0$), and we further assume that there is no
circular polarization ($V=0$). Isotropy also implies that intensity is
equal to its monopole $I=I_\emptyset$.

\subsection{Kompaneets equations}

Under these symmetries, it is shown that there is no
contribution from the Thomson terms, since scattering out and
scattering in term exactly cancel. Furthermore let us consider the
situation in the bulk frame of baryons ($\gr{v}=0$). The only
possible terms, when considering the expansion up to order
$\epsilon^2$, are the thermal and recoil terms.

The collision term reduces then to the celebrated Kompaneets equation
~\cite{Kompaneets1957}
\be\label{Kompaneets1}
\frac{\dd t}{\dd \tau} I^{\rm Kom}_C = \frac{1}{mE^2}\partial_E\left\{E^4
  \left[T_e\partial_E I_\emptyset + I_\emptyset \left(1+\frac{I_\emptyset}{2}\right)\right] \right\}\,.
\ee

We notice that if $f_\gamma \equiv I_\emptyset/2$ (as in
\S~\ref{SecGenCollTerm}, this is the distribution {\it per helicity state}) is a Planck distribution at temperature $T_\gamma$
\be
f_\gamma(E) = \nbb(E/T_\gamma)\,,\qquad \nbb(x) \equiv \frac{1}{{\rm e}^x-1}\,,
\ee
then it satisfies
\be\label{MagicBE}
T_\gamma \partial_E f_\gamma = -f_\gamma(1+f_\gamma)\,,
\ee 
and the Kompaneets collision term is recast as
\be\label{SZ1}
\frac{\dd t}{\dd \tau}  I^{\rm Kom}_C = (T_e-T_\gamma)\frac{1}{mE^2}\partial_E\left(E^4\partial_E I_\emptyset \right)\,.
\ee
The factor multiplying $T_e - T_\gamma$ (where $I_\emptyset/2$ is a Planck spectrum) is exactly the spectral shape of a $y$-type spectral
distortion. If the photon temperature is equal to the electron
temperature, it vanishes. Or said differently, the Planck spectrum at
the electron temperature is a fixed point of the associated Boltzmann equation. It is customary to define a {\it Compton} optical depth by
\be\label{tauK}
\frac{\dd \tau^{\rm C}}{\dd \tau} \equiv \frac{T_e}{m}\,,\qquad \frac{\dd  \tau^{\rm C}}{\dd t} \equiv \frac{n_e \sT T_e}{m}\,. 
\ee
The Boltzmann equation associated with the Kompaneets collision term
then takes the very simple form
\be\label{KompaneetsInx}
\frac{\dd f_\gamma}{\dd \tau^{\rm C}} =\frac{1}{E^2}\partial_E\left\{E^4
  \left[\partial_E f_\gamma + \frac{1}{T_e}f_\gamma \left(1+f_\gamma\right)\right] \right\}\,.
\ee

In fact, property (\ref{MagicBE}) holds even if the distribution is a Bose-Einstein distribution, that is with
a constant chemical potential $f_\gamma(E) = \nbb[(E+\mu)/T_\gamma]$. Indeed, it is obvious under the form (\ref{Kompaneets1}) that the Kompaneets
term conserves the number of photons [as it should since Compton
collisions of the type (\ref{DefComptonreaction}) do conserve in
general photons]. Since a Planck spectrum at a given temperature $T_\gamma$ has
a number density of photons uniquely determined by $T_\gamma$, a
chemical potential must develop when photons thermalize with
electrons, so as to ensure the conservation of photons, and the final
spectrum is a Bose-Einstein spectrum with non-vanishing chemical
potential, see \citet{Sunyaev1970}, \citet{Burigana1991} and \citet[\S 3.2.2]{HuThesis}.

Higher order effects in the Fokker-Planck expansion have been
considered in the isotropic case, either without electron bulk
velocity \citep{Challinor1997,Itoh:1997ks} or with a bulk velocity on top of
thermal effects \citep{Challinor1998,Nozawa:1998zu,Itoh:1998wv,Sazonov1998}. These corrections are
particularly relevant for the Sunyaev-Zel'dovich effect \citep{SZ1972} of hot galaxy
clusters on the CMB, whose lowest order description is Eq.~(\ref{SZ1})
restricted to the limit $T_\gamma \ll T_e$.

\subsection{Thermalization on electrons}

Let us discuss briefly the thermalization process from the Kompaneets
equations in two limiting cases. First we consider that electrons
are completely dominating the energy content, and then the case where
the photon energy density dominates over baryons. The first case
applies in the late universe, whereas the second case applies in the
radiation dominated era.

\subsubsection{Test distribution}

\begin{figure}[!htb]
     \includegraphics[width=\columnwidth,angle=0]{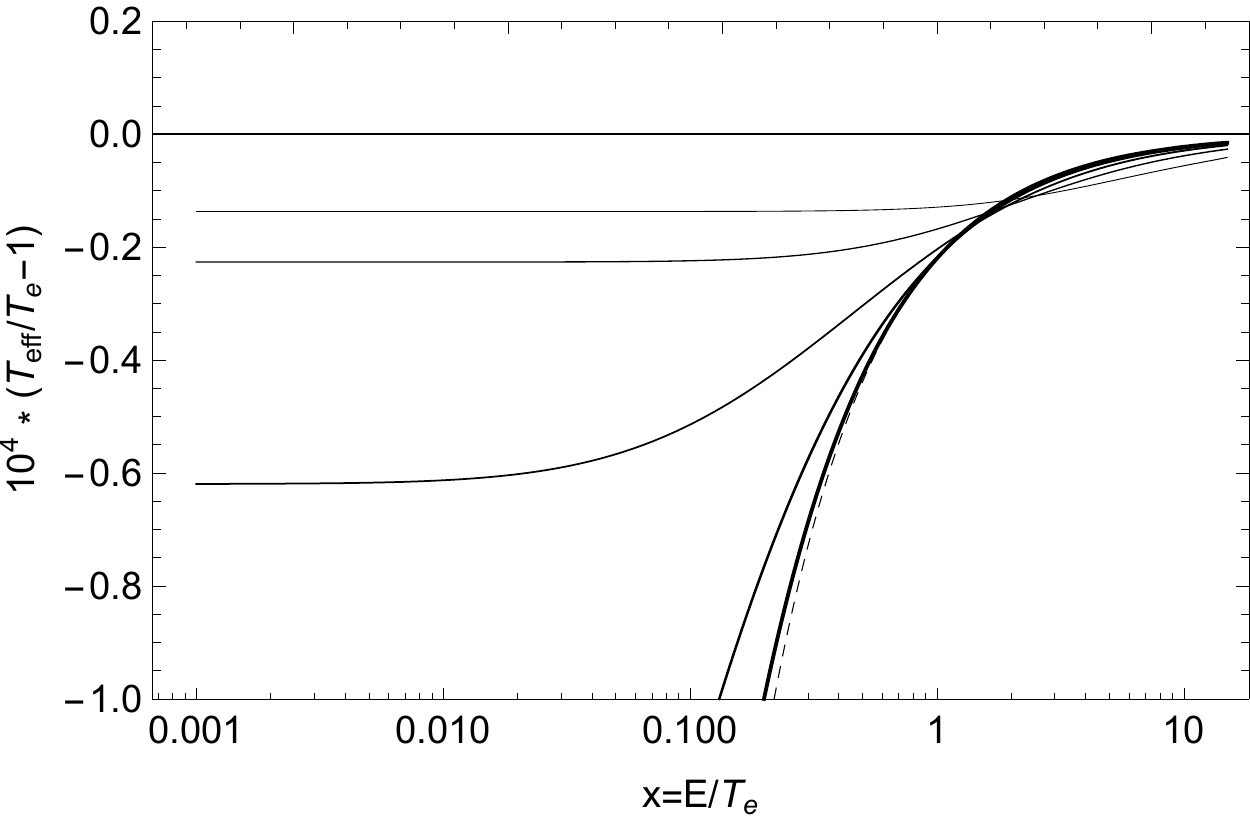}
     \caption{Effective temperature $T_{\rm eff} \equiv
       E/\ln[(1+f_\gamma)/f_\gamma] $ for various Compton optical depths. The
       curves of increasing thickness (and from top to bottom) correspond to  $\tau^{\rm C}=0, 0.25, 0.5, 1, 2,4$ and the dashed curve is the expected  Bose-Einstein final spectrum.}
\label{FigK1}
\end{figure}

In the case where the energy density of baryons dominate, their
temperature is not affected by the back-reaction of the Kompaneets
collision term on them. Electrons set the final value of the photons temperature. In Fig.~(\ref{FigK1}) we plot the response of
a Planck spectrum to a rapid increase of $0.01\%$ of the electron temperature.
We find that approximately for $\tau^{\rm C} \gtrsim 4$, the
spectrum is reasonably converged to a Bose-Einstein distribution with
the appropriate chemical potential. However, the case of dominant baryonic
energy density applies to the late universe, for instance in the
galaxy inter-cluster hot gas, for which the effect of the Kompaneets equation is
known as the Sunyaev Zel'dovich (SZ) effect \citep{SZ1972} and in such
cases the Compton optical depth is much lower than unity. Hence
the associated spectral distortions are expected to be of the
$y$-type, that is with a distortion given by the Kernel
(\ref{SZ1}). Note that distortions from all clusters should in
principle contribute collectively to a global monopolar distortion of the SZ type including relativistic temperature corrections~\citep{Refregier:2000xz,Chluba2015PRL}.

\subsubsection{Dominant distribution}

In the opposite case where the energy density of baryons is subdominant, the photons cannot gain energy from their interactions
with electrons. Hence the total energy density transfer rate must vanish, that is the electron temperature must be such that $\int 
I_C^{\rm Kom} E^3 \dd E= 0$. It is then given by
\be
T_e = \frac{1}{4}\frac{\int E^4 f_\gamma (1+f_\gamma)\dd E}{\int f_\gamma E^3 \dd E}\,.
\ee
The effect of the Kompaneets equation is to redistribute photons while
conserving the total energy density so as to approach a Bose-Einstein
spectrum. This case applies to the radiation dominated era. The photon
spectrum is redistributed via Compton interactions with
electrons, but there is no net creation of photons, nor energy gained
or lost. Note that Comptonization of electrons is much faster~\citep{Iwamoto1983}, precisely because their energy density is
subdominant and they always possess a thermal spectrum, that is a well
defined temperature. 

\begin{figure}[!htb]
     \includegraphics[width=\columnwidth,angle=0]{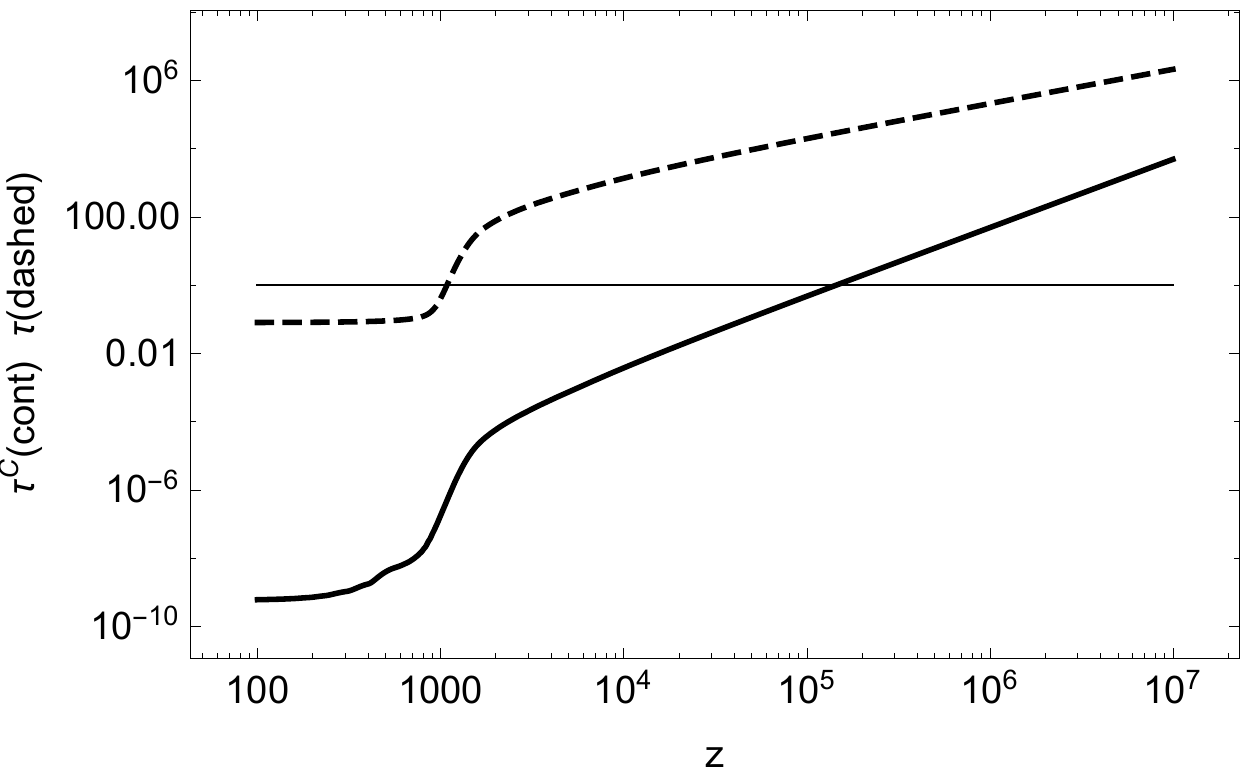}
     \caption{Optical depth (dashed line) and Compton optical
       depth (continuous line). The horizontal line depicts unity. The
       fact that the Compton optical depth $\tau^{\rm C}$ is lower than the optical
       depth $\tau$ is the very reason why spectral distortions open a window
       on earlier cosmological times than the anisotropies of the CMB temperature.}
\label{FigTau}
\end{figure}

One can still define the Compton optical depth via
Eq.~(\ref{tauK}). In the early universe, any energy injected at a time
corresponding to $\tau^{\rm C} \gtrsim 4$ results in a Bose-Einstein
spectrum, that is a spectral distortion of the $\mu$-type, whereas
energy injected later, and thus corresponding to a low value of
$\tau^{\rm C}$ (that is $\tau^{\rm C}\lesssim 0.1$) are of the
$y$-type. For the most recent cosmological
parameters~\cite{Planck2016}, the Compton optical depths scales for
redshifts $z$ belonging to the radiation era (that is for $z\gtrsim
3\times 10^3$) as\footnote{The scaling in $(1+z)^2$ is deduced from
  $\dd \tau^{\rm C}/\dd a = n_e T_e / (H a) \propto 1/a^3$, where $H$ is the cosmic time Hubble function and
  $a$ the scale factor.} 
\be
\tau^{\rm C} \propto 4.8\times 10^{-11}\,(1+z)^2\,,
\ee
as depicted on Fig.~(\ref{FigTau}). Hence it is found that $\tau^{\rm C}
\simeq 4$ corresponds to $z \simeq 3\times 10^5$, above which
distortions are of the $\mu$-type, whereas $\tau^{\rm C}\simeq 0.1$
corresponds to $z \simeq 5\times 10^{4}$ below which they are of the $y$-type.

However this simplistic picture is altered by the processes which do
not conserve the number of photons (double Compton scattering and
Bremsstrahlung). 
These create photons of low energy and force the low
energy part of the spectrum to stick to the Planck distribution at
$T_e$. Hence, these photon creating processes also solve the issue of negative chemical
potentials. Photons in this lower part of the spectrum are subsequently up-scattered by Compton
interactions, that is by the Kompaneets collision term, and this tends to decrease the chemical
potential. Eventually, the spectrum relaxes fully to a Planck
spectrum~\citep[\S 3.4.1]{HuThesis}. For very large redshifts ($z\gtrsim z_{\rm max} \simeq
2\times 10^6$), the distortions of the
$\mu$-type are no-longer visible~\citep{Chluba:2011hw} as they are erased by these photon
creating processes. More details about spectral distortions generated during the radiation
dominated era can be found in
\citet{HuThesis,Chluba2012,Chluba2016}, and numerical solutions are presented in \citet{Chluba:2011hw,Chluba2015Green}.

\subsubsection{Compton cooling}

The temperature of massive particles tends to decay like $1/a^2$ where
$a$ is the scale factor, whenever their temperature is much lower than
their mass, which is the case for $z\ll 10^{10}$. Hence, baryons (electrons and nuclei), tend to cool faster
than photons. From the Kompaneets term (\ref{SZ1}) it is seen that
colder baryons will tend to extract energy from photon, via the
Compton interactions between photons and electrons. Let us evaluate
briefly the energy extracted from the photons by the faster adiabatic
cooling of baryons, using basic thermodynamics.

The pressure of baryons is very well approximated by $P_B = n_B T$. Hence the
adiabatic evolution of baryons kinetic energy when the volume $V$
expands is
\be
\dd U_B = -P_B \dd V - \delta Q = -n_B T_B \dd V - \delta Q\,.
\ee
where $\delta Q$ is the energy brought to the photons by the electrons, which is expected to be negative.
If we use that for baryons (that is massive particles) $U_B \simeq 3/2 n_B V T_B$, and
if we assume that baryons are forced to follow nearly exactly the
photon temperature because of Compton interactions ($\dd T_B/T_B =\dd T_\gamma/T_\gamma=
-1/3 \dd V/V$), then we get 
\be
\delta Q = \frac{1}{2} n_B T_\gamma \dd \ln V\,.
\ee
This results in a total cooling of the photons 
\be
\frac{\delta (a^4\rho_\gamma)}{a^4 \rho_\gamma} = -\frac{1}{2} \left.\frac{n_B
  T_\gamma}{\rho_\gamma}\right|_{\rm today} \int \dd \ln V\,.
\ee
Estimating that  Compton cooling is efficient for $z \gtrsim z_{\rm min}
\simeq 200$, we can estimate the total energy extracted in the CMB spectrum by integrating
from $z_{\rm min}$ up to $z_{\rm max}$. In a Planck spectrum, the average energy of photons is about
$2.70 T_\gamma$, hence $\rho_\gamma \simeq 2.70 n_\gamma T_\gamma$. We
then use that the baryon-to-photon number density ratio is of order
$\eta \simeq 6.1 \times 10^{-10}$~\cite{PitrouBBN}. However this is the ratio of the
number of nucleons to the number of photons. Taking into account that
$24.7\%$ in mass is in the form of Helium nuclei~\cite{PitrouBBN}, and adding also the
contribution of electrons we estimate $n_B/n_\gamma \simeq 1.81 \eta$.
Hence the energy extracted from the CMB is estimated to be
\be
\frac{\delta \rho_\gamma}{\rho_\gamma} \simeq -6.1\times 10^{-10}\ln\left[\frac{1+z_{\rm
  max}}{1+z_{\rm min}}\right] \simeq -5.6\times 10^{-9} \,,
\ee
in very good agreement with the estimation performed in \citet[\S
3]{Chluba2012Cooling}, previously published in \citet{Chluba:2011hw}.
In fact only the contribution above roughly $z=10^5$ has the time to
generate a $\mu$-type distortion, as energy extracted at lower
redshift is mainly in the form of $y$-type distortions. Hence the
energy extracted and which results in a $\mu$-type distortion is
reduced to $\delta \rho_\gamma/\rho_\gamma \simeq -1.8 \times 10^{-9}
$. This translates into
\be
\frac{\mu}{T_\gamma} \simeq -2.5 \times 10^{-9}\,,
\ee
where we used $\mu/T_\gamma \simeq 1.4 \delta \rho_\gamma/\rho_\gamma $ \citep[Eq.~3.48]{HuThesis}. 
It should not be surprising that energy extraction by Compton cooling
results in a negative chemical potential, since it is also accompanied
by a small decrease of the photon temperature on top of the
cosmological redshifting $\propto 1/a$, and this chemical potential
ensures the conservation of photons.

\section{Anisotropic distribution functions}\label{SecAnisotropicCMB}

The Kompaneets equation~(\ref{Kompaneets1}) takes a simple form because of the
high symmetries of an isotropic distribution. However the general form
of the collision term at order $\epsilon^2$ in the Fokker-Planck
expansion is much more involved as we detail in this section, since it
involves the various angular moments of the distribution.

\subsection{Choice of frame}\label{SecChoiceFrame}

Given that we know how the distribution transforms from one frame (a
tetrad) to another frame, we are free to choose the one which is the
most adapted to the description of the photon spectrum. It is even
possible to use the boost operator approach of~\citet{Dai2014} to
relate harmonic space multipoles of different frames at all orders in
the boost velocity. It is customary in cosmology to consider the cosmological frame, which is
defined by the fact that the time-like vector of the tetrad
$\gr{e}_0$, is normal to constant coordinate time hyper-surfaces
\citep{Pitrou2008}. 

However it appears that this choice is not optimal, and it is simpler
and physically more transparent in many situations to work in the
baryons frame, that is the one in which the bulk velocity of electrons vanishes.
Of course this statement is arbitrary and one might prefer to work
with the cosmological frame. Let us list the advantages of the baryon
frame.
\begin{itemize}
\item The collision term remains at most quadratic in the distribution
  function, and even linear if we keep only the Thomson term. This
  would remain true even if we were to perform the Fokker-Planck
  expansion up to an arbitrary large power of $\epsilon$. However, if we were to transform the result to the
  cosmological frame, one would need to add terms with arbitrary high
  powers of the baryons velocity. One would thus needs another criteria to
  cut this expansion. This is discussed in appendix~\ref{SevBulkV}.
\item The variations of the electrons temperature is better analyzed
  in their frame. Since the energy transferred to electrons is deduced
  from the Compton collision term, it is easier to work directly in
  the baryons frame in order to have the energy transfer directly in
  the needed frame.
\item The number density of electrons appears in Compton collisions,
  and it is much more natural to define this density in the frame of
  electrons, in order to avoid extra Lorentz factors.
\item The effect of the baryons velocity cannot be fully removed by a
  change of frame. Once the bulk velocity is removed from the
  collision term, it reappears as a modification of the $\dd \ln E/\dd
  t$ in the Liouville term, that is it changes how free-streaming affects the energy of
  photons. However this can be seen as the effect of two boosts,
  one being at the last scattering event of the photon, and one at its
  reception. This is seen evidently in the line-of-sight
  reformulation \citep{Zaldarriaga:1996xe,Hu:1997hp} of the Boltzmann equation. Furthermore, a boost only aberrates the distribution, that is it changes directions, and shifts all $\ln E$ by the same quantity. Hence, if the spectrum is seen as a function of
  $\ln E$ instead of $E$, the effect of the boost related to the
  change of frame is simple. This idea is at the basis of our decomposition of spectra
  presented in \S~\ref{SecSpectralDistortions}, for which a shift in
  $\ln E$ is not a distortion, but simply a change of temperature. 
\item Eventually, working in the baryon frame allows for a neat
  separation between collisional effects, and special and general
  relativistic effects. What is unusual is not using the baryon frame, but
  instead using a different frame as it obfuscates this separation.
\item All cosmological predictions are given up to a boost due to our
  peculiar velocity with respect to the cosmological frame, so we
  should always use the one which makes computations easier, or
  possibly the one which is more closely related to our peculiar velocity.
\item Working in the baryons frame brings results which are valid up
  to order $\epsilon^3$ even though we performed an expansion up to
  order $\epsilon^2$, because there can be no odd powers. In the
  baryon frame one would need to keep terms which are cubic in the
  baryons velocity. 
\item More generally there is an inflation of terms when
  not working in the baryons frame, because they all appear as the
  effect of a boost on a simpler collision term computed in the baryon
  frame~\citep{Pitrou2008,Chluba2012}.
\item Finally, it has also been argued in \citet{Chluba2012MNRAS} that the optical depth
    \eqref{DefTau} is better defined in the baryon frame, so as to simplify the interpretation of the SZ signal.
\end{itemize}

If we make such an important case for working in the baryon frame, it
is because so far all CMB theoretical computations are formulated with
the photon distributions considered in the cosmological frame. It
appeared natural since numerical integrations are also associated to the
cosmological frame. Describing the photon spectrum in the baryon
frame, while computing its evolution following the cosmological time,
would not seem a natural choice at first. This reformulation would not require much work for linear
perturbation theory, as e.g. \citet{Ma:1995ey,Hu:1997hp,Zaldarriaga:1996xe}.
It would also be useless as our case of using the baryon frame is crucial only to
describe the spectrum correctly, and at first order in cosmological
perturbations no spectral distortions can arise, but only temperature
variations. It is rather easy to see that once expressed with the
line-of-sight method, the modification of the energy evolution $\dd \ln
E / \ln t \to \dd \ln E / \ln t - \dd (v^i n_i)/\dd t $ would
give the same contribution as a collision term expressed in the
cosmological frame, because perturbative effects at the observer's
position are always ignored.

However, using the baryon frame to describe the photon spectrum would
require to revise more substantially the literature (and its
associated numerical codes) on the second-order Boltzmann equation, among which \citet{BenekeFidler,Huang:2012ub,ZhiqiFilippo,Pettinari:2013he,Su:2012gt} but also \citet{Pitrou2008,PUB2010}.
For completeness, we give in appendix~\ref{SevBulkV} the collision term in the
baryon frame, but only up to second order in cosmological perturbations.

\subsection{Thomson term}\label{SecThomson}

The lowest order terms in the Fokker-Planck expansion are those of
order $\epsilon^0$. For an isotropic and unpolarized spectrum, they
vanish. However, as soon as we allow for an angular structure in the
spectrum, they lead to the Thomson contribution to the collision
term. They are compactly written in the form

\be
\frac{\dd t}{\dd \tau}  C^{\rm Tho}_{ij} =-f_{ij}+\myA[f]_{ij}\,.
\ee
The first term is the effect of scattering out events and is thus
proportional to the distribution function itself. The second term
accounts for the scattering in events. Its
general expression is
\bea
\myA[f]_{ij} &=&\frac{{\cal H}_{ij}}{2} \left[I_\emptyset +
\left(\frac{1}{10}I_{ij}-\frac{3}{5}E_{ij}\right) n^{ij}\right]\nonumber\\
&&+\left[-
\frac{1}{10}I_{ij}+\frac{3}{5}E_{ij}\right]^{\cal T}\nonumber\\
&&-\frac{\ii}{2}\epsilon_{ij}\left(\frac{1}{2}
V_k n^k\right)\,,
\eea
where the $\cal T$ indicates that the projected traceless part must be
taken thanks to the projector (\ref{DefTTproj}). We have clearly separated the intensity, linear
and polarization parts in a decomposition similar to
Eq.~(\ref{fmunuphotons}) and we note them $I_\myA[f] $,
$\polar_\myA[f]_{ij}$ and $V_\myA[f] $. With this notation, the various components of the Thomson collision term are
\beas
\frac{\dd t}{\dd \tau} I^{\rm Tho} &=& -I + I_\myA[f]\,, \\
\frac{\dd t}{\dd \tau}  \polar^{\rm Tho}_{ij} & =& -\polar_{ij} + \polar_\myA[f]_{ij}\,,\\
\frac{\dd t}{\dd \tau}  V^{\rm Tho} & =& -V + V_\myA[f]\,.
\eeas
\subsection{Thermal effects}\label{SecThermal}

The thermal effects are of order $\epsilon^2$ and arise from the integration~(\ref{Moment2}).
They also take a compact form with appropriate definitions. We find
\be
\frac{\dd t}{\dd \tau}  C^{T_e/m}_{ij} = \frac{T_e}{m}\left[\frac{1}{E^2}\partial_E^2(E^2 \myB[f]_{ij}) + \myD[f]_{ij}\right],
\ee
where we defined
\bea
\myB[f]_{ij} &=& \frac{{\cal H}_{ij}}{2}\left[I_\emptyset-\frac{2}{5}
  I_i n^i+\left(\frac{1}{10}I_{ij}-\frac{3}{5}E_{ij}
  \right)n^{ij}\right.\nonumber\\
&&\left.\qquad+\left(\frac{1}{7}E_{ijk}-\frac{3}{70}I_{ijk}\right)n^{ijk}\right] 
\nonumber\\
&&+\left[-\frac{1}{10}I_{ij}+\frac{3}{70}I_{ijk}n^k\right.\nonumber\\
&&\left.\quad+
  \frac{3}{5}E_{ij}-\frac{1}{7}E_{ijk}n^k-\frac{2}{5}B_{c(i}\epsilon_{j)}^{\,\,\,\,c}\right]^{\cal
T}\nonumber\\
&&-\frac{\ii}{2}\epsilon_{ij}\left(-\frac{1}{2}V_\emptyset+\frac{1}{2}V_i
  n^i-\frac{2}{5}V_{ij}n^i n^j\right)\,,
\eea

\bea
\myD[f]_{ij} &=& \frac{{\cal H}_{ij}}{2}\left[-2
  I_\emptyset+\frac{2}{5}I_i
  n^i+\left(\frac{18}{5}E_{ij}-\frac{8}{10}I_{ij}\right)n^{ij}\nonumber\right.\\&&\qquad
\left.+\left(\frac{9}{35}I_{ijk}-\frac{6}{7}E_{ijk}\right)n^{ijk}\right]\nonumber\\
&&+\left[\frac{6}{10}I_{ij}-\frac{12}{5}E_{ij}-\frac{9}{35}I_{ijk}n^k+\frac{6}{7}E_{ijk}n^k\right]^{\cal
T}\nonumber\\
&&-\frac{\ii}{2}\epsilon_{ij}\left(V_\emptyset -2 V_i n^i
  +\frac{2}{5}V_{ij}n^i n^j\right)\,.
\eea
Again, we have written these expressions such that their intensity
($I_\myB[f]$ and $I_\myD[f]$), circular polarization ($V_\myB[f]$ and $V_\myD[f]$), and linear polarization ($\polar_\myB[f]_{ij}$ and
$\polar_\myD[f]_{ij}$) components can be read directly. Hence the components of the collision term due to thermal effects are

\beas
\frac{\dd t}{\dd \tau}  I^{T_e/m} &=& \frac{T_e}{m}\left[\partial_E^2(E^2
  I_\myB[f]) + I_\myD[f]\right],\slabel{IThermal}\\
\frac{\dd t}{\dd \tau}  \polar^{T_e/m}_{ij} &=&\frac{T_e}{m}\left[\partial_E^2(E^2 \polar_\myB[f]_{ij})
  + \polar_\myD[f]_{ij}\right],\\
\frac{\dd t}{\dd \tau}  V^{T_e/m} &=& \frac{T_e}{m}\left[\partial_E^2(E^2 V_\myB[f]) + V_\myD[f]\right]\,.
\eeas
Eq. (\ref{IThermal}) matches exactly the thermal terms inside
Eq. (C19) of \citet{Chluba2012} when ignoring linear and circular polarization.

\subsection{Recoil effects}\label{SecRecoil}

The contribution of the electron recoil to the collision term also
takes a very compact form and requires no further definition. It reads
simply as
\bea
\frac{\dd t}{\dd \tau}  C^{E/m}_{ij} &=&  \frac{E}{m}\left[\left(f_{ik}+\frac{{\cal
      H}_{ik}}{2}\right)\frac{1}{E}\partial_E (E^2
\myB[f]_{kj})\right]\nonumber\\
&+&\frac{E}{m}\left[\left(f_{kj}+\frac{{\cal
      H}_{kj}}{2}\right)\frac{1}{E}\partial_E
(E^2\myB[f]_{ik})\right]\nonumber\\
&+&2\frac{E}{m}f_{ij}\,.
\eea
The last term corresponds to a reduction of the scattering out contribution. 
Contrary to the other contributions, there are both linear and quadratic terms in the distribution functions due to the stimulated
emission effects. The components of the recoil term are easily read. The intensity component is
\bea
\frac{\dd t}{\dd \tau}  I^{E/m} &=& \frac{E}{m}(2+E\partial_E) I_\myB[f]+2\frac{E}{m} I\\
&+&2\frac{E}{m}f^{ji} (2+E\partial_E) \myB[f]_{ij}\,,\nonumber
\eea
and it matches the recoil terms inside Eq.~(C19) of \citet{Chluba2012}, when ignoring linear and circular polarization.
The polarization components of the recoil term are
\bea
&&\frac{\dd t}{\dd \tau} \polar^{E/m}_{ij} = \frac{E}{m}(2+E\partial_E)\polar_\myB[f]_{ij} +2\frac{E}{m} \polar_{ij}\\
&&+\frac{E}{m}\left[f_i^{\,\,k}(2+E\partial_E)
  \myB[f]_{kj}+f^k_{\,\,\,j}(2+E\partial_E) \myB[f]_{ik}\right]^{\cal T}\,,\nonumber
\eea\bea
&&\frac{\dd t}{\dd \tau} V^{E/m} = \frac{E}{m}(2+E\partial_E) V_\myB[f]+2\frac{E}{m} V\\
&&+\frac{\ii \epsilon^{ij}E}{m}\left[f_i^{\,\,k}(2+E\partial_E) \myB[f]_{kj}+f^k_{\,\,j}(2+E\partial_E) \myB[f]_{ik}\right]\,.\nonumber
\eea
We did not yet open explicitly the terms which are quadratic in the
distribution function and which were on the second lines of the three
previous equations. These purely quadratic terms are
\bea\label{Irecoilquad}
\frac{\dd t}{\dd \tau}  I^{E/m}_{\rm quad} &=& \frac{E}{m} I
(2+E\partial_E)I_\myB[f]\\
&+&2 \frac{E}{m}\polar^{ij}(2+E\partial_E)\polar_\myB[f]_{ij}\nonumber\\
&+&\frac{E}{m}V(2+E\partial_E)V_\myB[f]\,,\nonumber
\eea
\bea\label{Precoilquad}
\frac{\dd t}{\dd \tau} \polar^{E/m}_{{\rm quad}\,ij} &=& \frac{E}{m}I
  (2+E\partial_E)\polar_\myB[f]_{ij}\\
&+&\frac{E}{m}\polar_{ij}(2+E\partial_E)I_\myB[f]\nonumber\\
&+&2\frac{E}{m} \left[\polar^k_{\,\,(i} (2+  E\partial_E)\polar_\myB[f]_{j)k}\right]^{\cal T}\,,\nonumber
\eea
\be\label{Vrecoilquad}
\frac{\dd t}{\dd \tau}  V^{E/m}_{\rm quad} = \frac{E}{m}\left[I  (2+E\partial_E)V_\myB[f]+V(2+E\partial_E)I_\myB[f]\right].
\ee
We remark that linear polarization does not affect circular polarization in
the recoil term, nor does circular polarization affect linear polarization.

In practice, since the recoil terms are already small, since they are
reduced by $E/m$, they might be linearized around an isotropic background
distribution. In that case only the contributions from the first term
of Eq.~(\ref{Irecoilquad}) or the two first terms of Eq.~(\ref{Precoilquad}) survive,
and the quadratic contributions reduce to
\be\label{Irecoilquad2}
\frac{\dd t}{\dd \tau}  I^{E/m}_{\rm quad} \simeq \frac{E}{m}\left[I_\emptyset (2+E\partial_E)I_\myB[f]+(I-I_\emptyset) (2+E\partial_E)I_\emptyset\right],
\ee
\be\label{Precoilquad2}
\frac{\dd t}{\dd \tau}  \polar^{E/m}_{{\rm quad}\,ij} \simeq \frac{E}{m}\left[I_\emptyset (2+E\partial_E)\polar_\myB[f]_{ij}+\polar_{ij}(2+E\partial_E)I_\emptyset\right],
\ee
\be\label{Vrecoilquad2}
\frac{\dd t}{\dd \tau}  V^{E/m}_{\rm quad} \simeq \frac{E}{m}\left[I_\emptyset  (2+E\partial_E)V_\myB[f]+V(2+E\partial_E)I_\emptyset\right].
\ee
 In \citet[Eq.~6.18]{Pitrou2008}, a linearization of the
effects coming from stimulated emission was used, hence matching only
these linearized quadratic terms.

The thermal contributions of \S~\ref{SecThermal} and the recoil terms of
this section, once added, constitute the {\it generalized
Kompaneets equation}, whose derivation is original in the case of
polarized radiation, and extends Eq.~(C19) of \citet{Chluba2012}.
Combined with the Thomson term of \S~\ref{SecThomson}, it rules the
thermalization of an anisotropic and polarized distribution over an
electron distribution. The total collision term, valid up to
order $\epsilon^2$ is the sum of the Thomson contributions and the
extended Kompaneets equations, that is
\be
C_{ij} =C^{\rm Tho}_{ij} + C^{E/m}_{ij}+C^{T_e/m}_{ij}\,.
\ee
Since we worked in the baryon frame, there are in fact no contribution
at order $\epsilon^3$ so this collision term is only corrected by
order $\epsilon^4$ contributions which have factors $E T_e/m^2$,
$E^2/m^2$ or $T_e^2/m^2$. The decomposition of the angular dependence
in spherical harmonics rather than in STF tensor is easily obtained,
especially if the quadratic terms are linearized. Indeed, in that case
one needs only the relations (\ref{MagicSTFYlm}) and (\ref{Magiceplusns}) to express the result with spherical harmonics.

The modifications of the Thomson contribution when considered in a general frame instead of the
baryon frame are gathered for completeness in appendix~\ref{SevBulkV}, even
though as argued in \S~\ref{SecChoiceFrame} it is preferable to work
in the baryon frame.  When decomposing the result in spherical
harmonics, the procedure is much more involved and one must use
various relations of \S~(\ref{SecProdYlmThorne}), yet another reason
for not working in the baryon frame.

Finally let us comment that if circular polarization is initially
vanishing, it is not generated by Compton collisions since
$V_\myA[f]$, $V_\myB[f]$ and $V_\myD[f]$ depend only on the circular
polarization multipoles, and the quadratic terms in the recoil term~(\ref{Vrecoilquad}) are
linear in circular polarization. It is therefore customary to ignore
circular polarization, unless in contexts where Faraday rotation
sources it from linear polarization thanks to birefringence [see
e.g. \citet{Montero-Camacho:2018vgs,Kamionkowski:2018syl}]. In such a
case, the circular part of the collision term presented in this part
should be used to describe properly its subsequent evolution under Compton scattering.

%%%%%%%%%%%%%%%%%%%%%%%%%%%%%%%%%%%%%
\mypart{1}{Spectral distortions}
%%%%%%%%%%%%%%%%%%%%%%%%%%%%%%%%%%%%%

The angular dependence of the distribution function is expanded in
moments, either with STF tensors or with spherical harmonics. However,
it would be convenient to find an expansion of the dependence in the
photon energies $E$ in suitable moments so as to reduce the number of
degrees of freedom. In the next section we review the proposition of \citet{Stebbins2007,PitrouStebbins} for
such a decomposition, and we argue that when restricting to the Thomson
terms, only the first few spectral moments are necessary to describe
the spectrum. The dynamical evolution of the spectral moments
deduced from the Thomson collision term is detailed in \S~\ref{SecDynStebbins}.

\section{Spectrum parameterization}\label{SecSpectralDistortions}

\subsection{Distribution of Planck spectra}
\subsubsection{Temperature transform} 

Restricting first to unpolarized radiation, the distribution of photons is characterized only by its intensity $I$. It is a function of the position in
space-time, the direction of propagation $\gr{n}$ and the energy $E$ of
radiation, hence it is of the form $I(E,\dots)$, where dots indicate all
the non-spectral dependence which we omit in most cases. In previous literature~\cite{1972JETP...35..643Z,1975ApJ...198..245C,1992ApJ...385..288S,Chluba:2004cn} the starting point for the description of the spectral dependence  is
to consider that $I$ is a superposition of Planck spectra with
different temperatures, given by the distribution $p(T,\dots)$, such that
\be\label{DefTransformsT}
I(E,\dots)=2\int_{0}^\infty \dd T p(T,\dots) \nbb\left(E\over T\right) ,
%=\int_{0}^\infty \dd T p(T) \nbb\left(\frac{E}{T}\right) 
\ee
with $\nbb (x) \equiv1/(\exp(x)-1)$.  If $\int_0^\infty p(T) dT\ne1$ the distribution is said to be ``gray''.
\citet{Stebbins2007} gives a full treatment of grayness and there it is shown that an initially
non-gray  distribution with only Compton-type interactions will remain non-gray.  Henceforth we 
consider only non-gray distributions (see \citet{Ellis:2013cu} for an example of a process inducing grayness). 
One can characterize the shape of the spectrum by the moments of the distribution $p(T)$. One thus defines
\be\label{Powerlawmoments}
\bar T_{(p)}\equiv \left(\int_0^\infty T^p p(T)\, \dd T\right)^{1\over p}
\,.
\ee
Different authors have concentrated on following only specific
moments.  The most commonly used are the Rayleigh-Jeans temperature
(which usually coincides with the electron temperature), 
$\bar T_{\rm RJ} \equiv \bar T_{(1)}$~\cite{Chluba:2004cn};
the number density temperature, 
$\bar T_{\rm n}    \equiv\bar T_{(3)}$~\cite{Pitrouysky,Naruko2013,PitrouyEBsky};
and the bolometric temperature 
$\bar T_{\rm b}    \equiv \bar T_{(4)}$~\cite{PUB2010,Creminelli:2011sq,Huang:2012ub};
giving respectively the low frequency brightness, the number density of photons, and the energy 
density in photons.
Indeed, using Eq. (\ref{DefTransformsT}) we find
\be
{T_{(p)}}^p \propto \int_0^\infty I(E) E^{p-1}\dd E\,,
\ee 
for $p\ge
2$. Note that if the distribution
function has a chemical potential, as in the case of a general Bose-Einstein
distribution, the low energy limit is then a constant ($[\exp(\mu/T) -1]^{-1}$), and it is thus impossible to describe such distribution as a
superposition of Planck spectra like in Eq.~(\ref{DefTransformsT})  whose low
energy limit is $\propto \bar T_{\rm
  RJ}/E$. 

However in \citet{Stebbins2007} an alternative description of
the spectrum based on different moments was  proposed. At the basis of the formalism,
is the use of the variable $\cT\equiv\ln T$ (where a reference unit of
temperature is implicit) whose distribution is $q(\cT)\equiv T p(T)$. The logarithmically averaged temperature 
is then simply defined by
\be
\bar \cT \equiv \langle \cT \rangle
 \equiv \ln \bar T,\,\,\,\,{\rm with}\,\,\,\,\langle f  \rangle \equiv \int_{-\infty}^\infty \dd \cT
f(\cT)q(\cT).
\ee

\subsubsection{Spectral moments} 

The spectral distortions are characterized by the logarithmically averaged moments (LAM) of $q(\cT)$:  the moments about 0, $\{\eta_p \}$; the central moments, $\{u_p\}$; and the  moments about a reference temperature, $\{d_p\}$, {\it i.e}
\be
\eta_p \equiv \langle \cT^p \rangle\,,\quad u_p \equiv
\langle (\cT-\bar \cT)^p \rangle\,,\quad d_p \equiv
\langle (\cT-\cT_0)^p \rangle 
\label{moments}
\ee
where $\cT_0 \equiv \ln T_0$ and $T_0$ is an arbitrary reference
temperature, usually chosen close to the mean. By construction,
$u_1=0$, and since the spectrum is non-gray $\eta_0=d_0=u_0=1$.  

Using $\cT = (\cT- \bar \cT) + \bar\cT$ and  $\cT = (\cT- \cT_0) +\cT_0$, the moments~(\ref{moments}) are related by Leibniz-type 
relations
\begin{align}
&u_p = \mystar{p}{-\bar \cT}{\{\eta_k\}}= \mystar{p}{-d_1}{\{d_k\}},\label{ufromdandeta}\\
&\eta_p = \mystar{p}{\bar \cT}{\{u_k\}}\,,\qquad d_p = \mystar{p}{d_1}{\{u_k\}}\,, \label{dandetafromu}
\end{align}
where 
\be
\mystar{p}{X}{\{Y_k\}}\equiv \sum_{m=0}^p{p\choose m} X^{p-m} Y_m\,.
\ee 
The meaning of the moments is clear as one can reconstruct the spectrum by
\be\label{RebuildI}
I(E) = \sum_{m=0}^\infty \frac{d_m}{m!} D^m \nbb\left(\frac{E}{T_0}\right) =\sum_{m=0}^\infty \frac{u_m}{m!} D^m \nbb\left(\frac{E}{\bar T}\right),
\ee
where 
\be
D^m \nbb(x) \equiv (-1)^m \frac{\dd^m \nbb(x)}{\dd \ln (x)^m}\,.
\ee 
Thus $\{d_m\}$ and $\{u_m\}$ are the coefficients of a generalized Fokker-Planck expansion around  $T_0$ and $\bar T$, respectively.  
The $u_p$ are frame independent, but this is not the case for the other types of 
moments~\cite{Stebbins2007}.  
The  observed spectrum as a function of frequency and direction requires knowledge of the observer frame  because of the 
Doppler effect and associated aberration, so one must also know $\bar \cT$ and thus 
\be\label{d1RelTT0}
d_1=\bar \cT -\cT_0 = \ln \bar T -\ln T_0.
\ee  
This is the first moment and it is directly related to the ``temperature relative
perturbation'' which is $\exp(d_1)-1$ since Eq.~(\ref{d1RelTT0}) is also
\be\label{TfromT0}
\bar T= T_0 {\rm e}^{d_1}\,.
\ee
The second moment gives the Compton $y$ distortion, 
\be\label{Defy}
y \equiv {1\over2}u_2={1\over2}(d_2-d_1^2).
\ee  
These two moments are the ones most relevant for current observations.
 
\subsubsection{Spectral moment of a polarized spectrum} 

Linear polarization will be generated by Compton scattering and the previous formalism can be extended to describe the polarization
spectrum~\cite{Stebbins2007,PitrouStebbins}, as we only need to
consider the tensor-valued distribution function $f_{ij}(E,\dots)$ to describe
both intensity and polarization. In practice, one would only consider
the components in the  two-dimensional sub-space, transverse
to the photon direction and the observer velocity, that is one would
use the $2\times2$ matrix (\ref{MatrixStokes}) noted $f_{ab}$, where
the indices $a,b$ refer to a basis in this subspace. For simplicity we
ignore circular polarization so $f_{ab}$ reduces to a trace ($I$) and
a STF part $f_{\langle ab \rangle}$ for linear polarization.

A tensor-valued distribution of Planck spectra $q^{ab}(\cT)$ is defined by\be\label{DefTransformsPolarization}
f^{ab}(E)=\int_{-\infty}^\infty \dd \cT q^{ab}(\cT)\ \nbb\left(E\,e^{-\cT}\right) ,
\ee
and its matrix-valued moments $\{d^{ab}_p\}, \{\eta^{ab}_p\}, \{u^{ab}_p\}$ can be generalized from 
the $\{d_p\}, \{\eta_p\}, \{u_p\}$, for which a trace and a STF part can be defined. 
The relations (\ref{ufromdandeta},\ref{dandetafromu}) are then straightforwardly extended for linear 
polarization.

From the structure of the Compton collision term, it can be shown~\cite{Stebbins2007} that 
$d_0^{\stf{ab}} =\eta_0^{\stf{ab}} = 0$ if initially so, but $u_1^{\stf{ab}}\neq 0$.  The set of variables 
for the polarized part is thus simply the set of $\{u_p^\stf{ab}\}_{p \ge 1}$ as they are frame 
independent. Compared to the intensity, the main difference is that there is no temperature to be 
defined for polarization, but there is the non-vanishing moment $u_1^\stf{ab}$ which is the dominant 
one.  A common misstatement or misunderstanding consists in treating this moment as a 
temperature perturbation, and to use the definition $\Theta^\stf{ab} \equiv u_1^\stf{ab}$, but strictly 
speaking, it is a pure spectral distortion, and as such frame independent. In 
\citet{Naruko2013}, it is called the {\it ``temperature part''} of the polarization, as opposed 
to the primary spectral distortion 
\be
u_2^\stf{ab}=d_2^\stf{ab} - 2d_1 d_1^\stf{ab}\,.
\ee

\subsubsection{Discussion on the choice of a set of variables} 

It is clear that since the $\{u_p\}_{p\ge 2}$ are frame invariant they
are good candidates to describe the spectral distortions. 
The use of $d_1$ for the temperature perturbation is then natural as it fits into this formalism. However, one might wonder
if this is the only set of variables with such appealing properties. Starting from the moments defined 
in (\ref{Powerlawmoments}), we can relate these to the $\{d_p\}$ and $\{u_p\}$ by
\be\label{Tpowern}
\langle T^p\rangle = (\bar T_{(p)})^p=T_0^p \sum_{m} \frac{p^m d_m}{m!}
= \bar T^p \sum_{m} \frac{p^m u_m}{m!} \,.
\ee
It appears clearly that, for a given $p$, the temperature $\bar T_{(p)}$ can be used to define a 
temperature perturbation and the moments
\be\label{MomentMn}
\Theta_{(p)} \equiv \frac{\bar T_{(p)}}{T_0}-1,  \,\,\, M_{(p),m} \equiv \frac{\langle \left(T-\bar T_{(p)}\right)^m \rangle}{\bar T_{(p)}^{m}}\,.
\ee

To illustrate how these temperature perturbations are related to
$d_1$, let us keep only the moments $m\leq 2$ to express the bolometric temperature perturbation
$\Theta_{\rm b}$ ($p=4$), and the number density temperature
$\Theta_n$ ($p=3$). We find they are related to $(d_1,d_2)$ or $(d_1,u_2)$ by
\beas
\Theta_{\rm b} &\simeq& d_1-\frac{3}{2}d_1^2 +2 d_2 = d_1+ \frac{1}{2} d_1^2 + 2 u_2\,,\\
\Theta_{\rm n} &\simeq&d_1 - d_1^2 + \frac{3}{2} d_2= d_1+ \frac{1}{2} d_1^2 + \frac{3}{2} u_2\slabel{Thetan}\,,
\eeas 
and in particular 
\be\label{Thetanb}
\Theta_{\rm b} \simeq \Theta_{\rm n} + y\,.
\ee  

The $\{M_{(p),m}\}_{m\ge 2}$ would be as good as the $\{u_m\}_{m\ge 2}$ to describe
the spectral distortions, since they are obviously frame invariant as
they involve only an (infinite) sum of products of the $\{u_p\}$.
In the next section, we argue that to decide which set of variables
should be used, one should examine the dynamical evolution, and choose the one which
has the simplest structure, and for which numerical integration is simplified.

\subsection{Spectral moments evolution}\label{SecEvolutionMoments}

\subsubsection{General form of the Boltzmann equation} 

The general form of the Boltzmann equation is (again we omit the
dependence in $(E,\dots)$ for brevity)
\be\label{GenBoltzmann}
L^{ab}[f] \equiv \frac{{\cal D} f^{ab} }{{\cal D} t} + \frac{\dd \ln E}{\dd  t}\frac{\partial f^{ab}}{\partial \ln E} = C^{ab}[f]\,,
\ee
where the convective derivative ${\cal D}/{\cal D}t$ acts on all the dependence except the spectral dependence, and accounts for the
effect of free streaming. 

The collision term can also be described by its moments $\{\eta_p^{C,\,ab}\}, \{u_p^{C,\,ab}\}, \{d_p^{C,\,ab}\}$ which are
related by relations similar to (\ref{ufromdandeta}) and
(\ref{dandetafromu}), that is
\begin{align}
&u^{C,ab}_p = \mystar{p}{-d_1}{\{d^{C,ab}_k\}}\,,\label{ufromdandetaC}\\
&d^{C,ab}_p = \mystar{p}{d_1}{\{u^{C,ab}_k\}}\,.\label{dandetafromuC}
\end{align}
In order to find the evolution of the $\{u_p^{ab}\}$, it proves simpler to first derive from~(\ref{GenBoltzmann}) the evolution
of the $\{d_p^{ab}\}$, and we get
\be\label{Liouvilledab}
\frac{{\cal D} d^{ab}_m}{{\cal D} t} = m d^{ab}_{m-1} \frac{\dd \ln E}{\dd t} + d^{C, \,ab}_m\,.
\ee 
So for the temperature perturbation ($d_1$), the trace of $m=1$ gives
\be\label{Eqd1}
\frac{{\cal D} d_1}{{\cal D} t} =  \frac{\dd \ln E}{\dd t} + d^C_1\,.
\ee
If the spectrum is initially non-gray, and radiation is only subject
to Compton scattering, it remains so and this property translates to
$d_0^{C,\,ab} =0$. In that case, this implies
\be\label{LiouvilledabNonGray}
\frac{{\cal D} d^{\stf{ab}}_1}{{\cal D} t} = d^{C, \,\stf{ab}}_1\,,
\ee 
and the first $u^{C, \,ab}_m$ are related to the first $d^{C, \,ab}_m$ by
\be\label{uC2dC2}
u^{C, \,ab}_1 = d^{C, \,ab}_1\,,\qquad u^{C, \,ab}_2 = d^{C, \,ab}_2 - 2 d_1 d^{C, \,ab}_1\,.
\ee

The moments $\{d_p^{C,ab}\}$ can be read off the collision
term as long as we do not consider recoil terms. Indeed the Thomson
and thermal terms involve only $\partial^n_{\ln E}$, which is exactly
what is used in the expansion (\ref{RebuildI}). In fact if thermal
effects are ignored, and if we work in the baryon frame, the Thomson term is
extremely simple, whereas one must use the general frame expression of
appendix~\ref{SevBulkV} if not working in the baryon frame. What is
crucial is that the moments $\{d_p^{C,ab}\}$  are linear in the variables $\{d^{ab}_p\}$ which describe the radiation spectrum. However, they
still couple non-linearly to the baryons bulk velocity if one insists
in not working in the baryon frame.

From the relations~(\ref{ufromdandetaC}), one infers that
\bea\label{Boltzmannu}
\frac{{\cal D} u^{ab}_p}{{\cal D} t} &=& \sum_{m=1}^p {m \choose p} \left(-d_1\right)^{p-m}\left[d^{C,\,ab}_m - m d^{ab}_{m-1}
d^C_1\right]\nonumber\\
&=&u^{C,ab}_p - p  u_{p-1}^{ab} d_1^C\,.
%&=&u^{C,\,ab}_p
\eea
This system of equation is closed at any order $p$, since the
equation-of-motion for  $u^{ab}_p$ depends only on  $u^{ab}_{p'}$ for
$p'\le p$.  One can truncate this system of equations at any order,
but one must bear in mind that we have neglected recoil terms.

It was crucial in these derivations that $\dd \ln E / \dd t $ does not
depend on $E$ but only on the metric and the direction of propagation
$\gr{n}$. Seen as a function of $\ln E$ instead of $E$, the spectrum
is only shifted by free-streaming but the overall shape remains
unchanged. Since the temperature transform (\ref{DefTransformsT}) depends on $E/T$, that is
on $\ln E - \ln T$, this property of global shifting is transferred to
the distribution $q(\cT)$ of superimposed Planck spectra. Centered
moments are thus very well adapted since only the center ($d_1$) is affected
by a global shift in Eq.~(\ref{Eqd1}), but not the centered moments $u_n^{ab}$
in Eq.~(\ref{Boltzmannu}). The structure is exactly similar for the
  effect of a boost, if we let aside the aberration which affects directions. Indeed it also
  shifts $\ln E$ by a constant quantity, and therefore $d_1$ is
  affected by a boost but not the centered moments $u^{ab}_n$.

\subsubsection{Doppler, SZ effect and $y$-type distortion} 

At first order one needs only the temperature perturbation $d_1$ and
$u_1^\stf{ab}=d_1^\stf{ab}$.   At second order, one adds the spectral distortions $u_2$ and 
$u_2^\stf{ab}$, and this distortion, known in this context as the non-linear kinetic SZ
effect~\cite{Pitrouysky,PitrouyEBsky}, is generated by the
r.h.s. of~(\ref{Boltzmannu}) with $p=2$.

The distortion generated by the thermal SZ effect~\cite{1969ApSS...4..301Z} is also captured by 
$u_2$ and the usual $y$ parameter associated with it is related by the
relation (\ref{Defy}). Note that it is apparent on Eq.~(\ref{SZ1})
when compared to the expansion (\ref{RebuildI}) that the thermal SZ
effect also induces a shift in $d_1$, but we also check from Eq.~(\ref{Thetan}) that it does not affect $\Theta_{\rm n}$ since
Compton collisions conserve the number of photons.

A polarized $y$-type distortion can also be defined~\cite{1980MNRAS.190..413S,Naruko2013,PitrouyEBsky} and is related to the moments by
\be\label{DefY}
Y^{\stf{ab}}\equiv{1\over2}u^{\stf{ab}}_2 ={1\over2}(d_2^\stf{ab}- 2
d_1 d_1^\stf{ab})\,.
\ee

\subsubsection{Structure of the numerics} \label{DefStructureNumerics}

Eq.~(\ref{Boltzmannu}) shows that
\begin{enumerate}
\item spectral distortions are affected only by the collision term, as they remain unaffected by metric
perturbations [see also \citet{Stebbins2007,Pitrouysky,Naruko2013}];
\item metric perturbations, which enter through the redshifting term $ \dd \ln E/\dd t $ affect only 
the evolution of the temperature perturbation $d_1$, and more importantly do not couple
non-linearly with $d_1$ [Eq. (\ref{Eqd1})];
\item the
collision term for the evolution of $u_p^{ab}$ [the r.h.s of~(\ref{Boltzmannu})], contains only terms of 
the form $d_1^{p-k} u_k^{ab}$ with $k\leq p$ (see \citet{Stebbins2007} for more details) 
multiplied by powers of the baryons bulk velocity. Therefore it restricts the non-linearities to products 
of at most $p$ factors of spectral moments, when considering the evolution of the moment of order 
$p$. N.B. for $p=1$ the collision term ($d_1^{C,ab}$) is linear in the moments.
\end{enumerate}
Any other parameterization of the distortion based on the $M_{(p),n}$ defined in (\ref{MomentMn}) 
would conserve property (1). However, property (3) would be lost with the $M_{(p),n}$. The 
loss of this property is, in principle, not a serious problem for the numerical integration, since 
interactions are localized in time by the visibility function. However, this would lead to unnecessary 
complications when going to higher orders of perturbations and thus higher moments. Our first 
argument here is that the simplest is the best.

Our second argument is that property (2) is crucial for the numerical integration since redshifting 
effects are not localized in time. Indeed, by avoiding a non-linear coupling between the temperature 
perturbations and the metric perturbations, the numerical integration is made possible even at the 
non-linear level as it avoids coupling between the angular moments of the temperature 
perturbations with the metric perturbation~\cite{Huang:2012ub}. Finding a form of the 
Boltzmann equation that satisfies this property, was the key to a successful numerical integration at 
second order~\cite{Huang:2012ub,Pettinari:2013he}. With the present formalism, this property 
arises  naturally for the variable $d_1$. Metric perturbations would also affect the geodesic and lead 
to  time-delay and  lensing effects, but these can be treated 
separately~\cite{Hu:2001yq,ZhiqiFilippo}.   There would be 
of course other variables for which property (2) holds. For instance, defining 
\be
\tilde \Theta_{(p)} \equiv \ln (1+\Theta_{(p)})\,,
\ee 
one obtains from~(\ref{Tpowern}) that the  variables
\be
\widetilde{\Theta}_{(p)} = d_1+\frac{1}{p} \ln\Big(1 + \sum_{m\ge 2} \frac{p^m u_m}{m!}\Big)
\ee
obviously satisfy property (2) but not property (3). Up to second order in cosmological 
perturbations (neglecting $\{u_p\}_{p\ge 3}$) the definitions for the most common temperatures are 
related by 
\be
d_1\simeq \widetilde \Theta_{\rm n}-\frac{3}{2} u_2 \simeq  
\widetilde \Theta_{\rm b}-2 u_2\simeq\widetilde \Theta_{\rm RJ}-
\frac{1}{2}u_2\,.
\ee
This motivated the use of $\widetilde{\Theta}_{\rm b}$ instead of $\Theta_{\rm  b}$ in the final
output of \citet{Huang:2012ub}, since property (2) is satisfied for
the former and not for the latter. 

Similarly, for the fractional perturbation to the energy density, one finds up to second order in cosmological perturbations 
\be
\Delta \simeq 4 [d_1 + 2d_1^2 + 2 u_2] \,,
\ee
 and using $\tilde \Delta \equiv \ln(1+\Delta)$, we find 
 
\be
\tilde \Delta \simeq 4 (d_1 + 2 u_2)= 4 \tilde \Theta_{\rm b}\,.
\ee 
Again this motivated the use of $\tilde \Delta$ instead of $\Delta$ in the intermediate numerics of \citet{Huang:2012ub}, so as 
to keep property (2) satisfied. A final example can be made with the fractional
energy density perturbation of linear polarization. One finds 
\be
\Delta^\stf{ab} \simeq 4 [d_1^\stf{ab}(1+4 d_1) + 2 u_2^\stf{ab}],
\ee 
and the non-linear term $d_1^\stf{ab}d_1$ will induce a non-linear coupling of the type
$d_1^\stf{ab} \dd \ln E / \dd t$ in the evolution equation of $\Delta^\stf{ab}$. However, using 
\be
\tilde \Delta^\stf{ab} \equiv \Delta^\stf{ab} (1- 4 d_1)\,,
\ee 
this non-linear coupling disappears~\cite{GuidoChristian} and property (2) is recovered. 
In all these three examples, property (2) can be restored with an
ad-hoc change of variable, but property (3) is not satisfied, due to the term in $u_2$ for the
first two examples, and due to the term $u_2^\stf{ab}$ for the last
one. It implies in particular that the evolution equation for the
lowest order moment in this description, {\it i.e.} their temperature perturbation, has a collision term which is not linear in the moments of radiation.

\subsection{Summary and notation} 

The essential properties described above for the structure of dynamical equations 
are only met with the set of variables made of $d_1$, $\{u_p\}_{p \ge 2}$ and
$\{u_p^\stf{ab}\}_{p\ge 1}$. Furthermore, the moments which characterize the spectral distortions 
are frame independent and thus do not depend on our local
velocity. Only the angular dependence is affected by the choice of
frame due to aberration effects. We strongly recommend that 
these moments should be used to parameterize the CMB spectrum when
recoil effects are neglected. However we shall use names which are
more reminiscent of temperature and for the next section we define
\be\label{DefTheta}
\Theta \equiv d_1\,,\qquad \theta^{\stf{ab}} \equiv d_1^{\stf{ab}}=u_1^{\stf{ab}}\,.
\ee
We must remember that the temperature associated with intensity is
recovered from Eq.~(\ref{TfromT0}), and that $\theta^{\stf{ab}}$ is strictly
speaking not a temperature, but rather the lowest spectral distortion of polarization.
Similarly we also do not work directly with $u_2$ nor $u_2^{\stf{ab}}$
for the spectral distortions, but rather with their halves, the $y$ and $Y^{\stf{ab}}$ variables
defined in Eqs.~(\ref{Defy}) and (\ref{DefY}). We also restore spatial
tetrad indices $i,j\dots$ instead of indices $a,b$ referring two the
screen projected space, that is we use $\theta_{ij}$ and $Y_{ij}$. 
%Say it is transverse projected traceless ?
We now restrict the expansion (\ref{RebuildI}) of the spectrum to these moments, that is to second
order effects only. Hence our spectrum parameterization is
\beas\label{SummarizeExpansion}
I(E) &=& -\Theta \dloge \nbb +  \left(y +
  \frac{1}{2}\Theta^2\right)\dloge^2\nbb,\\
\polar_{ij}(E) &=& -\theta_{ij} \dloge
\nbb+  \left(Y_{ij} +\Theta \theta_{ij}\right)\dloge^2\nbb.
\eeas
where the argument of the Planck spectrum is $\nbb(E/T_0)$ and with the logarithmic derivative
\be\label{Defdloge}
\dloge \equiv \partial_{\ln E}=E\frac{\partial}{\partial E}\,.
\ee
Note that the temperature defined in \citet{Pitrouysky,Naruko2013} is
exactly $\Theta_n$. The expansions (2.24) and (2.26) of
\citet{Naruko2013} appear at first sight different, but using the
relation (\ref{Thetan}) we check that they agree with Eqs.~(\ref{SummarizeExpansion}).

Finally,  the moments of $\Theta$ and $y$ in an expansion of the type (\ref{Ilm}) are noted $\Theta_{L}$ and
$y_L$. As for $\theta^{ij}$ and $Y^{ij}$, we decompose
them as in Eq.~(\ref{DefEBTsagas}), that is in $E$-modes (noted $E^\theta_L$ and $E^Y_L$)
and $B$-modes (noted $B^\theta_L$ and $B^Y_L$).

\section{Angular correlations of spectral distortions}\label{SecDynStebbins}

\subsection{Collision term of spectral moments}

We consider only the effect of the Thomson collision term and we work
in the baryon frame. One should use the collision terms of appendix
(\ref{SevBulkV}) if one wishes to rephrase these results in a general
frame. However we argued that distortions are better analyzed in the
baryon frame. This description with only the Thomson terms applies to the reionization epoch or around
recombination when dealing with the dissipation of baryon acoustic
oscillations, since as a first approximation the effect of the
extended Kompaneets collision term can be ignored on anisotropies~\cite{Chluba2012}.

From the discussion in \S~\ref{SecEvolutionMoments} the evolution of
$\Theta$ and $\theta_{ij}$ is extremely simple. It is
just read from the Thomson term exposed in \S~\ref{SecThomson} where
we use the replacement rules 
\be
I \to \Theta\,,\qquad \polar_{ij} \to \theta_{ij}\,,
\ee
and the same rules for the associated STF multipoles. For completeness
we repeat the result here which is
\bea
&&\frac{\dd t}{\dd \tau}C[\Theta] = - \Theta + \Theta_\emptyset +
\left(\frac{1}{10}\Theta_{ij}-\frac{3}{5}E^\theta_{ij}\right)
n^{\langle ij \rangle}\,,\slabel{CTheta}\nonumber\\
&&\frac{\dd t}{\dd \tau}C[\theta_{ij}] = -
\theta_{ij} +\left(\frac{3}{5}E^\theta_{ij}-\frac{1}{10}\Theta_{ij}\right)^{\cal T} . \slabel{CThetaij}
\eea
There is no term quadratic in the spectrum entering the collision term for $\Theta$ nor
$\theta_{\stf{ij}}$ as already stressed in
\S~\ref{DefStructureNumerics}!  In \citet{Naruko2013} it is found that quadratic terms arise, but it is only because of the use of $\Theta_n$.
Furthermore, red-shifting effects (encoded formally by  $\dd \ln E/\dd
t$) do not affect $\theta_{ij}$, as seen on Eq. (\ref{LiouvilledabNonGray}). With our definitions, the
red-shifting of energy only affects $\Theta$. 

This shows once more the importance of choosing appropriate
variables. Of course, one must not forget that non-linearities arising from metric
perturbations enter the evolution of $\Theta$ through $\dd \ln E /\dd
t$, or from the transport operator ${\cal D}/{\cal D}t$, and
eventually the temperature is obtained by the non-linear relation (\ref{TfromT0}).

We now turn to the collision terms of $y$ and
$Y_{ij}$. These contain quadratic terms, even when
restricting to the Thomson collision contribution in the baryon frame,
and this can be seen from the relation ~(\ref{uC2dC2}).

From results of \S \ref{SecEvolutionMoments}, and using Eq.~(\ref{uC2dC2}) we find
\bea\label{BAOy}
&&\frac{\dd t}{\dd \tau}C[y] =-y + y_\emptyset
+\left(\frac{1}{10}y_{ij}-\frac{3}{5} E^Y_{ij}\right)n^{\langle ij
  \rangle}\\
&&+\frac{1}{2}\left\{\left[(\Theta-\Theta_\emptyset)^2\right]_\emptyset+(\Theta-\Theta_\emptyset)^2\right.\nonumber\\
&&\quad\,\,\left.+\frac{1}{10}\left[(\Theta-\Theta_\emptyset)^2\right]_{ij}n^{\langle
    ij \rangle}-\frac{1}{5}(\Theta-\Theta_\emptyset)
  \Theta_{ij}n^{\stf{ij}}\right\}\nonumber\\
&&+\frac{3}{5}\left\{(\Theta-\Theta_\emptyset)E^\theta_{ij}- E[(\Theta-\Theta_\emptyset)\theta_{kl}]_{ij} \right\}n^{\stf{ij}}\,,\nonumber
\eea
where the notation $[\dots]_\emptyset$ means that we must extract the
monopole when decomposing the angular dependence of the expression in
square brackets in STF tensors. Similarly $[\dots]_{ij}$ indicates
we must extract the quadrupole of the scalar quantity inside the
brackets and $E[\dots]_{ij}$ that we must extract the electric type
quadrupole of the tensorial quantity inside brackets. Physically,
Thomson scattering remaps directions and thus mixes Planck spectra of
different temperatures if the distribution is not isotropic, yielding a y-type distortion at lowest order.

The collision term associated with $Y_{ij}$ is 
\bea\label{BAOYij}
&&\frac{\dd t}{\dd \tau}C[Y_{ij}] = \left\{-Y_{ij}+\frac{3}{5}
E^Y_{ij}-\frac{1}{10}y_{ij}\right.\\\
&&+(\Theta-\Theta_\emptyset)
\left(\theta_{ij}+\frac{1}{10}\Theta_{ij}-\frac{3}{5}E^\theta_{ij}\right)-\frac{1}{20}[(\Theta-\Theta_\emptyset)^2]_{ij}\nonumber\\
&&\left.+\frac{3}{5}E[(\Theta-\Theta_\emptyset)\theta_{kl}]_{ij}+\theta_{ij}\left(\frac{3}{5}E^\theta_{kl}-\frac{1}{10}\Theta_{kl}\right)n^{kl}\right\}^{\cal
T}. \nonumber
\eea
Note that the dissipation of baryon acoustic oscillations originates
when the temperature $\Theta$ ceases to be equal to its monopole
$\Theta_\emptyset$, hence feeding the evolution of $y$ in
Eq. (\ref{BAOy}) but also sourcing the distortion of polarization
through Eq. (\ref{BAOYij}). A tight-coupling expansion~\citep{Pitrou2010TC} of Eqs.~(\ref{CTheta}) and
(\ref{CThetaij}) allows to obtain $\Theta-\Theta_0$ but also the quadrupoles which can be used to estimate the effect~\citep{Chluba2012}.

\subsection{Non-linear kSZ effect during reionization}

\begin{figure}[!htb]
  \center
  \includegraphics[width=\columnwidth]{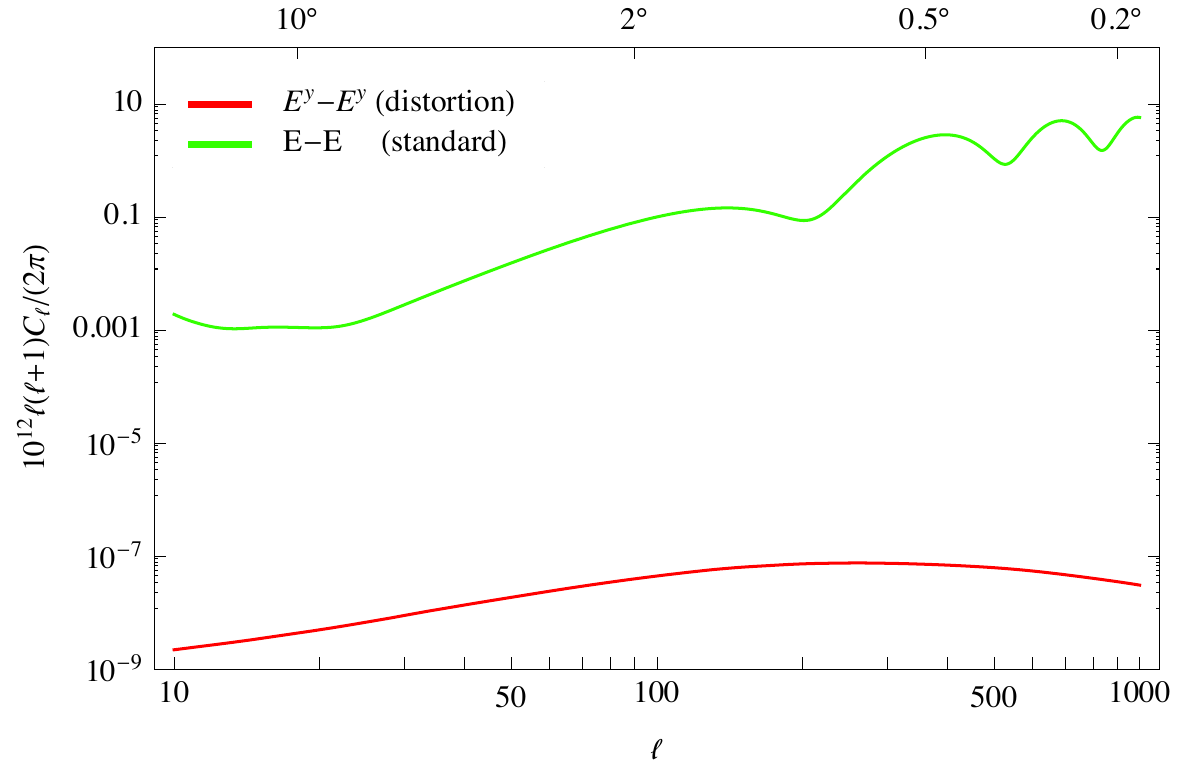}
      \caption{{\it Red}: $E$-modes multipoles of spectral distortions
        $Y_{ij}$. {\it Green}: $E$-mode multipoles of the temperature-like signal
      $\theta_{ij}$.} 
      \label{FigEE}
\end{figure}

\begin{figure}[!htb]
  \center
  \includegraphics[width=\columnwidth]{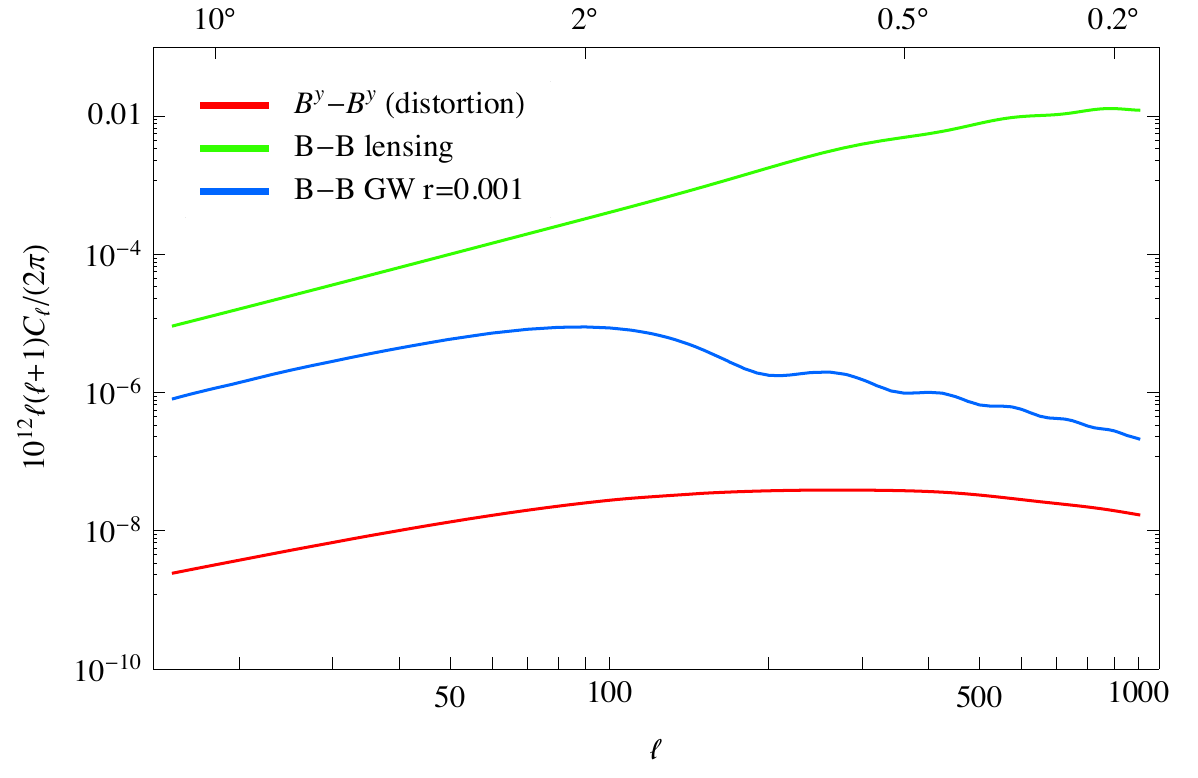}
      \caption{{\it Red}: $B$-modes multipoles of spectral distortions
        $Y_{ij}$. {\it Green}: $B$-modes multipoles of the temperature-like
        signal $\theta_{ij}$ generated by lensing of $E$-modes. {\it
          Blue}: $B$-modes multipoles of the temperature-like
        signal $\theta_{ij}$ from primordial tensor modes with
        tensor-to-scalar ratio $r=0.001$.} 
      \label{FigBB}
\end{figure}

\begin{figure}[!htb]
  \center
  \includegraphics[width=\columnwidth]{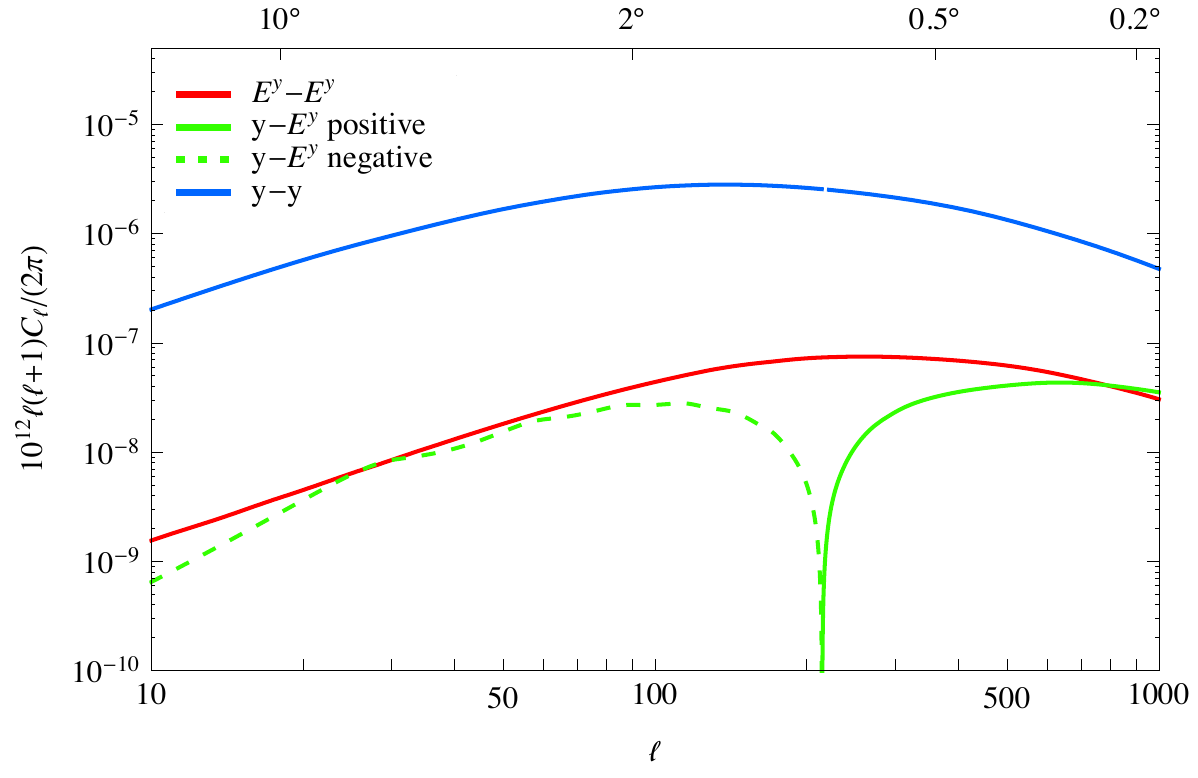}
      \caption{{\it Blue}: $y$-distortion multipoles. {\it Red}: $E$-type multipoles of the distortion signal $Y_{ij}$. {\it Green}: Cross-correlation of both signals.} 
      \label{FigEEyy}
\end{figure}

After recombination, that is below $z\simeq 10^3$, baryons start to decouple from photons and the velocity difference
between them and photons starts to grow. Since we performed
computations in the baryons frame, this velocity difference is hidden
in the dipole of the distribution. Hence we define
\be
 {\cal V}_i\equiv [\Theta-\Theta_\emptyset]_i\,,
\ee
and we assume that it is sufficient to characterize the angular
dependence of temperature ($\Theta \simeq \Theta_\emptyset + {\cal V}_i
n^i$). During the reionization era, ${\cal V}_i$ is growing because
matter collapses whereas radiation free-streams. The collision terms
of the previous section considerably simplify as they reduce to
\bea
\frac{\dd t}{\dd \tau}C[y] &\simeq& -y + y_\emptyset+\left(\frac{1}{10}y_{ij}-\frac{3}{5} E^Y_{ij}\right)n^{\langle ij  \rangle}\\
&&+\frac{1}{3} {\cal V}_i {\cal V}^i+ \frac{11}{20}{\cal V}_{\langle
  i} {\cal V}_{j\rangle}n^{\langle ij \rangle}\,,\nonumber \\
\frac{\dd t}{\dd \tau}C[Y_{ij}] &=& -Y_{ij}+\left[\frac{3}{5}
E^Y_{ij}-\frac{1}{10}y_{ij}-\frac{1}{20}{\cal V}_{\langle i}{\cal V}_{j\rangle}\right]^{\cal T}.
\eea
The velocity difference ${\cal V}_i$ sources the monopole and the
quadrupole of the $y$ spectral distortion via quadratic terms. As for the distortion in
polarization $Y_{ij}$, its quadrupolar electric type multipole is also
sourced by quadratic terms in the velocity. This effect is nothing but
the non-linear kinetic Sunyaev Zel'dovich (kSZ) effect. The angular
correlations of $y$ but also
of the $E$ and $B$ modes of $Y_{ij}$ generated during the
recombination era due to the large scale velocity of baryons in the
intergalactic medium has been computed in
\citet{Pitrouysky,PitrouyEBsky} with the line-of-sight method, and we
reproduce the figures. In Fig.~\ref{FigEE} we plot the $E$-type
multipoles (the $C_\ell$'s) associated
with the temperature-like signal ($\theta_{ij}$) along with those from
the distortion ($Y_{ij}$). Then in Fig.~\ref{FigBB} we also compare
the $B$-type multipoles of $\theta_{ij}$ and $Y_{ij}$, but also those
arising from primordial gravitational waves with tensor-to-scalar
ratio $r=0.001$. Finally $E$-type multipoles of distortions and its
correlations with $y$-type distortions are plotted in
Fig.~\ref{FigEEyy}. 
For the $y$-distortion the effect should be subdominant compared to the thermal $y$-distortions from all
unresolved clusters which has been already detected~\cite{Planck2016_22,ACT_SZ}.
However there is no thermal counterpart for the distortion in polarization $Y_{ij}$.


\section*{Conclusion}

%\part*{Conclusion}
%\addcontentsline{toc}{section}{{\large Conclusion}}

We have emphasized the similarities in the construction of
distribution functions for fermions and bosons. While polarization for
fermions is naturally described by a vector, we need a tensor to
describe the polarization state of photons. In the case of massless
fermions, there are even more similarities since linear polarization is described in a screen-projected
space and circular polarization is defined separately, exactly as for
a gas of photons. We can then anticipate the description of a gas of gravitons when considered as
spin-$2$ particles. Indeed, the stochastic background of gravitational
waves [see e.g. \citet{Cusin2017a,Cusin2017b,Cusin2018PRL}] could be
considered as a gas whose transfer could be deduced from a Liouville equation.
Once covariantized with the polarization tensors $\epsilon_{\pm}^{i}\epsilon_{\pm}^{j}$, its polarization state built as in
Eq.~(\ref{DefPolar}) would be a $4$-index symmetric transverse (to both
$p^\mu$ and $u^\mu$) traceless tensor, whose multipolar decomposition is performed as in Eq.~(\ref{PplusplusYlms}) but with spin-$4$ spherical harmonics~\citep{Cusin:2018avf}.
  
The derivation of the collision terms are also extremely similar
between the Fermi theory of weak interactions and the effective
$4$-vertex description of QED. They only differ in the statistical
factor of the final distributions, which are Pauli-blocking factors for
fermions in weak interactions but stimulated emission factors for
photons in Compton scattering. Apart from this change of sign
in the statistics, the structure of the effect of the final state
distribution [Eq.~(\ref{ColAss})], which cannot be guessed a priori,
is completely similar.

We then emphasized a third similarity in the treatment of weak
interactions exchanging neutrons and protons before BBN, and Compton
scattering. In both cases there is a massive particle in the initial
and final state and the collision term can be computed with
one-dimensional integrals using a Fokker-Planck expansion, that is an
expansion in the momentum transferred to the massive particle. For the
neutron-proton conversions, this allowed to compute the so-called
finite nucleon mass corrections with a method previously introduced in \citet{PitrouBBN}.
For Compton scattering, it allows to obtain the thermal and recoil corrections to the lowest order
approximation known as Thomson scattering. These corrections when
considered for an anisotropic photon distribution are only consistent
when polarization is included (since the quadrupole of the distribution
generates linear polarization) and lead to the {\it extended Kompaneets equation}
presented in \S~\ref{SecAnisotropicCMB}. It extends the results of
\citet[Eq.~C19]{Chluba2012} which were derived in the anisotropic but
unpolarized case. Furthermore, we argued that spectral
distortions should be computed in the baryon frame even though any
frame is in principle suitable since it is always possible to boost
the distribution functions.

Finally, we discussed that by remapping directions, the Thomson
collision term generates spectral distortions through the mixing of Planck
spectra when the distribution is not isotropic. We argued that a parameterization based on logarithmic and
centered moments of the distribution of Planck spectra should be
preferred as it leads to a simple separation of the collision term
into thermal and spectral contributions. Furthermore, apart from
the variable describing temperature fluctuations, all spectral moments are
frame invariant in the sense that only the directions are aberrated by
a Lorentz transformation. With this parameterization, no quadratic term arises in the
collision term governing the evolution of the quantity which
characterizes temperature fluctuations. We summarized the equations
governing the dissipation of baryon acoustic oscillations and the
non-linear kinetic SZ effect, and we stressed that distortions exist also in polarization.

%Perspectives. 

\ifmerci
\begin{acknowledgments}
It is a pleasure to thank Jean-Philippe Uzan, Francis Bernardeau, Thiago Pereira, Pierre
Fleury, Sébastien Renaux-Petel, Guillaume Faye, Giulia Cusin,
Christian Fidler, Atsushi Naruko and Julien Froustey for collaborations and continuous discussions on the topic. 
I thank particularly Guilhem Lavaux and Lucie Gastard for their long
standing encouragements to complete this work.
\end{acknowledgments}
\else\fi

\bibliography{BiblioHDR}

%%%%%%%%%%%%%%%%%%%%%%%%%%%%%%%%%
\mypart{4}{Appendices}
%%%%%%%%%%%%%%%%%%%%%%%%%%%%%%%%%

\appendix

\section{Spinor valued operators}\label{SecSpinorOperators}

A spinor valued operator has $16$ degrees of freedom and we thus need
a $16$-dimensional basis to decompose operators in spinor space. The
set (\ref{SetO}) is a complete basis for the space of operators in spinor space, where
we defined the matrices 
\bea\label{DefSigma}
\gamma^5 &\equiv&\frac{\ii}{4!} \epsilon_{\alpha\beta\mu\nu}\gamma^\alpha
\gamma^\beta \gamma^\mu \gamma^\nu = \ii \gamma^0 \gamma^1 \gamma^2 \gamma^3\,,\\
\Sigma^{\mu\nu} &\equiv& \frac{\ii}{4}[\gamma^\mu,\gamma^\nu],\,\,\widetilde{\Sigma}^{\mu\nu} \equiv
\frac{1}{2}{\epsilon^{\mu\nu}}_{\alpha\beta}\Sigma^{\alpha\beta} =\ii \gamma^5 \Sigma^{\mu\nu}.
\eea

The operators are orthogonal and we find for any two different operators $X$ and $Y$(${X_\mathfrak{a}}^\mathfrak{b}$ and
${Y_\mathfrak{a}}^\mathfrak{b}$ in spinor components) in the set ${\cal
  O}$ that ${\rm  Tr}[X \cdot Y]={X_\mathfrak{a}}^\mathfrak{b} {Y_\mathfrak{b}}^\mathfrak{a}
=0$.
Using this property, any operator can be decomposed onto
this basis with the help of the Fierz identity~\cite{Nishi:2004st} (see also Eq.~G.1.99
of \citet{BibleSpinors} taking into account a factor $2$ difference in the definition of $\Sigma^{\mu\nu}$)
\bea\label{MyFierz}
\delta_{\mathfrak{a}}^\mathfrak{b} \delta_{\mathfrak{c}}^{\mathfrak{d}} &=& \frac 1 4 \left[
  \delta_{\mathfrak{a}}^\mathfrak{d} \delta_{\mathfrak{c}}^{\mathfrak{b}}
  +(\gamma^{5})_{\mathfrak{a}}^{\,\,\mathfrak{d}}
  (\gamma^{5})_{\mathfrak{c}}^{\,\,\mathfrak{b}} \sgnzz
  (\gamma^{\mu})_{\mathfrak{a}}^{\,\,\mathfrak{d}}
  (\gamma_{\mu})_{\mathfrak{c}}^{\,\,\mathfrak{b}}\right. \\
&&\,\,\left.\sgnii (\gamma^\mu
  \gamma^5)_{\mathfrak{a}}^{\,\,\mathfrak{d}} (\gamma_\mu
  \gamma^5)_{\mathfrak{c}}^{\,\,\mathfrak{b}} + 2 (\Sigma^{\mu
    \nu})_{\mathfrak{a}}^{\,\,\mathfrak{d}} (\Sigma_{\mu\nu})_{\mathfrak{c}}^{\,\,\mathfrak{b}}
\right] \nonumber\\
&\equiv& \sum \limits_{X \in {\cal O}} c_X X_{\mathfrak{a}}^{\,\,\mathfrak{d}} X_{\mathfrak{c}}^{\,\,\mathfrak{b}}\,.\nonumber
\eea
The last equality defines the coefficients $c_X$ of the expansion which, by
construction, satisfy $c_X  = 1/{\rm Tr}[X\cdot X]$.
Note also that this identity can be used with the matrices $\widetilde{\Sigma}^{\mu\nu}$
instead of $\Sigma^{\mu\nu}$ by employing $ (\Sigma^{\mu
    \nu})_{\mathfrak{a}}^{\,\,\mathfrak{d}}
  (\Sigma_{\mu\nu})_{\mathfrak{c}}^{\,\,\mathfrak{b}} =-  (\widetilde\Sigma^{\mu
    \nu})_{\mathfrak{a}}^{\,\,\mathfrak{d}} (\widetilde
  \Sigma_{\mu\nu})_{\mathfrak{c}}^{\,\,\mathfrak{b}}$.

Any operator $O$ in spinor space is decomposed on the basis ${\cal O}$ thanks to Eq. (\ref{MyFierz}) as
\be\label{DecompositionFormula}
{O_\mathfrak{a}}^\mathfrak{b} = \sum_{c_X \in {\cal O}} c_X {\rm Tr}[O\cdot X] {X_\mathfrak{a}}^\mathfrak{b}\,.
\ee
In particular, any bilinear tensor product of the form $u_r(p)
\bar{u}_s(p)$ or $v_r(p) \bar{v}_s(p)$ [with the standard notation
$\bar u_s = u_s^\dagger \gamma^0$], is a spinor-space operator and can be decomposed as 
\bea\label{urus}
u_{r,\mathfrak{a}}(p) \bar{u}_{s}^\mathfrak{b}(p) &=& \sum \limits_{X
  \in {\cal O}} c_X
[\bar{u}_{s}(p) X u_r(p) ] {X_\mathfrak{a}}^\mathfrak{b}\nonumber\\
&=& \sum \limits_{X  \in {\cal O}} c_X [\bar{u}_{s}^\mathfrak{c}(p) {X_\mathfrak{c}}^\mathfrak{d} u_{r,\mathfrak{d}}(p) ] {X_\mathfrak{a}}^\mathfrak{b}  \,.
\eea
Hence, we only need to compute the $\bar{u}_{s}(p) X u_r(p) $ and
$\bar{v}_{s}(p) X v_r(p) $ [these expressions are computed in \citet[App.~D]{FidlerPitrou}] for all
$X \in {\cal O}$ to decompose the operator ${F_\mathfrak{a}}^\mathfrak{b}$, defined in Eq. (\ref{DefFparticles}).

\section{Construction of the classical collision term}\label{SecConstructionCollision}
We define the operator $N_{rs}^{(0)}$ characterising the ingoing states prior to the collision. To first order in the interaction we obtain
\be\label{eq:firstorder}
N_{rs}(t) = N_{rs}^{(0)} + i\int \limits_{0}^{t} \dd t' [H_{\rm I}(t'),N_{rs}^{(0)}]\,.
\ee
The interpretation of this equation is that at the time $t=0$ the system is starting to interact, but as the background does not yet contain any correlations between the interacting species we can still evaluate the collisions using the zeroth order number operator. This first order solution describes forward scatterings and we need to go to second order to find the first non-forward interactions.

We insert the first order solution~(\ref{eq:firstorder}) into Eq.~(\ref{dfdtfromdNdt}) and find to second order
\bea
\frac{\dd N_{rs}}{\dd t} &=& \ii [H_{\rm I}(t), N_{rs}(t)] \approx \ii
[H_{\rm I}(t), N_{rs}^{(0)}] \nonumber\\
&&- \int \limits_{0}^{t} \dd t' [H_{\rm I}(t), [H_{\rm I}(t'),N_{rs}^{(0)}]]\,.
\eea
The second order contribution describes an interaction which is active between the time $t'$ and $t$. We identify this timescale with our microscopic timescale $t_{\rm mic} = t-t'$, quantifying the timescale of individual particle interactions. The averaged fluid however does not change significantly on this timescale and evolves on the much larger mesoscopic time-scale $t_{\rm mes} \gg t_{\rm mic}$. Since we compute the derivative of the number operator $N$ with respect to the time $t$ we may identify the mesoscopic time $t_{\rm mes} = t$. 
Expressed in these parameters we obtain
\bea
\frac{\dd N_{rs}}{\dd t_{\rm mes}} &=& \ii [H_{\rm I}(t_{\rm mes}),
N_{rs}^{(0)}] \\
&-& \int \limits_{0}^{t_{\rm mes}} \dd t_{\rm mic} [H_{\rm I}(t_{\rm mes}), [H_{\rm I}(t_{\rm mes} - t_{\rm mic}),N_{rs}^{(0)}]].\nonumber
\eea
Ideally we would like to evaluate this equation at the initial time and set $t_{\rm mes} =0$ to compute the change of our initial states under the considered interactions. 
This choice however is mathematically inconsistent as we are mixing mesoscopic and microscopic timescales in the integration. Instead we average the resulting time-derivative, considering the time-reversal symmetry, over a box that is centered on the initial time and has a length of $2\epsilon$ which is chosen to be small compared to the scale of macroscopic evolution.
\bea
&&\frac{\dd N_{rs}(0)}{\dd t}\Big|_{\rm classical} \equiv\ii [H_{\rm
  I}(0), N_{rs}^{(0)}]\\
&&- \int \limits_{-\epsilon}^{\epsilon} \frac{\dd t_{\rm
    mes}}{2\epsilon}{\rm sgn}(t_{\rm mes})\int \limits_{0}^{t_{\rm
    mes}} \dd t_{\rm mic} [H_{\rm I}(t_{\rm mes}),\nonumber\\
&&\quad \times [H_{\rm I}(t_{\rm mes}-t_{\rm mic}),N_{rs}^{(0)}]]\,.\nonumber
\eea
We may split this integration into three regions. First, the central region $\epsilon \approx t_{\rm Compton}$, where $t_{\rm Compton}$ the typical Compton time scale of particles, is highly non-trivial, but this region is negligible compared to our entire integration volume. In the remaining positive and negative regions the integrand is constant in time. The reason is that the integral over the microscopic time already has sufficient support and is converged. The remaining time-dependence based on the mesoscopic time $t_{\rm mes}$ is not relevant as we have chosen the box small compared to the mesoscopic evolution and we may now set $t_{\rm mes} = 0$ yielding
\bea
&&\frac{\dd N_{rs}(0)}{\dd t}\Big|_{\rm classical} \equiv  \ii [H_{\rm
  I}(0), N_{rs}^{(0)}] \nonumber\\
&&- \frac{1}{2}\int \limits_{-\epsilon}^{\epsilon} \dd t_{\rm mic} [H_{\rm I}(0), [H_{\rm I}(t_{\rm mic}),N_{rs}^{(0)}]] \,.
\eea

Finally we may extend the integration limit $\epsilon$ to infinity compared to the microscopic evolution using a separation of scales.

We note that the interaction Hamiltonian appearing in this equation may always be evaluated based on the non-interacting field value as we only utilise times which are small compared to the mesoscopic time.
Our expression is equivalent to those used in \citet{1993NuPhB.406..423S,Kosowsky:1994cy,BenekeFidler}.

We finally deduce from Eq.~(\ref{dfdtfromdNdt}) that the classical evolution of the distribution function is dictated by
\bea\label{QuantumBoltzmann}
&&\deltarel(0){\frac{\dd f_{rs}(t)}{\dd t}} = \ii \langle \Psi(t) |
[H_I(t), N^{(0)}_{rs}] |\Psi(t)\rangle \\
&&-\frac{1}{2} \langle \Psi(t) | \int_{-\infty}^{\infty} \dd t_{\rm mic} [H_I(t),[H_I(t+t_{\rm mic}), N^{(0)}_{rs}]] |\Psi(t)\rangle,\nonumber
\eea
where the first term on the rhs is the forward scattering term. It is responsible for refractive effects or flavor oscillations in matter (see \citet{Lesgourgues:2006nd,Lesgourgues:2012uu} for neutrino oscillations in cosmology) such as the MSW effect~\cite{1978PhRvD..17.2369W,1986NCimC...9...17M,Marciano:2003eq,1993NuPhB.406..423S,Volpe:2015rla}). The second term is the collision term and we define
\bea\label{eq:defcollAppendix}
&&\deltarel(0) C[f_{rs}(t)]\equiv \\
&&-\frac{1}{2} \langle \Psi(t) | \int_{-\infty}^{\infty} \dd t_{\rm mic} [H_I(t),[H_I(t+t_{\rm mic}), N^{(0)}_{rs}]] |\Psi(t)\rangle\nonumber
\eea
such that the Boltzmann equation (\ref{QuantumBoltzmann}) (when neglecting forward scattering and restoring the notation of the momentum dependence) is written
\be\label{QuantumBoltzmann2Appendix}
\deltarel(0) \frac{\dd f_{ss'}(t,p)}{ \dd t } = \deltarel(0) C[f_{ss'}(t,p)]\,.
\ee

\section{Finite nucleon mass corrections}\label{AppFM}

It is not fully correct to consider that nucleons have an infinite mass. Indeed, the
typical energy transfer in weak interactions to electrons and
neutrinos is of the order of the mass gap $\Gap\simeq 1.29 \,{\rm MeV}$,
which is $1.4 \times 10^{-3}$ smaller than the
nucleon mass. It corresponds to a temperature of $1.5 \times 10^{10}
{\rm K}$ which is not much larger than the freeze-out temperature. In the infinite
nucleon mass approximation, we have thus neglected factors of the type
$E_\nu/m_N$, $E_e/m_N$ or $\Gap/m_N$ (where $m_N$ is the average
nucleon mass $m_N\equiv (m_p+m_n)/2$) which represent order $10^{-3}$
corrections with respect to the leading one around $10^{10}{\rm K}$ and
even larger corrections at higher temperature. Our method consists in expanding
the full reaction rate in power of a small parameter $\epsilon$
related to the momentum transfer. Given the relation between kinetic
energy and momenta, $T/m_N$ is of order $\epsilon^2$. Terms of the type $E_\nu/m_N$ or $E_e/m_N$ are also of
order $\epsilon^2$ and terms of the type $\Gap/m_N$ are also treated as
being of order $\epsilon^2$. Our implementation of the finite mass corrections consists
in including all the terms up to order $\epsilon^2$, but neglecting
terms of higher order. This means that we neglect terms whose
importance is of order $10^{-6}$.

If we ignore radiative corrections at null temperature, these corrections take the form
\beas\label{GammaFMnoRC}
\delta\Gamma^{\rm FM}_{n \to p} &=& \myk \int_0^\infty p^2 \dd p\\
&\times&[\chi^{\rm FM}_+(E,g_A) +\chi^{\rm FM}_+(-E,g_A)]\nonumber\\
\delta\Gamma^{\rm FM}_{p \to n} &=& \myk \int_0^\infty p^2 \dd p \\
&\times&[\chi^{\rm FM}_-(E,-g_A)  +\chi^{\rm FM}_-(-E,-g_A)]\,,\nonumber
\eeas
and the functions $\chi^{\rm FM}_\pm$ are
\begin{widetext}
\bea\label{chiFMFull}
\chi^{\rm FM}_\pm(E,g_A) &=& \tilde c_{LL}\frac{p^2}{m_N E}g^{(2,0)}_\nu(E_\nu^\mp) g(-E) -\tilde
c_{RR}\frac{E_\nu^\mp}{m_N} g^{(2,0)}_\nu(E_\nu^\mp) g(-E) \\
&&+\left(\tilde c_{LL}+\tilde c_{RR}\right)\frac{T}{m_N}\left(g^{(2,1)}_\nu(E_\nu^\mp)
  g(-E)\frac{p^2}{E}-g^{(3,1)}_\nu(E_\nu^\mp)
  g(-E)\right)\nonumber\\
&&+\left(\tilde c_{LL}+\tilde c_{RR}+\tilde
  c_{LR}\right)\left[\frac{T}{2 m_N}\left(g^{(4,2)}_\nu(E_\nu^\mp)
  g(-E)+g^{(2,2)}_\nu(E_\nu^\mp) g(-E) p^2\right) \right.\nonumber\\
&&\qquad\qquad\qquad \qquad\qquad\left.+ \frac{1}{2
  m_N}\left(g^{(4,1)}_\nu(E_\nu^\mp) g(-E)+g^{(2,1)}_\nu(E_\nu^\mp)
  g(-E)p^2\right)\right]\nonumber\\
&&-\left(\tilde c_{LL}+\tilde c_{RR}+\tilde
  c_{LR}\right)\frac{3 T}{2}\left[1-\left(\frac{m_n}{m_p}\right)^{\pm 1}\right] g^{(2,1)}_\nu(E_\nu^\mp)
  g(-E)\nonumber\\
&&+\tilde
  c_{LR}\left[-\frac{3 T}{m_N} g^{(2,0)}_\nu(E_\nu^\mp)  g(-E)+\frac{p^2}{3 m_N
      E} g^{(3,1)}_\nu(E_\nu^\mp) g(-E)+\frac{p^2 T}{3 m_N E}g^{(3,2)}_\nu(E_\nu^\mp) g(-E)  \right]\nonumber
\eea
\end{widetext}
where $p=\sqrt{E^2-m_e^2}$, $E_\nu^\mp=E\mp\Gap$. We defined the reduced couplings
\beas\label{gtildeLLRRLR}
\tilde c_{LL} &\equiv&\frac{4}{1+3 g_A^2} c_{LL}\,,\\
\tilde c_{RR} &\equiv&\frac{4}{1+3 g_A^2} c_{RR}\,,\\
\tilde c_{LR} &\equiv&\frac{4}{1+3 g_A^2} c_{LR}\,,
\eeas
and the functions [with the notation (\ref{DefgnuANDg})]
\be
g_\nu^{(n,p)}(E_\nu) \equiv \frac{\partial^p [(E_\nu)^n
    g_\nu(E_\nu)]}{\partial E_\nu^p}\,.
\ee
However, the finite nucleon mass corrections must be coupled with
radiative corrections, and one must also account for the
weak magnetism in the neutron/proton current as it is also of the same
order as finite nucleon mass corrections. The full set of corrections
is reported in \citet{PitrouBBN}.

\section{Symmetric trace-free (STF) tensors}\label{AppSTF}

\subsection{Notation}

We introduce the multi-index notation
\be\label{ProdnIl}
I_\ell \equiv i_1 \dots i_\ell\,,\qquad n^{I_\ell} \equiv n^{i_1}\dots n^{i_\ell}\,,
\ee
and when no ambiguity can arise we use $L$ instead of $I_\ell$. When
$\ell=0$ we use the notation $I_\emptyset$.

The symmetric trace-free part of a set of indices is noted
$\langle \dots  \rangle $ and it can be used with multi-index
notation, e.g. $n^{\langle I_\ell \rangle }$. General formula for
extracting symmetric and then traceless parts can be found in
e.g. \citet{Thorne1980,BlanchetDamour1986}.

The angular integration of a product of direction vectors is 
\be\label{Angularn}
\int \frac{\dd^2 \gr{n}}{4\pi} n^{I_\ell} = 
\begin{cases} 0&{\rm if}\,\, \ell \,\, {\rm odd}\,,\\ 
\frac{1}{\ell+1} \delta^{(i_1 i_2} \dots \delta^{i_{\ell-1} i_\ell)}&{\rm if}\,\,  \ell \,\,{\rm even} \,.\end{cases}
\ee

\subsection{Relation to spherical harmonics}

Let us define for functions $A(\gr{n})$ and $B(\gr{n})$ 
\be
\left\{ A | B \right\} = \left\{ B | A \right\}^\star \equiv \int \dd^2 \gr{n} A^\star(\gr{n}) B(\gr{n})\,.
\ee
It is possible to obtain the orthogonality relations
\beas
\left\{ n^{\langle I_\ell \rangle } | n_{\langle J_\ell \rangle }
\right\} &=& \Delta_\ell\delta_{j_1}^{\langle i_1} \dots
\delta_{j_\ell}^{i_\ell \rangle }\,,\slabel{OrthoSTF}\\
\left\{ Y_{\ell m} | Y_{\ell
  m'}\right\} &=& \delta_{m m'}\,,\slabel{OrthoYlm}
\eeas
where
\be
\Delta_\ell \equiv \frac{4\pi \ell!}{(2\ell+1)!!}\,.
\ee
Eq.~(\ref{OrthoSTF}) is a particular case of \citet[Eq.~C2]{FBI2014}. Defining
\be
{\cal Y}_{\ell m}^L \equiv \Delta_\ell^{-1}\left\{ n^{\langle L\rangle
} | Y_{\ell m}\right\}\,,
\ee
we can expand the directional dependence either on spherical harmonics
or $n^{\langle L\rangle}$ using
\be\label{YlmfromSTF}
Y_{\ell m}(\gr{n}) = \Delta_\ell^{-1}n_{\langle L\rangle } \left\{ n^{\langle L
  \rangle } | Y_{\ell m}\right\} = n_{\langle L\rangle} {\cal Y}_{\ell m}^L\,.
\ee
The inverse relation is
\bea\label{MagicSTFYlm}
n^{\langle L \rangle} &=& \sum_{m=-\ell}^\ell Y_{\ell
  m}(\gr{n})\left\{ Y_{\ell m} | n^{\langle L \rangle } \right\}\\
&=&\sum_{m=-\ell}^\ell\Delta_\ell Y_{\ell m}(\gr{n}) {\cal
  Y}_{\ell m}^{\star L}\,. \nonumber
\eea
From the closure relation
\be
\sum_{m=-\ell}^\ell Y_{\ell m}(\gr{n}) Y^\star_{\ell m}(\gr{n}) = \frac{2\ell+1}{4\pi}\,,
\ee
we get the closure relation
\be
\sum_{m=-\ell}^\ell {\cal Y}_{\ell m}^{I_\ell} {\cal Y}^{\ell m\,\star}_{J_\ell}=\Delta_\ell^{-1}\delta_{j_1}^{\langle i_1} \dots
\delta_{j_\ell}^{i_\ell \rangle }\,.
\ee
Explicitly the $ {\cal Y}_{\ell m}^L$ are given by
\bea
{\cal Y}_{\ell m}^{I_\ell} &=& C_{\ell m}
\sum_{j=0}^{[(\ell-m)/2]}a_{\ell m j}\left(\delta_1^{i_1}+\ii
  \delta_2^{i_1}\right)\dots \left(\delta_1^{i_m}+\ii
  \delta_2^{i_m}\right)\nonumber\\
&\times&\delta_3^{i_{m+1}}\dots\delta_3^{i_{\ell-2j}}\delta^{{\tiny\ell-2j+1}\,\,
{\tiny \ell-2j+2}}\dots\delta^{i_{\ell-1}\,i_\ell}\,,
\eea
where
\bea
C_{\ell m} &\equiv& (-1)^m \left[\frac{2\ell+1}{4\pi}\frac{(\ell-m)!}{(\ell+m)!}\right]^{1/2}\,,\\
a_{\ell m j} &\equiv& \frac{(-1)^j (2\ell - 2j) !}{2^\ell j! (\ell-j)!(\ell-m-2 j)!}\,.
\eea
Since we use a Cartesian or triad basis we also define
${\cal Y}_{\ell m}^{I_\ell} ={\cal Y}^{\ell m}_{I_\ell} $ and we have
property $ {\cal Y}_{\ell m}^{\star\,I_\ell} = (-1)^m {\cal Y}_{\ell
  \,-m}^{I_\ell} $. The ${\cal Y}_{\ell m}^L$  satisfy the orthogonality property
\be
{\cal Y}_{\ell m}^{I_\ell}{\cal Y}^{\ell m' \star}_{I_\ell}=\Delta_\ell^{-1}\delta_m^{m'}\,.
\ee

\subsection{Relation to spin-weighted spherical harmonics}

The  ${\cal Y}_{\ell m}^L$ are also related to spin-weighted spherical
harmonics. To that purpose, we use the polarization basis 
\be
\gr{e}_\pm \equiv\frac{1}{\sqrt{2}}\left(\gr{e}_\theta\mp \ii \gr{e}_\varphi\right)\,.
\ee
Let us define (for $s>0$) the compact notation \citep{Pitrou:2019ifq}
\be\label{Defns}
n_{\pm s}^{\langle I_\ell \rangle } \equiv e_\pm^{\langle i_1}\dots e_\pm^{i_s} n^{i_{s+1}}
n^{i_\ell \rangle}\,,%\\
%n_{-s}^{\langle I_\ell \rangle } \equiv e_-^{\langle i_1}\dots e_-^{i_s} n^{i_{s+1}}
%n^{i_\ell \rangle}\,,
\ee
which generalizes the products~(\ref{ProdnIl}). For $s>0$ the spin-weighted spherical harmonics are related by
\be\label{YlmsfromSTF}
Y_{\ell m}^{\pm s}(\gr{n})=(\mp 1)^s b_{\ell s}\,{\cal Y}^{\ell m}_{I_\ell}
n_{\mp s}^{\langle I_\ell\rangle }\,,%\\
%Y_{\ell m}^{-s}(\gr{n})&=&b_{\ell s}\,{\cal Y}^{\ell m}_{I_\ell}
%n_{s}^{\langle I_\ell\rangle }\,,
\ee
where
\be
b_{\ell s}\equiv (\sqrt{2})^s
\sqrt{\frac{(\ell!)^2}{(\ell+s)!(\ell-s)!}} \,.
\ee
The relations (\ref{YlmsfromSTF}) are inverted as
\be\label{Magiceplusns}
n_{\mp s}^{\langle I_\ell\rangle }=\frac{(\mp
  1)^s\Delta_\ell}{b_{\ell s}} \sum_{m=-\ell}^\ell Y_{\ell m}^{\pm s}(\gr{n})
{\cal Y}^{\star I_\ell}_{\ell m}\,,%\\
%n_{s}^{\langle I_\ell\rangle }&=&\frac{\Delta_\ell}{b_{\ell s}} \sum_{m=-\ell}^\ell Y_{\ell m}^{-s}(\gr{n})
%{\cal Y}^{\star I_\ell}_{\ell m}\,.
\ee

\subsection{Products and contractions of STF tensor}\label{SecProdYlmThorne}

The symmetrized products of the ${\cal Y}_{\ell m}^L$ are directly
related to the products of spherical harmonics. Indeed, it can be
shown that~\footnote{This formula, though appearing first in
  \citet{Pitrou2008}, has been derived by Guillaume Faye. It is
  obtained by contracting the l.h.s with $n^{I_{\ell_1+\ell_2}}$, and
  using Eq.~(\ref{YlmfromSTF}) to recognize the products of spherical
  harmonics whose expressions in terms of Clebsch-Gordan coefficients
  is known. The formula is then recovered by taking $\partial_{I_{\ell_1+\ell_2}}$ with some algebraic manipulations.}
\bea
&&{\cal Y}^{\ell_1 m_1}_{(i_1 \dots i_{\ell_1}} {\cal Y}^{\ell_2
  m_2}_{i_{\ell_1+1} \dots i_{\ell_1+\ell_2})}  =
\sum_{\ell_3=0\,}^{\ell_1+\ell_2}\sqrt{\frac{(2\ell_1
    +1)(2 \ell_2 +1)}{4 \pi (2 \ell_3 +1)}}\nonumber\\
&& C_{\ell_1 m_1 \ell_2
  m_2}^{\ell_3 m_3}C_{\ell_1 0 \ell_2 0}^{\ell_3 0}  {\cal Y}^{\ell_3 m_3}_{(i_1\dots i_{\ell_3}}\delta_{i_{\ell_3+1}i_{\ell_3+2}}\dots\delta_{i_{\ell_1+\ell_2-1}i_{\ell_1+\ell_2})}\nonumber
\eea
where $m_3=m_1+m_2$ and the sum runs only over even
$\ell_1+\ell_2-\ell_3$. The Clebsch-Gordan coefficients are related to Wigner-3j symbols by
\be
C_{\ell_1 m_1 \ell_2 m_2}^{\ell_3 m_3} =
\troisj{\ell_1}{\ell_2}{\ell_3}{m_1}{m_2}{-m_3} \sqrt{2\ell_3 + 1}(-1)^{m_3-\ell_1+\ell_2}\nonumber
\ee
with $\ell_3 = \ell_1+\ell_2$ and $m_3 = m_1+m_2$.

In particular, we deduce a relation similar to the Gaunt integral of
three  spherical harmonics, which is
\bea\label{Gaunt}
&&\Delta_{\ell_3}{\cal Y}^{\ell_1 m_1}_{I_{\ell_1}} {\cal Y}^{\ell_2
  m_2}_{J_{\ell_2}}{\cal Y}_{\ell_3 m_3}^{I_{\ell_1} J_{\ell_2}} =\troisj{\ell_1}{\ell_2}{\ell_3}{m_1}{m_2}{m_3}\\
&&\times\troisj{\ell_1}{\ell_2}{\ell_3}{0}{0}{0}\sqrt{\frac{(2\ell_1+1)(2
    \ell_2 +1)(2\ell_3+1)}{4\pi}}\,.\nonumber
\eea

As a first application, it allows to obtain the symmetrized and {\it trace-free} product (still for $\ell_3 = \ell_1+\ell_2$)
\bea
&&{\cal Y}^{\ell_1 m_1}_{\langle I_{\ell_1}} {\cal Y}^{\ell_2
  m_2}_{J_{\ell_2}\rangle}= (-1)^{m_3}\troisj{\ell_1}{\ell_2}{\ell_3}{m_1}{m_2}{-m_3}\nonumber\\
&&\troisj{\ell_1}{\ell_2}{\ell_3}{0}{0}{0}\sqrt{\frac{(2\ell_1+1)(2
    \ell_2 +1)(2\ell_3+1)}{4\pi}}{\cal Y}^{\ell_3 m_3}_{I_{\ell_1}J_{\ell_2}}\,.\nonumber
\eea
When considering functions $A(\gr{n})$ and $B(\gr{n})$, which are
expanded in spherical harmonics multipoles $A_{\ell m}$ and $B_{\ell m}$ as in Eq.~(\ref{Ilm}) or
in STF tensor $A_{L}$ and $B_{L}$  as in Eq.~(\ref{STFscalar}), this relation is the key to extract the multipoles (in terms of
spherical harmonics) of a product of the type $A_{\langle I_{\ell_1}}
B_{J_{\ell_2} \rangle }$ in terms of the $A_{\ell m}$ and $B_{\ell
  m}$. It allows to work entirely with STF tensors, and only convert
to spherical harmonics multipoles at the very end if desired.

The second application of Eq. (\ref{Gaunt}) are the contractions
\bea
&& \frac{\Delta_{\ell_3}}{\Delta_{\ell_1}} {\cal Y}^{\ell_2
  m_2}_{J_{\ell_2}}{\cal Y}_{\ell_3 m_3}^{I_{\ell_1}J_{\ell_2}} =(-1)^{m_1}\troisj{\ell_1}{\ell_2}{\ell_3}{-m_1}{m_2}{m_3}\nonumber\\
&&\times\troisj{\ell_1}{\ell_2}{\ell_3}{0}{0}{0}\sqrt{\frac{(2\ell_1+1)(2
    \ell_2 +1)(2\ell_3+1)}{4\pi}}{\cal Y}_{\ell_1 m_1}^{I_{\ell_1}}\,.\nonumber
\eea
Again this allows to get the spherical harmonics multipoles of a product of the type $A_{I_{\ell_1} J_{\ell_2}} B^{J_{\ell_2}}$ in terms of the $A_{\ell m}$ and $B_{\ell
  m}$. 

When the Levi-Civita is involved we must use
\be
\ii \sqrt{\Delta_1} {\cal Y}^{1 n}_c \epsilon^{bc}_{\phantom{cb}\langle i_1} {\cal
  Y}^{\ell\, m-n}_{i_2 \dots i_\ell \rangle b} = {}^n \lambda^{m}_\ell{\cal Y}^{\ell m}_{I_\ell}\,,
\ee
which is deduced from \citep[Eqs.~2.26c-e]{Thorne1980}
\be
\ii \Delta_\ell {\cal Y}_{\ell m}^{\star I_\ell} \epsilon^{bc}_{\phantom{cb}\langle i_1} {\cal
  Y}^{\ell\, m-n}_{i_2 \dots i_\ell \rangle b} = {}^n
\lambda^{m}_\ell \sqrt{\Delta_1} {\cal Y}_{1 n}^{\star c}\,,
\ee
where
\bea
{}^n \lambda^{m}_\ell &\equiv&  (-1)^{m+1}\frac{(\ell+1)(2 \ell+1)}{2} \\
&&\times\troisj{\ell}{1}{\ell}{-m-n}{n}{m}\troisj{\ell}{1}{\ell}{2}{0}{-2}\,.\nonumber
\eea
Their explicit expression is \citep[Eq. 7.33]{Pitrou2008}
\be
{}^0\lambda^m_\ell \equiv \frac{-m}{\ell},\quad {}^n\lambda^m_\ell \equiv \frac{n}{\ell}\sqrt{\frac{(\ell+nm)(\ell+1-nm)}{2}}\,.\nonumber
\ee

Finally let us report the useful relation
\citep[Eq.~A.22a]{BlanchetDamour1986}
\be\label{UsefulBlanchet}
n^{\langle I_\ell \rangle} n^j = n^{\langle I_\ell j \rangle} +
\frac{\ell}{2\ell+1}n^{\langle I_{\ell-1}}\delta^{i_\ell \rangle j} \,.
\ee
We extend it for products of the type (\ref{Defns}) and we
find~\citep[App. C.2]{Pitrou:2019ifq}
\bea\label{UsefulBlanchetPitrou}
n_s^{\langle I_\ell \rangle} n^j &=& n_s^{\langle I_\ell j \rangle} +
\frac{(\ell-s)(\ell+s)}{\ell(2\ell+1)}n_s^{\langle
  I_{\ell-1}}\delta^{i_\ell \rangle j} \nonumber\\
&&+\frac{\ii s}{\ell+1} \epsilon_k^{\,\,\,j \langle i_\ell}
n_s^{I_{\ell-1} \rangle k}\,.
\eea

\subsection{The ${\cal Y}_{\ell m}^L$ in the literature}

The ${\cal Y}_{\ell m}^L$ are often present in the literature even
though not under the present notation nor with the same normalizations. For instance, the $Q_i^{(m)}$
and $Q_{ij}^{(m)}$ of \citet{Hu:1997hp} and \citet{Pitrou2008} or the
$\xi^i_m$ and $\chi_{2m}^{ij}$ of \citet{BenekeFidler} are
proportional to the ${\cal Y}^{1m}_i$ and ${\cal Y}^{2m}_{ij}$ when
evaluated at vanishing radial distance. In
particular, the relations (66) of \citet{BenekeFidler} are particular
cases of Eqs.~(\ref{Magiceplusns}).

\section{Collision term in a different frame}\label{SevBulkV}

In this part we report the collision term when expressed in a general
frame (that is not in the baryon frame) such that baryons possess a
spatial bulk velocity $\gr{v}$. We consider that this bulk velocity is
a factor of order $\epsilon$ in the Fokker-Planck expansion, even
though this is not really the case as it was initially an expansion in
the momentum transferred to the electron. But this choice allows a
first set of simplifications, since we ignore the coupling  of recoil and
thermal terms with the baryon bulk velocity as these would be of
order $T_e/m |\gr{v}|$ or $E/m |\gr{v}|$ and thus of order
$\epsilon^3$. That is restricting to second order in $\epsilon^2$, one
would use the thermal and recoil terms of \S~\ref{SecThermal} and
\ref{SecRecoil}. Hence the modifications introduced by the bulk baryon
velocity arise only from the Thomson term.

Furthermore, as expressions can still be very sizable, we also perform
a secondary expansion in which all multipoles (except the monopole of
intensity $I_\emptyset$) and the baryon bulk velocity, are considered
as first order quantities. In the cosmological context this amounts to
an expansion in cosmological perturbations. We choose to restrict to
second order in this expansion.

The intensity, linear and circular polarization parts of the collision
term arising from bulk baryon velocity in the Thomson term are
summarized in the next sections, in which we use the short-hand
notation (\ref{Defdloge}) and we omit to write the dependence of the multipoles on $E$.

\subsection{Intensity}

The intensity part of the Thomson contribution in a general frame is
\be\label{GeneralIThomsonVelocity}
\frac{\dd t}{\dd \tau}I^{\rm Tho,v}=-I (1- v_i n^i) + \sum_\ell \myZ^I_{I_\ell}
n^{\langle I_\ell \rangle}\,.
\ee
The non-vanishing multipoles in the right hand side are
\bea
 \myZ^I_\emptyset &=& I_\emptyset+\frac{1}{3}(\dloge^2 +3\dloge)I_\emptyset v^2
  +\frac{1}{3}(2+\dloge)I_i v^i\,,\nonumber\\
\myZ^I_i &=&-(1+\dloge)I_\emptyset v_i -\frac{1}{25}(4+\dloge) I_{ij} v^j
 \nonumber\\
&& +\frac{6}{25}(4+\dloge) E_{ij} v^j\,,\nonumber
\eea
\bea
\myZ^I_{ij} &=& \frac{1}{10}I_{ij}-\frac{3}{5}E_{ij}+\left(\frac{11}{20}\dloge^2+\frac{9}{20}\dloge\right)I_\emptyset
  v_{\langle i} v_{j\rangle} \nonumber\\
&&+\frac{1}{10}(\dloge-1)I_{\langle i}
  v_{j\rangle}-\frac{1}{7}(4+\dloge)E_{ijk} v^k\nonumber\\
&&+\frac{3}{70}(4+\dloge)I_{ijk}
  v^k+\frac{2}{5}(1+\dloge) B^l_{\,\,\langle i }\epsilon_{j\rangle  kl}v^k\,,\nonumber\\
\myZ^I_{ijk}&=& \frac{1}{10}(1-\dloge)I_{\langle ij} v_{k\rangle} -\frac{3}{5}(1-\dloge)E_{\langle ij} v_{k\rangle} \,.\nonumber
\eea
The full multipolar decomposition is then obtained by decomposing the
first term in Eq.~(\ref{GeneralIThomsonVelocity}), using
\be\label{MagicInivi}
[I v_i n^i]_{J_\ell } = \frac{\ell+1}{2\ell+3} I_{J_\ell k} v^k
+I_{\langle J_{\ell-1}} v_{j_\ell \rangle}\,,
\ee
where in the left hand side is meant the STF components of $I v_i n^i$ in an
expansion of the type (\ref{STFscalar}). This relation is easily shown
using Eq. (\ref{UsefulBlanchet}). Once Eq.~(\ref{GeneralIThomsonVelocity}) is integrated over $E^3 \dd
E$ so as to get a collision term for the brightness only, we can check that we recover Eq. (6.24) of \citet{Pitrou2008}.

\subsection{Linear polarization}

The linear polarization part of the Thomson contribution in a general frame is
\bea\label{GeneralPijThomsonVelocity}
\frac{\dd t}{\dd \tau}\polar^{\rm Tho,v}_{ij}&=&-\polar_{ij}(1- v_i n^i)\\
&&+\sum_\ell
\left[(\myZ^E_{ijK_\ell}-\epsilon^p_{\,\,(i}
\myZ^B_{j)p K_\ell}) n^{\langle K_\ell \rangle}\right]^{\cal T}\,.\nonumber
\eea
The non vanishing multipoles in the right hand side are
\bea
\myZ^E_{ij} &=& \frac{3}{5}E_{ij}
-\frac{1}{10}I_{ij}-\frac{1}{20}(\dloge^2-\dloge)I_\emptyset v_{\langle
  i}v_{j\rangle}\nonumber\\
&&-\frac{1}{10}(\dloge-1)I_{\langle i} v_{j\rangle
}+\frac{1}{7}(4+\dloge)E_{ijk}v^k\nonumber\\
&&-\frac{3}{70}(4+\dloge)I_{ijk}v^k
-\frac{2}{5}(1+\dloge) B^l_{\,\,\langle i }\epsilon_{j\rangle
  kl}v^k\,,\nonumber\\
\myZ^E_{ijk}&=&(1-\dloge)\left(\frac{3}{5} E_{\langle ij} v_{k\rangle}-\frac{1}{10}I_{\langle ij}v_{k\rangle}\right)\,,\nonumber
\eea
\bea
\myZ^B_{ij} &=& \frac{1}{15}(2+\dloge) I^l_{\,\,\langle i }\epsilon_{j\rangle  kl}v^k-\frac{2}{5}(2+\dloge) E^l_{\,\,\langle i }\epsilon_{j\rangle  kl}v^k\,.\nonumber
\eea
The full multipolar decomposition of the form (\ref{DefEBTsagas}) is
then obtained by decomposing the first term of
Eq. (\ref{GeneralPijThomsonVelocity}) using
\bea
E[\polar_{ij} v_k n^k]_{J_\ell } &=&
\frac{(\ell-1)(\ell+3)}{(2\ell+3)(\ell+1)} E_{J_\ell k} v^k
+E_{\langle J_{\ell-1}} v_{j_\ell \rangle} \nonumber\\
&&- \frac{2}{\ell+1}B^q_{\,\,\langle J_{\ell-1}}\epsilon_{j_\ell\rangle
  pq} v^p\,,
\eea
\bea
B[\polar_{ij} v_k n^k]_{J_\ell } &=&
\frac{(\ell-1)(\ell+3)}{(2\ell+3)(\ell+1)} B_{J_\ell k} v^k
+B_{\langle J_{\ell-1}} v_{j_\ell \rangle} \nonumber\\
&&+ \frac{2}{\ell+1}E^q_{\,\,\langle J_{\ell-1}}\epsilon_{j_\ell\rangle pq} v^p\,,
\eea
where in the left hand sides are meant the STF components of $E$ and
$B$ type of $\polar_{ij} v_k n^k$ in an expansion of the type
(\ref{DefEBTsagas}). The proof of these identities follows from the
use of Eq.~(\ref{UsefulBlanchetPitrou}) with
Eq. (\ref{PolarplusplusEB}). Once Eq.~(\ref{GeneralPijThomsonVelocity}) integrated over $E^3 \dd E$ so as to get a collision term for the brightness only, we can check that we recover Eqs.~(6.25-6.26) of \citet{Pitrou2008}.\\

\subsection{Circular polarization}

The circular polarization part of the Thomson contribution in a general frame is
\be\label{GeneralVThomsonVelocity}
\frac{\dd t}{\dd \tau}V^{\rm Tho,v}=-V (1- v_i n^i) + \sum_\ell \myZ^V_{I_\ell}
n^{\langle I_\ell \rangle}\,.
\ee
The non-vanishing multipoles in the right hand side are
\bea
\myZ^V_\emptyset &=&-\frac{1}{6}(3+\dloge) V_i v^i\,,\nonumber\\
\myZ^V_i &=& \frac{1}{2} V_i+\frac{1}{2}\dloge V_\emptyset v^i+\frac{1}{5}(3+\dloge)V_{ij} v^j\,,\nonumber\\
\myZ^V_{ij}&=& -\frac{1}{2} \dloge V_{\langle i} v_{j\rangle}\nonumber\,,
\eea
and one should use a relation of the form (\ref{MagicInivi}) to obtain
the decomposition of the first term of
Eq.~(\ref{GeneralVThomsonVelocity}) in STF tensors.

\end{document}